\documentclass[submission, Phys]{SciPost}

\usepackage[utf8]{inputenc} 
\usepackage[T1]{fontenc} 	
\usepackage[english]{babel} 

\usepackage[bitstream-charter]{mathdesign}
\usepackage{geometry} 		
\usepackage{amsmath} 		
\usepackage{mathtools} 		
\usepackage{float} 			
\usepackage{graphicx} 		
\usepackage{tabularx} 		
\usepackage{booktabs} 		
\usepackage{color, xcolor} 	
\usepackage{pdfpages} 		
\usepackage{extarrows} 		
\usepackage{multirow} 		
\usepackage{multicol} 		
\usepackage{enumitem} 		
\usepackage{xspace} 		
\usepackage{stackrel} 		
\usepackage{tikz} 			
\usepackage{braket} 		
\usepackage{bm} 			
\usepackage{tensor} 		
\usepackage{slashed} 		
\usepackage{siunitx} 		
\usepackage{lastpage} 		
\usepackage{cite} 			
\usepackage[normalem]{ulem} 
\usepackage{fontawesome} 	
\usepackage{tocloft} 		
\usepackage{titlesec} 		
\usepackage{doi} 			
\usepackage{hyperref} 		
\usepackage[most]{tcolorbox} 					
\usepackage[nameinlink, capitalize]{cleveref} 	
\usepackage[nottoc, notlot, notlof]{tocbibind} 	
\usepackage[ruled, vlined]{algorithm2e} 		
\usepackage{makecell}
\usepackage[makeroom]{cancel}
\usepackage{feynmf}
\usepackage{cancel}

\makeatletter
\def\BState{\State\hskip-\ALG@thistlm}
\makeatother

\makeatletter
\@ifundefined{pdfoutput}{}{\DeclareGraphicsRule{*}{mps}{*}{}}
\makeatother

\makeatletter
\DeclareRobustCommand*{\bfseries}{%
   \not@math@alphabet\bfseries\mathbf
   \fontseries\bfdefault\selectfont
   \boldmath
}
\makeatother

\hypersetup{
	pdftitle={CMSUnfolding},
	pdfauthor={},
	colorlinks=true, 			
	linkcolor={red!50!black}, 	
	citecolor={blue!50!black}, 	
	urlcolor={blue!80!black} 	
} 

\DeclareSymbolFont{usualmathcal}{OMS}{cmsy}{m}{n}
\DeclareSymbolFontAlphabet{\mathcal}{usualmathcal}

\SetArgSty{textnormal}
\SetKwComment{Comment}{{\small\#}~}{}
\SetCommentSty{mycommfont}

\setitemize{itemsep=0pt, parsep=0pt} 				
\setenumerate{itemsep=0pt, parsep=0pt} 				
\setitemize{itemsep=2pt,topsep=2pt,parsep=0pt,partopsep=0pt,leftmargin=*}
\setenumerate{itemsep=0pt,topsep=2pt,parsep=0pt,partopsep=0pt,labelindent=3pt,leftmargin=*}
\usepackage{amsmath}
 
\usepackage{amsthm} 		
\theoremstyle{definition}

\usetikzlibrary{arrows, shapes.geometric, arrows.meta, shapes, decorations.pathreplacing, fit, patterns, patterns.meta}

\definecolor{Rcolor}{HTML}{E99595}
\definecolor{Gcolor}{HTML}{C5E0B4}
\definecolor{Bcolor}{HTML}{9DC3E6}
\definecolor{Ycolor}{HTML}{FFE699}
\definecolor{Ycolor_light}{HTML}{FFF7DE}
\definecolor{Gcolor_light}{HTML}{F1F8ED}

\tikzstyle{expr} = [circle, minimum width=1.8cm, minimum height=1.8cm, text centered, align=center, inner sep=0, draw,font=\LARGE]
\tikzstyle{txt_huge} = [align=center, font=\Huge, scale=2]
\tikzstyle{txt} = [align=center, font=\LARGE, minimum height=1cm]
\tikzstyle{cinn} = [double arrow, double arrow head extend=0cm, double arrow tip angle=130, shape border rotate=90, inner sep=0, align=center, minimum width=2.1cm, minimum height=2.3cm, fill=Bcolor, draw,font=\LARGE]
\tikzstyle{cinn_black} = [cinn, minimum height=2.5cm, fill=black]
\tikzstyle{arrow} = [thick,-{Latex[scale=1.0]}, line width=0.2mm, color=black]
\tikzstyle{loss} = [rectangle, align=center,  minimum width=1.8cm, minimum height=1.5cm,fill=Rcolor,font=\LARGE, rounded corners]
\tikzstyle{xt} = [rectangle, align=center,  minimum width=5cm, minimum height=1.5cm,fill=Gcolor,font=\LARGE, rounded corners]
\tikzstyle{xts} = [rectangle, align=center,  minimum width=1cm, minimum height=1.5cm,fill=Gcolor,font=\Large, rounded corners]

\tikzstyle{embed} = [rectangle, rounded corners=0.3ex, minimum width=1.5cm, minimum height=1cm, text centered, align=center, inner sep=0, fill=Ycolor, font=\large, draw]

\tikzstyle{small_cinn} = [double arrow, double arrow head extend=0cm, double arrow tip angle=130, inner sep=0, align=center, minimum width=1.1cm, minimum height=0.5cm, fill=Bcolor, draw]

\tikzstyle{transformer} = [rectangle, rounded corners, minimum width=6cm, minimum height=2.4cm, font=\large, fill=Gcolor_light, draw]

\tikzstyle{attention} = [rectangle, rounded corners=0.3ex, minimum width=5.5cm, minimum height=1.2cm, align=center, fill=Gcolor, draw, font=\large] 

\definecolor{red_cb}{HTML}{e41a1c}
\definecolor{blue_cb}{HTML}{377eb8}
\definecolor{green_cb}{HTML}{4daf4a}
\definecolor{purple_cb}{HTML}{984ea3}
\definecolor{orange_cb}{HTML}{ff7f00}

\definecolor{EmeraldGreen}{HTML}{1ea78d}
\definecolor{EnglishRed}{HTML}{b02427}
\hypersetup{colorlinks=true,urlcolor=EmeraldGreen,citecolor=EmeraldGreen,linkcolor=EnglishRed}

\newcommand{\xg}{x_\text{gen}}
\newcommand{\xr}{x_\text{reco}}

\newcommand{\pl}{p_\text{latent}}
\newcommand{\pd}{p_\text{data}}
\newcommand{\pmd}{p_\text{model}}
\newcommand{\psim}{p_\text{sim}}
\newcommand{\punf}{p_\text{unfold}}

\newcommand{\qqquad}{\qquad\quad}

\newcommand\one{\leavevmode\hbox{\small1\normalsize\kern-.33em1}}
\newcommand{\normal}{\mathcal{N}} 			

\newcommand{\really}{\stackrel{!}{=}}

\newcommand{\loss}{\mathcal{L}} 	

\newcommand{\arXiv}[2][]{%
	\ifthenelse{\equal{#1}{}}%
	{\href{http://arxiv.org/abs/#2}{arXiv:#2}}%
	{\href{http://arxiv.org/abs/#2}{arXiv:#2~[#1]}}}

\def\slashchar#1{\setbox0=\hbox{$#1$}           
   \dimen0=\wd0                                 
   \setbox1=\hbox{/} \dimen1=\wd1               
   \ifdim\dimen0>\dimen1                        
      \rlap{\hbox to \dimen0{\hfil/\hfil}}      
      #1                                        
   \else                                        
      \rlap{\hbox to \dimen1{\hfil$#1$\hfil}}   
      /                                         
   \fi}

\newcommand{\tikznode}[2]{%
\ifmmode%
\tikz[remember picture,baseline=(#1.base),inner sep=0pt] \node (#1) {$#2$};%
\else
\tikz[remember picture,baseline=(#1.base),inner sep=0pt] \node (#1) {#2};%
\fi}

\def\mathswitchr#1{\relax\ifmmode{\text{#1}}\else$\text{#1}$\xspace\fi}
\def\mathswitch#1{\relax\ifmmode#1\else$#1$\xspace\fi}

\graphicspath{{./figs/}}

\begin{document}

\vspace*{-2.5em}
\hfill{\small DESY-25-012 }
\vspace*{0em}

\begin{center}{\Large \textbf{
How to Unfold Top Decays
}}\end{center}

\begin{center}
  Luigi Favaro\textsuperscript{1,2}, 
  Roman Kogler\textsuperscript{3},
  Alexander Paasch\textsuperscript{4}, \\
  Sofia Palacios Schweitzer\textsuperscript{1},
  Tilman Plehn\textsuperscript{1,5}
  and Dennis Schwarz\textsuperscript{6}
\end{center}

\begin{center}
{\bf 1} Institut f\"ur Theoretische Physik, Universit\"at Heidelberg, Germany\\
{\bf 2} CP3, Universit\'e catholique de Louvain, Louvain-la-Neuve, Belgium\\
{\bf 3} Deutsches Elektronen-Synchrotron DESY, Germany \\
{\bf 4} Institut f\"ur Experimentalphysik, Universit\"at Hamburg, Germany\\
{\bf 5} Interdisciplinary Center for Scientific Computing (IWR), Universit\"at Heidelberg, Germany
{\bf 6} Institute for High Energy Physics, Austrian Academy of Sciences, Austria \\
\end{center}

\begin{center}
\today
\end{center}


\section*{Abstract}
{\bf Using unfolded top-quark decay data we can measure the top quark mass, as well as search for unexpected kinematic
  effects. We present a new generative unfolding method for 
  the two tasks and show how they both benefit from unbinned,
  high-dimensional unfolding. Unlike weight-based or iterative generative methods we include a targeted unbiasing 
  with respect to the training data. This shows significant
  advantages over standard, iterative methods, in terms of applicability, flexibility and
  accuracy.}

\vspace{10pt}
\noindent\rule{\textwidth}{1pt}
\tableofcontents\thispagestyle{fancy}
\noindent\rule{\textwidth}{1pt}
\vspace{10pt}

\clearpage
\section{Introduction}
\label{sec:intro}

Particle physics studies the fundamental properties of particles and
their interactions, with the goal to discover physics beyond the
Standard Model. The methodology is defined by the interplay
between precision predictions and precision measurements. A key
challenge is that perturbative quantum field theory makes predictions
for partons, while experiments observe particles through
their detector signatures.  First-principle simulations link these two
regimes~\cite{Campbell:2022qmc}.  They start with predictions for the
hard process from a Lagrangian, and then add parton decays, QCD
radiation, hadronization, and the detector response, to eventually
compare with experimental data. This forward-simulation inference is
the basis of, essentially, all LHC analyses.

The first problem with forward inference is that it requires access to
the data and the entire simulation chain; neither of them
are available outside the experimental
collaborations. Second, it is not guaranteed that the best
theory predictions are implemented in the forward simulation
chain. Finally, in view of the high-luminosity LHC, hypothesis-driven
forward analyses will overwhelm our computing resources for 
precision theory predictions and detector simulations.
All three problems motivate alternative analysis
techniques.

An exciting alternative analysis method is based on inverse
simulations or unfolding.  Instead of simulating detector effects
for each predicted event, we can correct the observed events, for example,
for detector effects. Then, we perform inference on particles before
the detector or even partons and their hard scattering.
Because the forward simulations are based on quantum physics and are 
stochastic,  unfolding poses an incomplete inverse problem
on a statistical basis. Still, in this way
\begin{enumerate}
\item analyses can be done outside the experimental collaborations;
\item theory predictions can be updated and improved easily;
\item and BSM hypotheses can be tested without full simulations.
\end{enumerate}

Machine learning (ML) methods are revolutionizing not only our daily
lives, but also LHC physics~\cite{Butter:2022rso}. While classical
unfolding methods are severely limited in many ways, ML-unfolding
allows us to unfold unbinned events in many
dimensions~\cite{Huetsch:2024quz}.  A reweighting-based ML-based
unfolding method is MultiFold or OmniFold~\cite{Andreassen:2019cjw}, applied to H1~\cite{H1:2021wkz, H1:2023fzk, H1:2024mox},
LHCb~\cite{LHCb:2022rky} and, recently, ATLAS~\cite{ATLAS:2024xxl}
data.  
Generative ML-unfolding either maps
distributions~\cite{Datta:2018mwd,Howard:2021pos,Diefenbacher:2023wec,Butter:2023ira, Butter:2024vbx}
or learns the underlying conditional
probabilities~\cite{Bellagente:2019uyp,Bellagente:2020piv,Vandegar:2020yvw,Backes:2022vmn,Leigh:2022lpn,Ackerschott:2023nax,Shmakov:2023kjj,Shmakov:2024gkd}. Which
of these complementary methods one would want to use depends on the specific
task.  Learning conditional probabilities to invert the forward
simulation chain gives us access to per-event probabilities
smoothly over phase space~\cite{Butter:2020qhk}, guaranteeing the
correct event migration.  Its success rests on sufficiently precise
generative
networks~\cite{Butter:2019cae,Butter:2021csz,Butter:2023fov,Das:2023ktd},
which are developed and benchmarked also for fast forward
simulations~\cite{Danziger:2021eeg,Janssen:2023lgz,Heimel:2022wyj,Heimel:2023ngj, Heimel:2024wph}.
In this paper we present a novel direction in ML-unfolding:
\begin{itemize}
\item we target an especially challenging task, mass
measurement and unfolding of strongly peaked kinematics. Here, established methods, weight-based as well iterative generative unfolding, fail;
\item we show the first unfolding results related to a CMS analysis~\cite{CMS:2019fak, CMS:2022kqg}. While this paper shows fast simulation results only, even more promising results for full CMS simulations can be obtained from the CMS members on our team.
\end{itemize}
This analysis also marks the first application of generative unfolding to properly simulated data by an LHC experiment.
In Sec.~\ref{sec:ana} we describe
the goal of the analysis, show the results from the classic CMS analysis, 
introduce the dataset, and sketch the basic features and the implementation of 
generative unfolding. In Sec.~\ref{sec:gen} we see how the top
mass appears in the unfolded dataset. We find that a major problem 
is the uncontrolled bias induced by the training data. It
can be solved as described in Sec.~\ref{sec:gen_bias}. Next,
we show in Sec.~\ref{sec:gen_mass} how the top mass can be measured
from the unfolded distributions, and in Sec.~\ref{sec:gen_full} we show how to 
then unfold the
entire top decay phase space for re-analysis. 

In App.~\ref{app:bias} we illustrate how iterative 
bias removal methods do not work for peaked phase space
distributions.
The goal of this paper is to 
show that decay kinematics can be unfolded and to provide a
blueprint for an LHC analysis using generative unfolding.

\section{Goal and method}
\label{sec:ana}

If we want to unfold top-quark decay events, the main challenge is the 
model dependence and resulting bias when the top masses assumed for the 
simulated training data and the actual top mass differ. We  could 
attempt this with iterative improvements of the unfolding network~\cite{Backes:2022sph}, but we will see that
this approach is numerically extremely challenging.
We follow a slightly different strategy:
\begin{enumerate}
\item we ensure that the bias from the 
top mass assumed in the simulated training data is small;
\item we infer the correct top mass from the data, using 
a reduced unfolded phase space;
\item we produce training data with the inferred top mass and
unfold the full phase space.
\end{enumerate}

\subsection{Top mass measurement}
\label{sec:ana_mass}

The extraction of the top mass from the invariant jet mass of 
highly boosted hadronic top quark decays can shed light on possible ambiguities in 
top mass measurements using simulated parton showers. 
The ultimate goal is to compare the measured jet mass distribution to predictions 
from analytic calculations.
For that, it is convenient to unfold detector effects.

Unfolding uses simulated data, biasing the unfolded data towards the model 
used in the simulation. 
In particular, the choice of the top mass in the simulation leads to a
significant uncertainty~\cite{CMS:2022kqg}.
These modelling biases can be reduced by including more information and 
granularity into the unfolding process, motivating the use of 
ML-unfolding methods.

In the existing CMS measurement this is done by also unfolding differentially in 
the top-jet transverse momentum and by including various sideband regions 
close to the measurement phase space.
Using ML-unfolding, the data can be unfolded in a larger number of phase
space dimensions, providing ways to reduce the model bias.

\begin{figure}[t]
    \centering
    \includegraphics[width=0.45\textwidth]{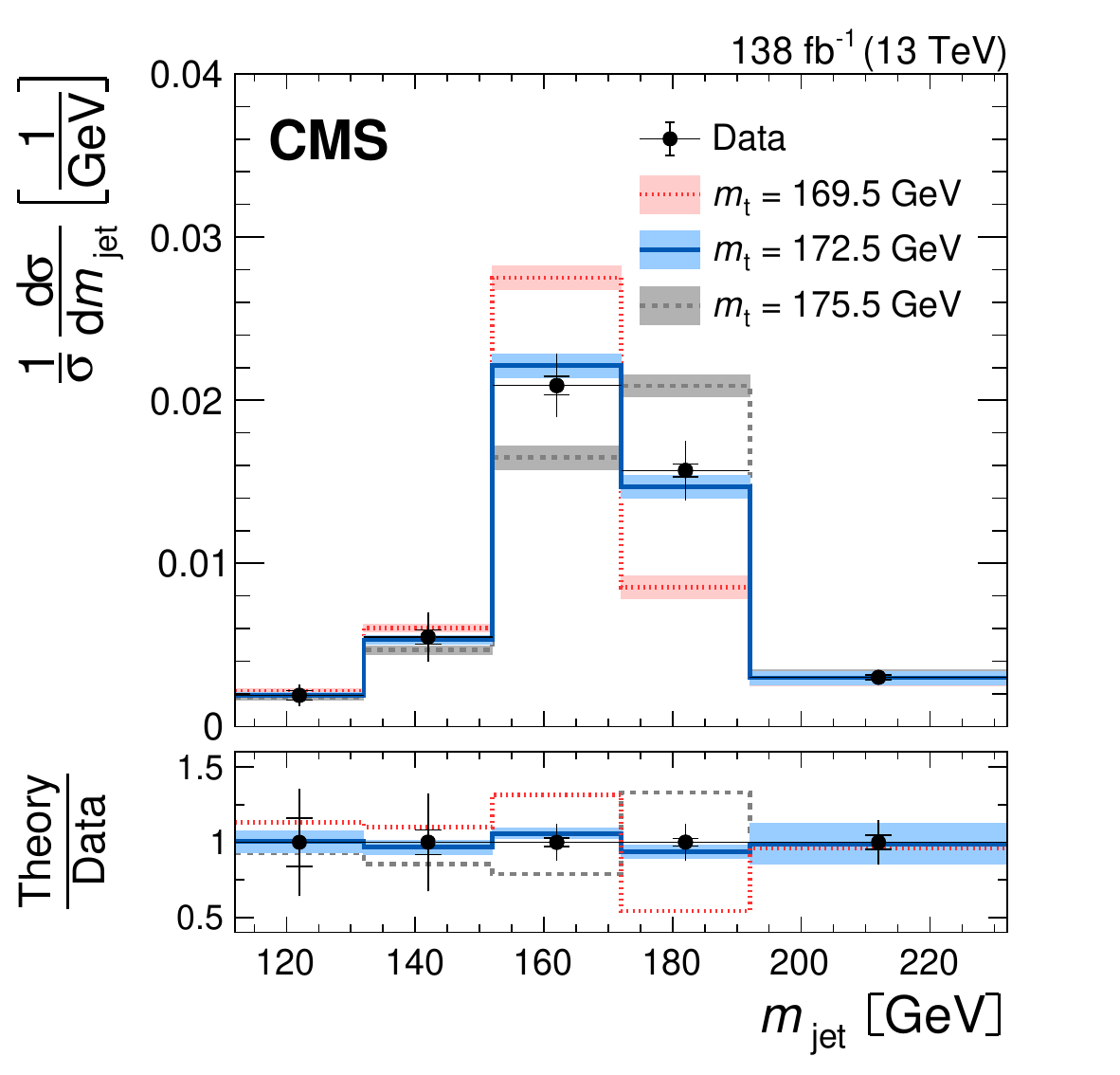}
    \caption{CMS benchmark result from Ref.~\cite{CMS:2022kqg}.
    It shows the differential top pair cross section as a function of the 
     top-jet invariant mass, compared to theory predictions for different
     top masses. The vertical bars represent the total uncertainties,  statistical uncertainties are shown as short horizontal bars, and 
     theoretical uncertainties as shaded bands.}
    \label{fig:classic}
\end{figure}

The result from our CMS benchmark analysis~\cite{CMS:2022kqg} 
is shown in Fig.~\ref{fig:classic}. This analysis unfolds 
the reconstructed 3-subjet mass $M_{jjj}$ and the corresponding reconstructed 
transverse momentum, $p_{T,jjj}$ to measure the top mass. 
The three subjets are obtained using a two-step clustering 
with the eXclusive Cone (XCone) algorithm~\cite{Stewart:2015waa}. 
In the first step, the event is clustered into two large-radius jets 
with a distance parameter $R=1.2$ to capture the decay products of 
the top quark and antiquark. In a second step, the two 
large-radius jets from the first step are each reclustered into three XCone subjets 
with $R=0.4$, where the subjets capture the dynamics of the 
hadronic top quark decay. 
Before the unfolding, the jet mass scale is calibrated
by reconstructing the $W$-boson from the two light-quark subjets 
and fitting the subjet energy scales to the resulting $W$ peak. 
The $W$ boson decay is identified with the help of $b$-tagging 
information, which is obtained for the XCone subjets by matching 
these in angular distance to small-$R$ anti-$k_T$ jets. 
This matching is needed because the $b$-tagging information 
is not calculated for XCone subjets in CMS.
The uncertainty in the unfolding from the modeling of 
final state radiation is reduced with the help of 
another auxiliary measurement of $N$-subjettiness 
ratios~\cite{Thaler:2010tr} on large-$R$ anti-$k_T$ jets, 
matched by angular closeness to the large-$R$ XCone jets. 
The matching procedures and auxiliary measurements add considerable 
complications to the measurement and come with non-negligible 
uncertainties. Because of the finite efficiency of the $b$ tagging 
and the associated mis-identification rate, the information from 
the $W$ reconstruction cannot be used in the unfolding because it 
breaks the permutation invariance among the jets. 
The leading systematic uncertainties in this measurement 
originate from the jet energy scale, jet mass scale, jet 
mass resolution, the $b$-jet response and the unfolding bias 
from the choice of the top mass in the simulation. Non-negligible 
uncertainties also arise from the modeling of non-perturbative 
effects. 
Ideally, unfolding enough phase space dimensions to capture the 
$W$ decay and the salient features of the jet substructure 
should allow us to constrain the dominating uncertainties 
in-situ and remove the top-mass bias in the unfolding.

Once we have measured the jet mass in an event sample 
and consequently the top mass, we can further analyze
the unfolded dataset. For instance, we can look for effects from 
higher-dimensional SMEFT operators on the decay of boosted tops, or we 
can search for anomalous 
kinematic distributions from new particles, modified interactions,
or enhanced QCD effects at the subjet level. While the unfolding 
for the top mass measurement has to include a sufficiently large number of 
dimensions, as discussed above, we now need to unfold the full, 12-dimensional
phase space. Three of these dimensions are 
finite jet masses, generated by QCD effects.

\subsection{Dataset }
\label{sec:ana_data}

We use simulated events for top pair production, similar to the
one used for a CMS measurement~\cite{CMS:2022kqg}.
We generate the events with Madgraph~5~\cite{Alwall:2014hca}. Hadronization, parton showers, and
multiple parton interactions are simulated with Pythia~8.230~\cite{Sjostrand:2014zea} with the underlying event tune CP5~\cite{CMS:2019csb}. The samples include a simulation of the detector response implemented in Delphes~3.5.0~\cite{deFavereau:2013fsa} using the default CMS card with pile-up, and the e-flow algorithm.
The pile-up subtraction only removes charged tracks associated to pile-up vertices. This simulation is a Delphes version of the CMS simulation for Ref.~\cite{CMS:2022kqg}.

In the simulated data, we have access to three stages of the simulation chain. 
The parton level includes the hard interactions of the top
quarks, that decay into a $b$-quark and a $W$-boson, that subsequently
decays into two quarks or lepton and neutrino.  The particle level
refers to all stable particles with lifetimes longer than
$10^{-8}$~s after parton shower and hadronization. Finally, the detector level
describes particle candidates after the detector simulation.
At this point, we limit ourselves to events which appear at all three stages,
our results show that the treatment of efficiency effects is sub-leading 
and beyond the scope of this study.

Event selections are applied at the particle and detector level. All
events that do not pass either of the selections are rejected from 
further analysis. For the signal or measurement region,
we only consider $t\bar{t}$ pairs in the lepton+jets decay at the parton level,
\begin{align}
    pp 
    \to t\bar{t} 
    \to (b q \bar{q}') \; (\bar{b} \ell^- \bar{\nu}) + \text{c.c.}
    \quad \text{with} \quad \ell = e,\mu \; ,
\end{align}
with the lepton acceptance
\begin{align}
 p_{T,\ell} > 60\;\text{GeV}
 \quad \text{and} \quad 
 |\eta_\ell|<2.4 \; .
\end{align}
The top jet
is constructed using XCone clustering and 
identified by the larger angular distance to the lepton. It must
fulfill
\begin{align}
 p_{T,J} > 400\,\unit{GeV}
 \qquad \text{and} \qquad 
 p_{T,j_{1,2,3}} > 30\,\unit{GeV} 
 \qquad 
 |\eta_{j_{1,2,3}}| < 2.5 \; ,
\end{align}
for the large-$R$ jet $J$ and three subjets $j_i$. In the following, we will 
refer to these subjets as jets. The second large-$R$ jet has to have
$p_{T,J} > 10\,\unit{GeV}$ to reject poorly reconstructed events 
where only the lepton and not the $b$ quark is reconstructed 
in the second large-$R$ jet.
To reduce the contribution from events where the full top quark
decay is not reconstructed within the top jet, 
we require the invariant
mass of the three jets, $M_{jjj}$, to exceed the invariant mass of
the lepton and the large-$R$ jet close to it.

At the detector level, in addition to the above requirements, the 
missing transverse momentum has to be larger than $50\,\unit{GeV}$ and at least one $b$-tagged jet must be present. 

The measurement-region selection criteria leave us with approximately 800,000 events simulated with a top mass of $m_t = 172.5\,\unit{GeV}$,
of which we use 75\% for the training.  
To be consistent with the amount of events available in CMS with 
the full detector simulations for the reference analysis, 
we choose samples with different top masses to 
have less events. All events contain the full 
generator (gen) and reconstruction (reco) level information. 
The XCone algorithm clusters the jets separately for 
reco-level jets and gen-level jets.
The clustered jets are sorted according to $p_T$. 

We only consider paired events in our signal, i.e. events that passed both reco- and gen-level cuts. Non-paired events can be treated as background if they are selected at the reco-level but are not part of the measurement’s fiducial phase space at the gen-level. On the other hand, events that were generated in the fiducial phase space at gen-level but were not reconstructed because of the detector’s acceptance or an inefficiency will need to be accounted for by an efficiency correction. 
This can be done through weights, as for example done 
in the Iterative Bayesian unfolding
method~\cite{1974AJ.....79..745L,Richardson:72,Lucy:1974yx,DAgostini:1994fjx} as implemented in RooUnfold~\cite{Adye:2011gm} and in 
TUnfold~\cite{Schmitt:2012kp}, and successfully applied in several jet substructure analyses at the LHC, see for example Refs.~\cite{ATLAS:2019kwg, ATLAS:2024dua, CMS:2022kqg, CMS:2023lpp}.
Another way to include efficiency and acceptance effects is through a classifier~\cite{Heimel:2023mvw}, but we leave the details of such a study to future work as these are closely related to the actual implementation of the data analysis. 

The CMS analysis~\cite{CMS:2022kqg} 
shows that continuum backgrounds, like $W$+jets production, can be subtracted bin-wise to the level where
they are no longer relevant for in the analysis. 
The normalization uncertainties in the different backgrounds introduce a shape uncertainty when changing the normalization of single processes. While the background normalizations vary between 20--100\% in the CMS analysis, the overall background uncertainty was estimated to be only 0.01 GeV in the extraction of the top quark mass and is thus negligible compared to other uncertainties. 
The method of bin-wise background subtraction can be generalized to the unbinned case with the help of a classifier~\cite{Andreassen:2021zzk}, which suggests that background uncertainties will remain small compared to other systematic uncertainties in this measurement. Therefore, we neglect these in our study and consider signal events only. 

\begin{figure}[t]
    \includegraphics[width=0.495\textwidth, page=16]{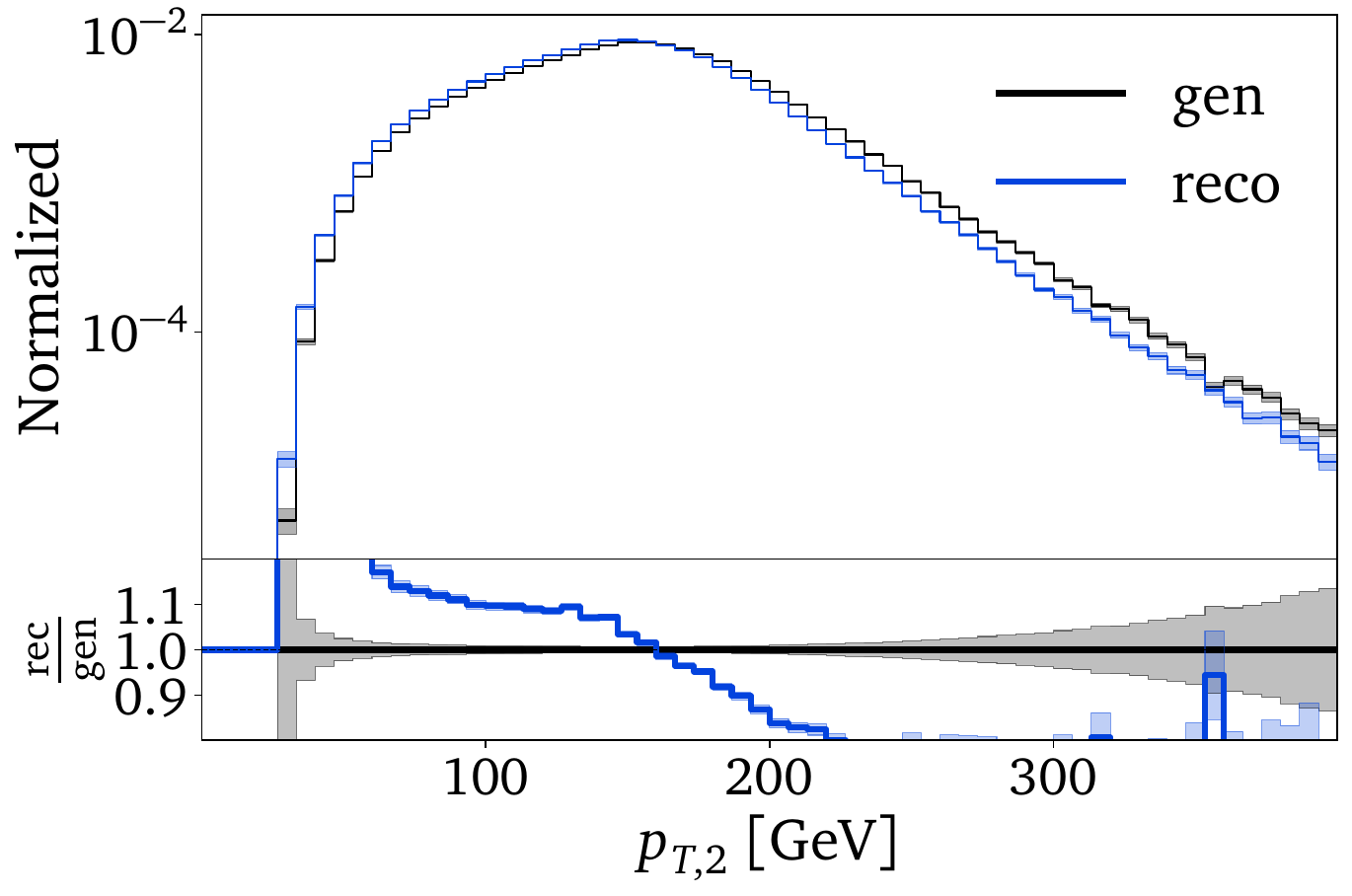} 
    \includegraphics[width=0.495\textwidth, page=19]{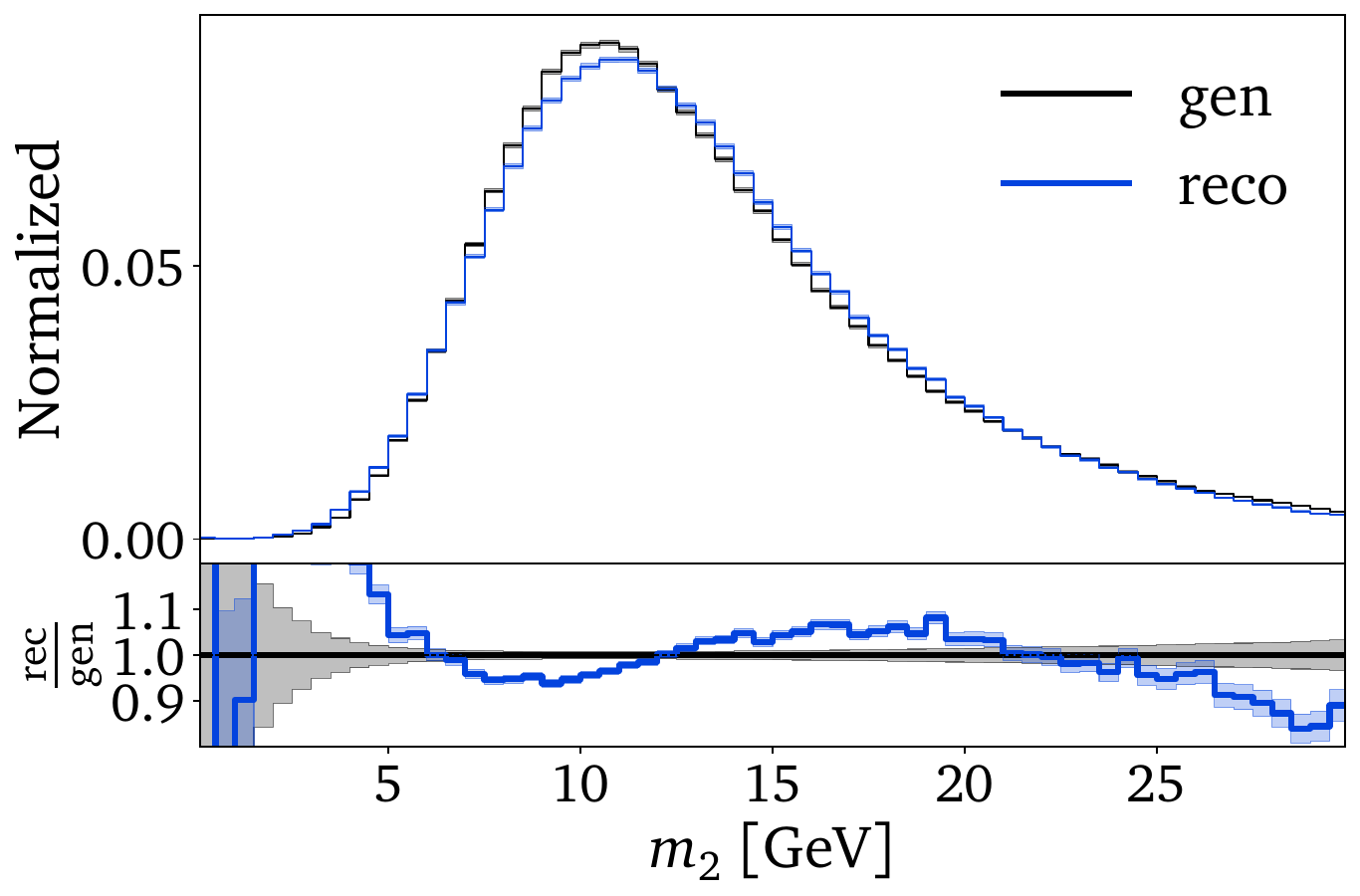} \\
    \includegraphics[width=0.495\textwidth, page=24]{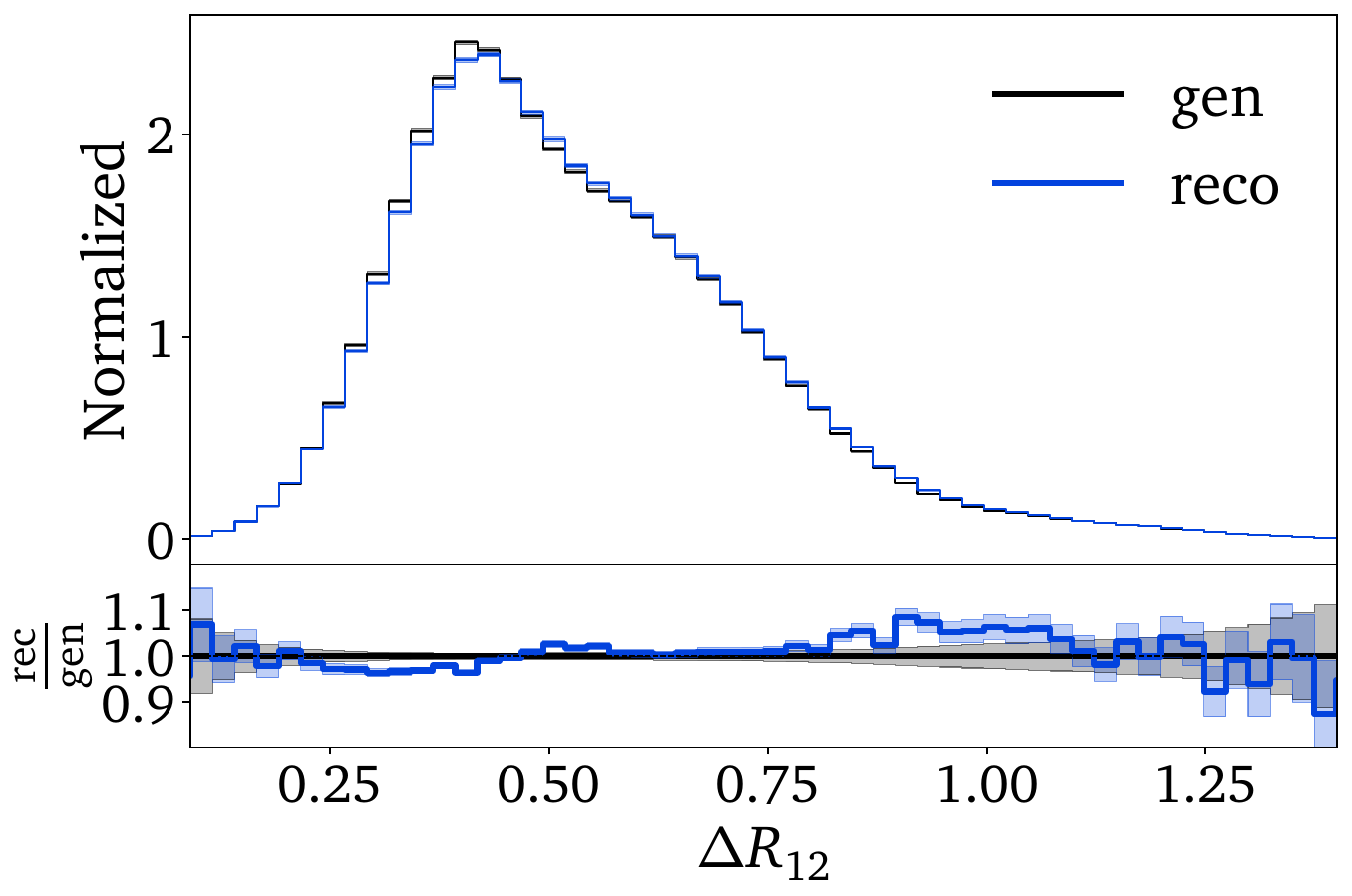}
    \includegraphics[width=0.495\textwidth, page=28]{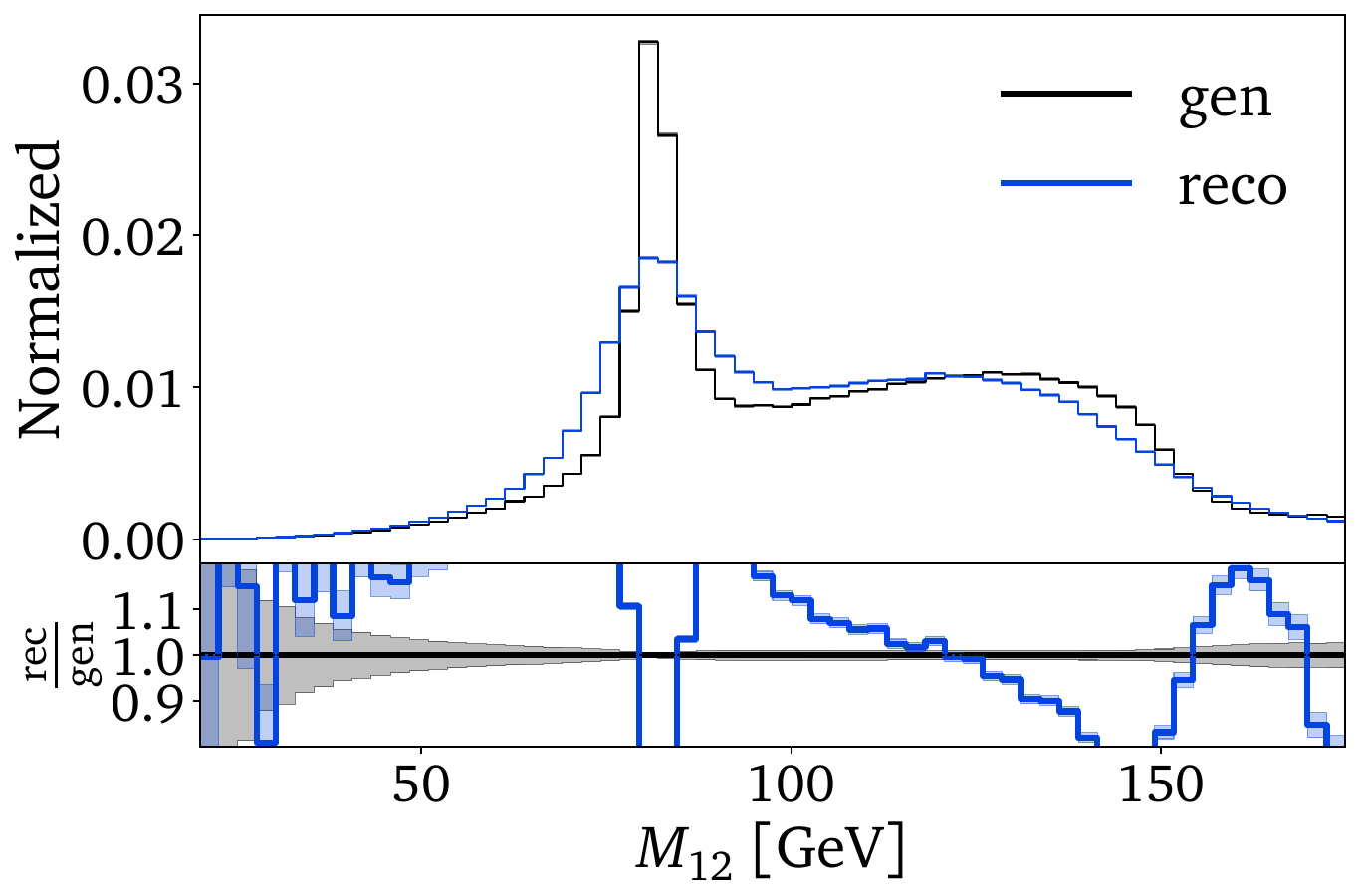} \\
    \includegraphics[width=0.495\textwidth, page=41]{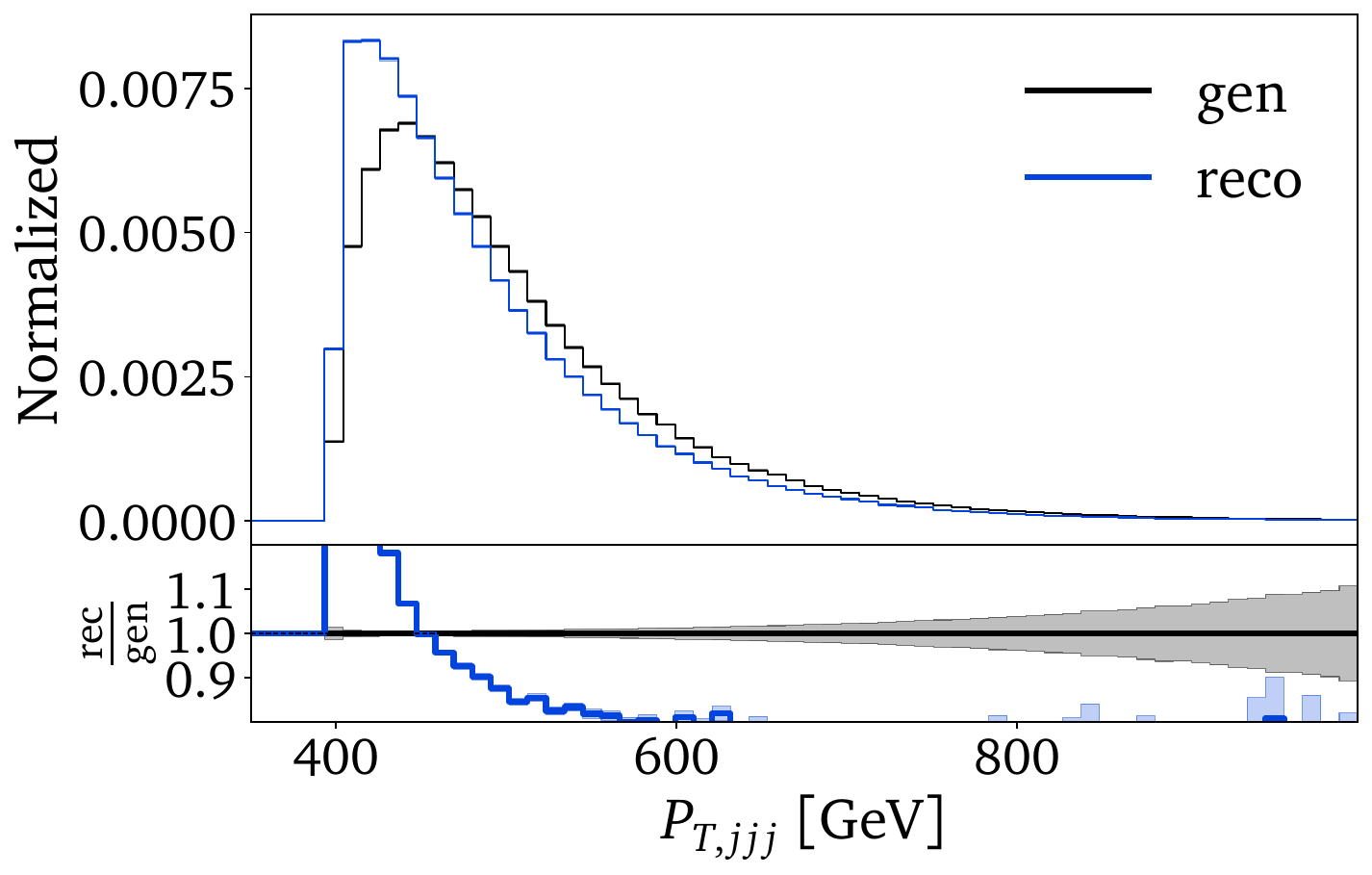}
    \includegraphics[width=0.495\textwidth, page=27]{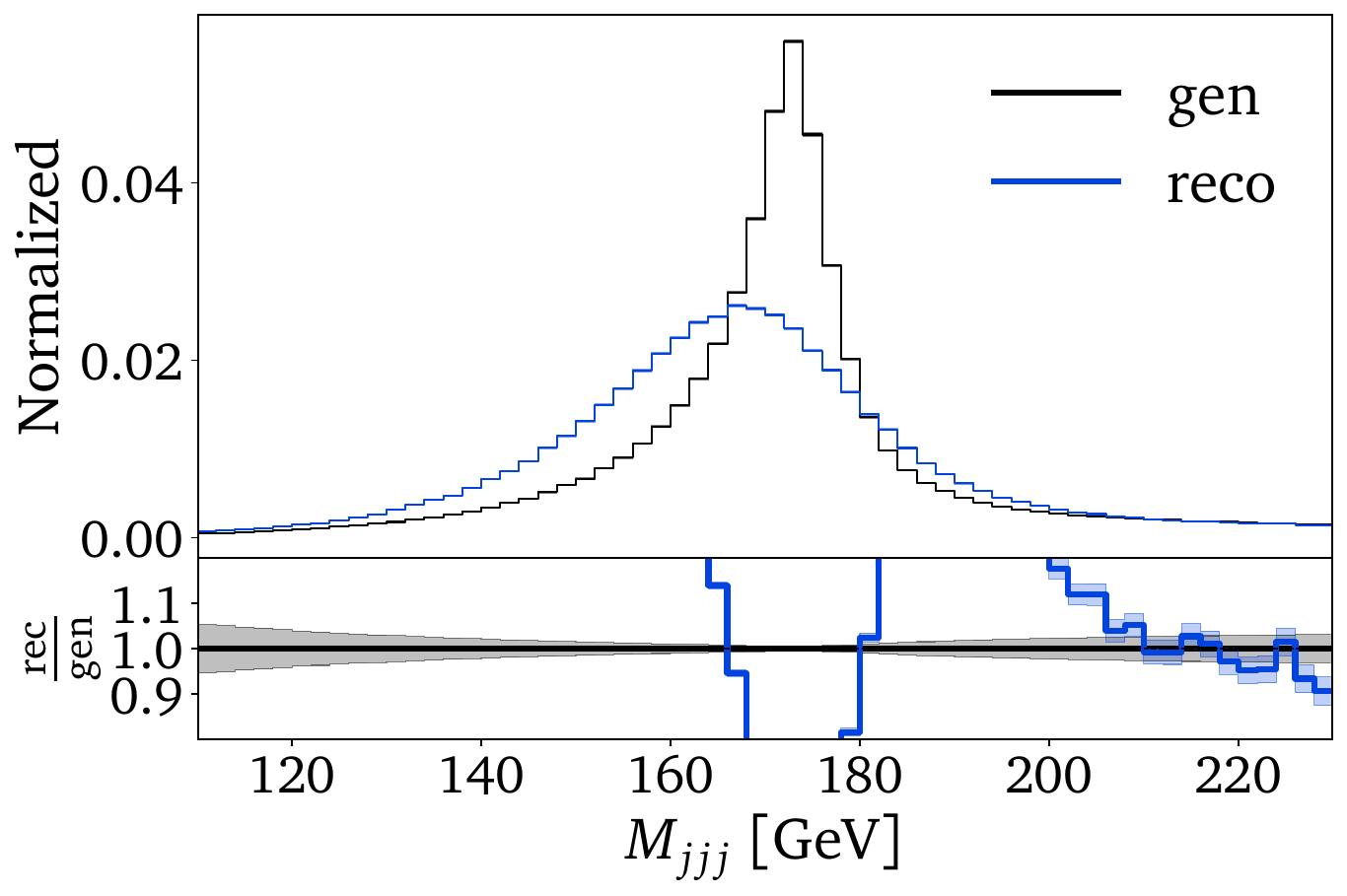}
    \caption{Kinematic distributions at reco-level and gen-level for
      the second jet (top), combining two jets (center), and
      combining three jets (bottom).}
    \label{fig:data}
\end{figure}

\subsection{Jet-mass features}
\label{sec:ana_masses}

For the generative unfolding algorithm a perfect matching
between reco-level and gen-level jets is not critical, as the
reco-level is used only as a condition.  We have checked that when
permuting the ordering of the reco-level jets randomly, we observe no
difference in performance.  Once we switch to the 4-momentum
representation ($m,p_T, \phi, \eta$), we see small differences
between reco-level and gen-level, for instance in the $p_T$ and individual jet masses shown in Fig.~\ref{fig:data} (top row).

Differences in the jet masses are mostly due to pile-up in our
simulation, which is added at the reco-level, and to a lesser degree 
from inefficiencies and mis-reconstructions in the reconstruction 
of photons, charged and neutral hadrons. 
Pile-up contributions are reduced by removing tracks
originating from pile-up vertices. The remaining difference in the jet
mass mostly comes from photons and neutral hadrons in the pile-up.This positive contribution to the jet masses is largest for the leading jet
because of its larger $p_T$ compared to the other jets.  
Figure~\ref{fig:data} implies that
unfolding detector effects includes correcting for these pile-up
effects. As Delphes assumes an idealized vertex reconstruction, 
we expect 
those differences to be larger when including full detector effects 
with GEANT4~\cite{Brun:1994aa}.

Going beyond single-jet observables, we need to understand and
eventually unfold detector effects on jet-jet correlations. In
Fig.~\ref{fig:data} (middle row) we show two examples.  
The distribution in the angular separation between the two leading jets shows a characteristic peak,  originating from the boosted decay 
kinematics combined with mass effects and the
detector acceptance. The 2-jet masses have a peculiar distribution,
owed to the fact that out of the three jets two come from the
$W$ decay. Because of the $p_T$-ordering, any of the three
combinations
\begin{align}
    M_{ik}^2 = m_i^2 + m_k^2 
    + 2 \; \left(
    m_{T,i} m_{T,k} \cosh{\Delta y_{ik}} - p_{T,i} p_{T,k} \cos{\Delta \phi_{ik}} \right) 
    \label{eq:dijetmass}
\end{align}
can reconstruct $m_W$. 
This is an exact equation for the three 2-jet masses, 
where $\Delta y_{ik}$ represents the difference in jet rapidities. 
Of the three 2-jet masses in
a top decay, two tend to be similarly close to $M_{ik} \sim
m_W$~\cite{Plehn:2010st}.  In Fig.~\ref{fig:data} (middle right), we also
observe the upper endpoint in the top decay kinematics at
gen-level~\cite{Plehn:2011nx}
\begin{align}
    m_{bj}^\text{max} < \sqrt{m_t^2 - m_W^2} \approx 155\,\unit{GeV} \; .
\end{align}

Following Eq.\eqref{eq:dijetmass}, we can improve the training of the
unfolding network by including the 2-jet masses as explicit features.
Each of the 2-jet masses then substitutes an angular variable.  With
this basis transformation we sacrifice access to the individual
azimuthal angles and are left with their absolute differences.

Next, we see in Fig.~\ref{fig:data} (bottom row) that the transverse
top quark momentum is not affected
significantly by detector effects, and the 3-jet mass peaks around the
top mass value. In our phase space parametrization we can calculate
the 3-jet mass as
\begin{align}
    M_{jjj}^2 = M_{12}^2 + M_{23}^2 + M_{13}^2 - m_1^2 - m_2^2 - m_3^2.
    \label{eq:trijetmass}
\end{align}
By using all these jet masses as training features, we can greatly
improve the learning and unfolding of the 3-jet mass. The
no-free-lunch theorem, however, tells us that this gain will lead to a
mismodelling of other correlations.  In particular, we will see that
there is no guarantee that $\cos{\Delta \phi} \in [0,1]$ anymore, 
leading to the generation of
unphysical event kinematics in some cases.

\subsection{Generative unfolding}
\label{sec:ana_cfm}

Traditional unfolding
algorithms~\cite{Cowan:2002in,Spano:2013nca,Arratia:2021otl} have been
used to unfold simple differential cross section measurements.  Widely used
methods include Iterative Bayesian
Unfolding~\cite{1974AJ.....79..745L,Richardson:72,Lucy:1974yx,DAgostini:1994fjx},
Singular Value Decomposition~\cite{Hocker:1995kb}, and
TUnfold~\cite{Schmitt:2012kp}.  Their limitation is the need for
binned data in a low-dimensional phase space. 
This also means that we have
to preselect the observables we want to unfold and decide on 
their binning before the unfolding.

To use ML-methods for high-dimensional and unbinned unfolding, 
we invert the forward simulation using Bayes' theorem
\begin{align}
    \label{eq:posterior}
    p(\xg|\xr)  = p(\xr|\xg) \; \frac{w(\xg) p(\xg)}{w(\xr)p(\xr)} \; ,
\end{align}
where $\xg$ is a point in the weighted gen-level phase space and $\xr$ a point in the weighted 
reco-level phase space. The gen-level and reco-level weights are encoded by $w(\xg)$ and $w(\xr)$ respectively. To unfold reco-level data, we need to learn 
\begin{align}
  \pmd(\xg|\xr) \approx p(\xg|\xr) 
\end{align}
as the statistical basis of an inverse simulation.  
The network endoding this conditional probability can be a 
GAN~\cite{Bellagente:2019uyp}, an INN version of a 
normalizing flow~\cite{Bellagente:2020piv}, or a diffusion 
network~\cite{Huetsch:2024quz}.
Once a
generative neural network encodes $\pmd(\xg|\xr)$, we calculate
\begin{align}
    \label{eq:unfolded_distribution}
    \punf(\xg) = \int d \xr \; \pmd(\xg|\xr) w(\xr) p(\xr) \; .
\end{align}
At the event level, this integral can easily be evaluated by 
marginalizing the corresponding joint probability.
Our method can be summarized as 
\begin{alignat}{9}
  & \psim(\xg)
  && \punf(\xg)
  \notag \\
  & \hspace*{-9mm} \text{\footnotesize paired data} \Bigg\updownarrow
  && \hspace*{+6mm} \Bigg\uparrow {\scriptstyle \pmd(\xg|\xr)}
  \notag \\
  & \psim(\xr) 
  \quad \xleftrightarrow{\text{\; correspondence \;}} \quad 
  && \pd(\xr)
\label{eq:scheme1}
\,.
\end{alignat}
The two distributions $\psim(\xr)$ and $\psim(\xg)$ are encoded in one
set of simulated events, before and after detector effects, or
at the parton- and the reco-level.

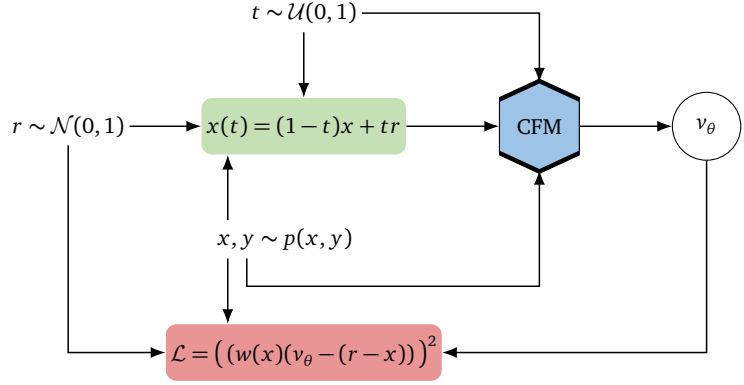
\begin{figure}[t]
    \centering
    \begin{tikzpicture}[node distance=2cm, scale=0.5, every node/.style={transform shape}]

\node (z) [txt] {$ r \sim \mathcal{N}(0,1)$};
\node (xt) [xt, right of= z, xshift=4.2cm]{$x(t) = (1-t) x + t r$};
\node (t) [txt, above of = xt, yshift=1cm]{$t \sim \mathcal{U}(0,1)$};
\node (xgen) [txt, below of = xt, yshift=-1cm, xshift=-0.5cm] {$x, y \sim p(x, y)$};
\node (model) [cinn_black, right of=xt, xshift=4.2cm] {};
\node (model) [cinn, right of=xt, xshift=4.2cm, fill=Bcolor] {CFM};

\node (loss) [loss, below of= xt , yshift = -4cm]{$\loss=\big( \left(w(x)(v_{\theta} - (r-x)\right) \big)^2$};
\node(vtheta)[expr, right of = model, xshift=2.4cm]{$v_{\theta}$};

\draw [arrow, color=black] (model.east) -- (vtheta.west);
\draw [arrow, color=black] (t.south) -- (xt.north);
\draw [arrow, color=black] (z.east) -- (xt.west);
\draw [arrow, color=black] (xt.east) -- (model.west);
\draw [arrow, color=black] (t.east) -- (model.north |- t.east) -- (model.north);

\draw [arrow, color=black] (z.south) -- (z.south |- loss.west) -- (loss.west);
\draw [arrow, color=black] (vtheta.south) -- (vtheta.south |- loss.east) -- (loss.east);
\draw [arrow, color=black] ([xshift=-1.5cm]xgen.south) -- ([xshift=-2cm]loss.north);
\draw [arrow, color=black] ([xshift=-1.5cm]xgen.north) -- ([xshift=-2cm]xt.south);
\draw [arrow, color=black] ([xshift=-1cm]xgen.south) -- ([xshift=-1cm, yshift=1cm]xgen.south |- loss.north) -- ([yshift=1cm]model.south |- loss.north) -- (model.south);

\end{tikzpicture}
    \caption{Schematic representation of generative unfolding with a
      CFM network.}
    \label{fig:schematic}
\end{figure}

The generative network we employ to learn $\pmd(\xg|\xr)$ is
Conditional Flow Matching (CFM). The generative CFM network is the leading
architecture for precision-LHC
simulations~\cite{Butter:2023fov}. Mathematically, CFM is based on two
equivalent ways of describing a diffusion process using an 
ordinary differential equation (ODE) or a
continuity equation~\cite{Plehn:2022ftl}
\begin{align}
  \frac{dx(t)}{dt} = v(x(t),t)
  \qquad \text{or} \qquad 
\frac{\partial p(x,t)}{\partial t} 
= - \nabla_\theta \left[ v(x(t),t) p(x(t),t) \right] \; ,
\label{eq:ode}
\end{align}
both with the same velocity field $v(x(t),t)$. The diffusion process
described by $t \in [0,1]$ relates a latent Gaussian distribution $\pl(r)$ to the
physical phase space $\pd(x)$,
\begin{align}
 p(x,t) \to 
 \begin{cases}
  \pd(x) \qqquad & t \to 0 \\
  \pl(r) = \normal(r;0,1) \qqquad & t \to 1  \; .
 \end{cases} 
\label{eq:fm_limits}
\end{align}
We employ a simple linear interpolation
\begin{align}
    x(t) 
    &= (1-t) x + t r 
    \to \begin{cases}
      x \qqquad & t \to 0 \\
      r \sim \normal(0,1) \qqquad & t \to 1  \; .
    \end{cases} 
\end{align}
Using this approximation, we train the network to learn
\begin{align}
  v_\theta(x(t),t) \approx v(x(t),t) 
  \label{eq:velocity_field}
\end{align}
using the continuity equation and then generate phase space
configurations using a fast ODE solver. Even though the corresponding MSE
loss function 
\begin{align}
  \loss_\text{CFM} = \left[ w(x) (v_\theta - (r-x)) \right]^2
  \label{eq:loss}
\end{align}
is not a likelihood loss, a Bayesian version of the CFM
generative network can learn uncertainties on the underlying phase
space density together with the central values underlying its
sampling~\cite{Butter:2023fov}.

The CFM setup is illustrated in Fig.~\ref{fig:schematic}. Its
conditional extension is straightforward, in complete analogy to
the conditional GANs~\cite{Bellagente:2019uyp} and conditional
INNs~\cite{Bellagente:2020piv} developed for unfolding. While the naive
GAN setup does not learn the event-wise (inverse) migration correctly
and therefore does not encode physical, calibrated conditional
probabilities, the cINN with its likelihood loss does exactly that.
The CFM succeeds because of its mathematical foundation,
Eq.\eqref{eq:ode}~\cite{Huetsch:2024quz}.

\subsubsection*{Training bias}

In Eq.\eqref{eq:scheme1} we describe the structure of generative unfolding, but we are missing
a critical complication --- the simulated reco-level data $\psim(\xr)$ might not agree with 
the actual reco-level data $\pd(\xr)$. 

Let us assume a simple case where  the simulation depends on 
a simulation parameter $m_s$ which we can tune to describe the actual data. This can be a 
physics parameter we eventually infer, or a nuisance parameter which we profile over. 
The dependencies of the four datasets on $m_s$ and its `correct' value in the data, $m_d$,
turn Eq.\eqref{eq:scheme1} into
\begin{alignat}{9}
  & \psim(\xg|m_s)
  && \punf(\xg|m_s,m_d)
  \notag \\
  & \hspace*{-9mm} {\scriptstyle p(\xr|\xg)} \Bigg\downarrow 
  && \hspace*{+6mm} \Bigg\uparrow {\scriptstyle \pmd(\xg|\xr,m_s)}
  \notag \\
  & \psim(\xr|m_s) 
  \quad \xleftrightarrow{\text{\; correspondence \;}} \quad 
  && \pd(\xr|m_d)
  \label{eq:scheme2}
  \; .
\end{alignat}
In the forward direction, 
$p(\xr|\xg)$ does not have an explicit $m_s$-dependence, but both 
simulated datasets follow $\psim(\xg|m_s)$ and $\psim(\xr|m_s)$ induced 
by the generator settings. By assumption, $m_s = m_d$ 
ensures that the simulated and actual data agree at the reco-level,
\begin{align}
  \psim(\xr|m_s=m_d) \really \pd(\xr|m_d) \; .
\end{align}
We then use this relation to infer $m_d$ at the reco-level. 

Alternatively, 
we can do the same inference at the gen-level, requiring
\begin{align}
  \psim(\xg|m_s=m_d) \really \punf(\xg|m_s=m_d,m_d) \; .
\end{align}
The problem with this unfolded inference is the dual dependence of 
$\punf(\xg|m_s,m_d)$ through the reco-level data and the learned conditional probability.
This dual dependence is automatically resolved if $\punf(\xg)$ 
only depends on $m_d$ through the reco-level data, so the bias from 
$\pmd(\xg|\xr,m_s)$ can be neglected. 
It is important to emphasize that such a bias from the training data
would lead to an uncontrolled
systematic shift and a wrongly measured mass value.

An established way to remove the bias is through iteratively re-
weighting the training dataset. This IcINN method~\cite{Backes:2022sph}
can of course applied to any conditional generative network.
It relies on a learned classifier over $\xg$ which 
reweights $\psim$ to $\punf$ including the $m_s$-dependencies and 
serves as a basis for re-training the unfolding network. 
It implicitly assumes that
$\punf(\xg|m_s,m_d)$ depends mostly on $m_d$ and at a reduced level on $m_s$. 
In that  case the endpoint of the Bayesian iteration
is reached when the two dependencies coincide at the level of the 
remaining statistical uncertainty. In App.~\ref{app:bias}
we show results for top decays and discuss the reasons for them not working.

\section{Unbinned top quark decay unfolding}
\label{sec:gen}

Unfolding top decays is technically challenging, because the top
mass and the $W$ mass are dominant features of an 
altogether 12-dimensional
phase space. We start with a naive unfolding in
Sec.~\ref{sec:gen_naive}, using our appropriate phase space
parametrization with reduced
dimensionality~\cite{Ackerschott:2023nax}. In Sec.~\ref{sec:gen_bias},
we show how the model dependence from the top mass in the training
data can be controlled. With this enhancement, we show in
Sec.~\ref{sec:gen_mass} how the high-dimensional unfolding 
improves the existing top mass measurement based on classic
unfolding. Finally, we show how to unfold the entire
12-dimensional phase space using the measured top mass in
Sec.~\ref{sec:gen_full}.

\subsection{Lower-dimensional unfolding}
\label{sec:gen_naive}

We know that the precision of learned phase space distribution using
neural networks scales unfavorably with the phase space
dimension~\cite{Maitre:2021uaa,Badger:2022hwf}.\footnote{For a possible
  improvement see Ref.~\cite{Spinner:2024hjm, Brehmer:2024yqw}.} 
The full 12-dimensional phase space will not be the optimal
representation to measure the top mass. Instead, we only use a
lower-dimensional phase space representation for the top mass
measurement, finding a balance between relevant kinematic information
and dimensionality. We postpone the full kinematic unfolding to the point
where we need to access the 
full kinematics and benefit from the measured top mass. 

For the traditional CMS analysis~\cite{CMS:2022kqg}, two phase space
dimensions were unfolded, $M_{jjj}$ and $p_{T,jjj}$, 
where the $p_{T,jjj}$ was integrated over in the final measurement. 
The jet mass calibration relies on the reconstructed $W$ boson. Identifying the $W$-decay jets in the top jet ideally requires $b$-tagging information, but because of the inefficiency
not all jets from the $W$ decay can be identified. Instead, 
the jet mass can be calibrated by using all possible 2-jet combinations, 
where each of the three resulting distributions feature 
a sharp $W$-mass peak (see Fig.~\ref{fig:data}). 
Therefore, we unfold those for the top mass measurement such that a reliable 
calibration can be performed at a later stage. 

\begin{figure}[t]
    \includegraphics[width=0.495\textwidth, page=1]{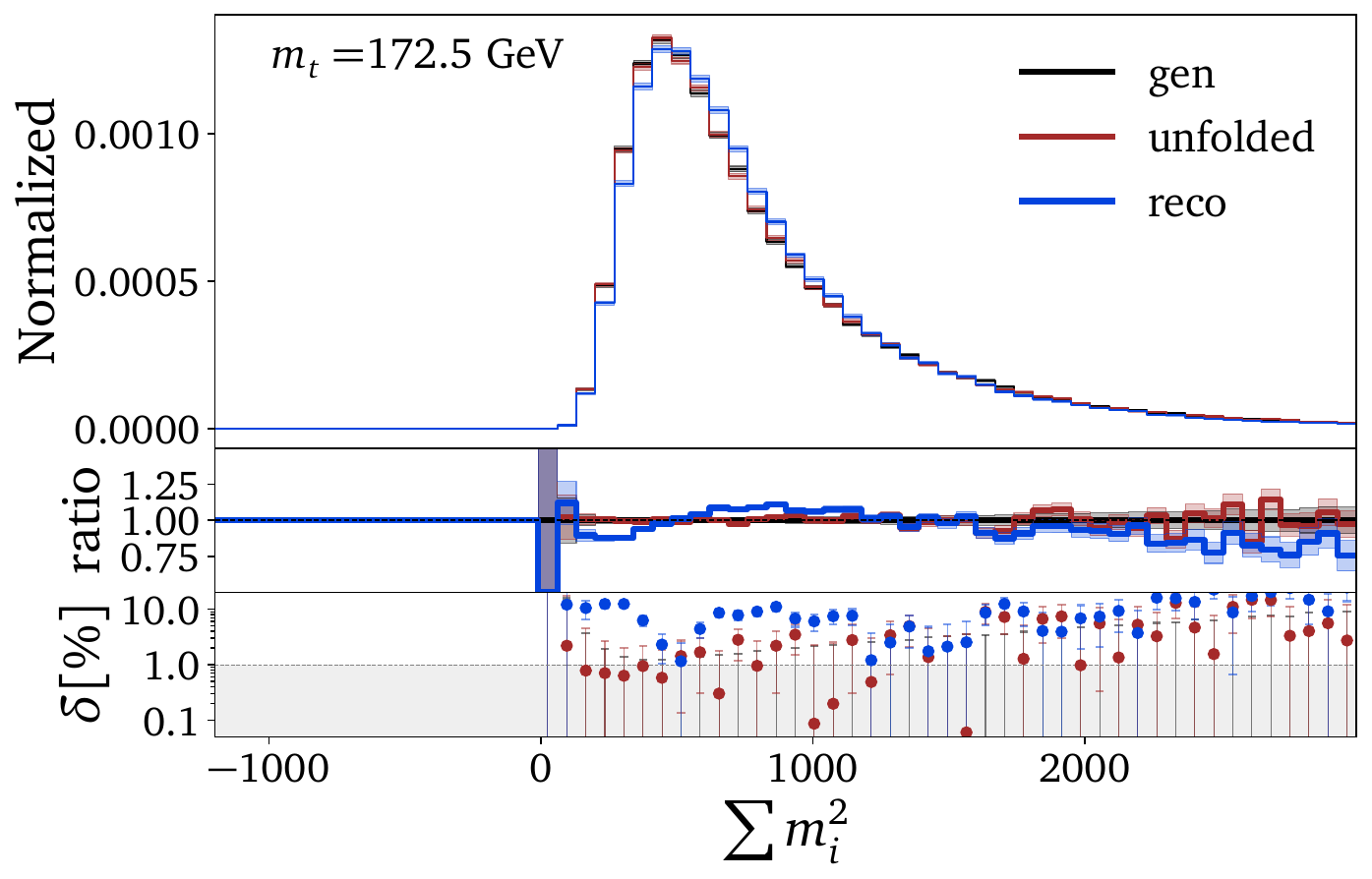}
    \includegraphics[width=0.495\textwidth, page=2]{figs/biased_results/4d/4d_biased_1725_delphes.pdf}\\
    \includegraphics[width=0.495\textwidth, page=5]{figs/biased_results/4d/4d_biased_1725_delphes.pdf}
    \includegraphics[width=0.495\textwidth, page=5]{figs/biased_results/4d/4d_biased_1735_delphes.pdf}
    \caption{Kinematic distributions from the 4-dimensional unfolding. We
      also show the reco-level and the gen-level truth for
      $m_t=172.5\,\unit{GeV}$. In the bottom-right panel we compare $M_{jjj}$
      for $m_t=172.5\,\unit{GeV}$ to generated unfolding for $m_t=173.5\,\unit{GeV}$,
      not seen during training.}
    \label{fig:4d_biased}
\end{figure}

Our unfolding setup follows 
Sec.~\ref{sec:ana_masses}. From Eq.\eqref{eq:trijetmass} we know that
we can extract the 3-jet mass as a proxy for the top mass from the set
of single-jet and 2-jet masses. Because the single-jet masses are
largely universal and not a good handle on the jet energy
calibration, our first choice is to measure the top mass from
a 4-dimensional unfolding of
\begin{align}
  \Big\{ \; M_{j1j2}, M_{j2j3}, M_{j1j3}, \sum_i m_i \; \Big\} \; .
  \label{eq:basis_4d}
\end{align}
The results are shown in Fig.~\ref{fig:4d_biased}. First, we see that
we can unfold the sum of the single jet masses extremely well, with
deviations of the unfolded data from the generator truth at the
per-cent level. This means that we expect to be able to extract the
3-jet mass essentially from the sum of all 2-jet masses with a
known and controlled offset.

\begin{figure}[t]
    \includegraphics[width=0.495\textwidth, page=7]{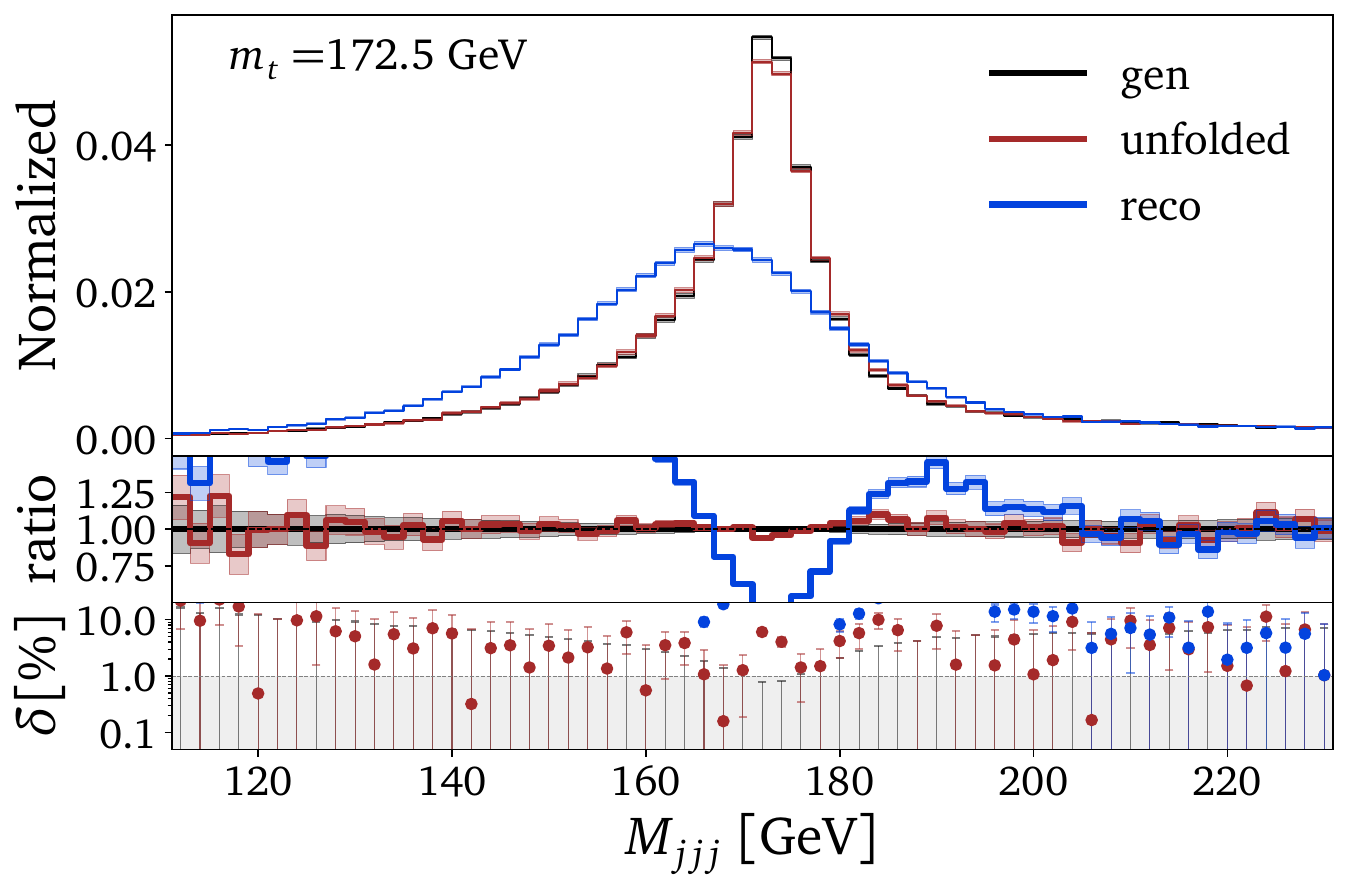}
    \includegraphics[width=0.495\textwidth, page=7]{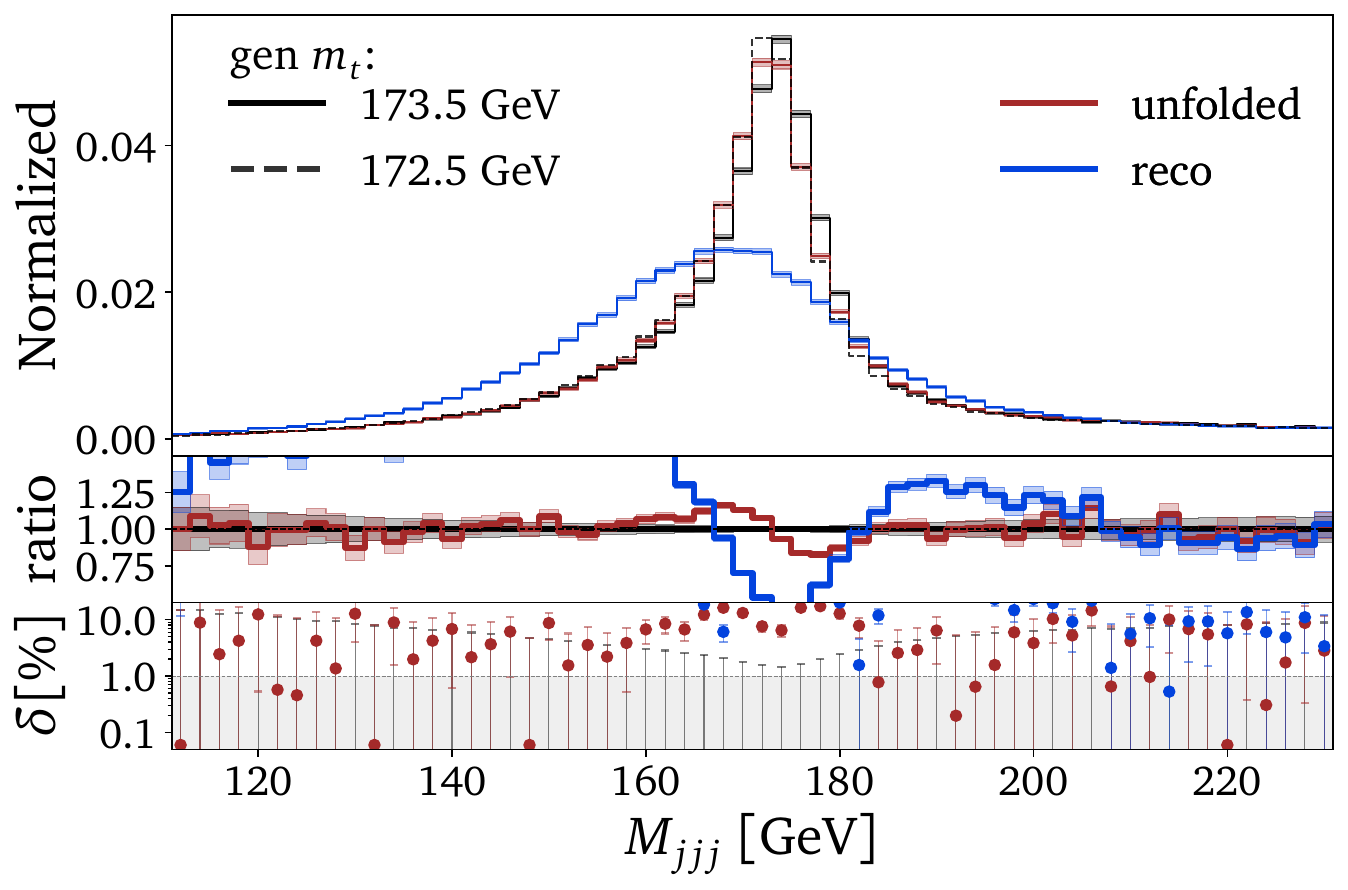}
    \caption{Kinematic distributions from 6-dimensional unfolding.  In
      the right panel we compare $M_{jjj}$ for $m_t=172.5\,\unit{GeV}$ to
      generated unfolding for $m_t=173.5\,\unit{GeV}$, not seen during
      training.}
    \label{fig:6d_biased}
\end{figure}

Next, we show a 2-jet mass, with the characteristic
$W$ peak and the shoulder at $m_{bj}^\text{max}$. The $W$ peak 
is washed out at the reco-level, but the generative
unfolding reproduces the gen-level extremely well. The relative deviation of the
unfolded to the truth 2-jet mass distributions is at most a few
per-cent, with no visible shift around the $W$ peak. The same quality
of the unfolding can be observed in the $M_{jjj}$ distribution,
perfectly reproducing the top mass at $m_t = 172.5\,\unit{GeV}$, the correct
value in the training data and in the data which gets unfolded.

The problem with measuring the top mass from unfolded data appears
when we unfold data simulated with a different top mass. In the
lower-right panel of Fig.~\ref{fig:4d_biased} we show the unfolded
$M_{jjj}$ distribution for reco-level data generated with $m_t=
173.5\,\unit{GeV}$, unfolded with generative networks trained on $m_t=
172.5\,\unit{GeV}$. We see that the top peak in the unfolded data
is dominated by the training bias of the network, specifically
a maximum at $M_{jjj} = (172 \pm 1)\,\unit{GeV}$. This means the top peak 
is entirely determined by the training
bias and hardly impacted by the reco-level data which we unfold.
 
From the 4-dimensional unfolding we know that the network
learns the $W$ peak in the 2-jet masses and the top peak in
the 3-jet mass at a precision much below the physical particle
widths. The problem is that the bias from the network training
completely determines the position of these mass peaks in the unfolded
data. To confirm that these findings are not an artifact of our
reduced phase space dimensionality, we repeat the same analysis for
the 6-dimensional phase space
\begin{align}
  \Big\{ \; M_{j1j2}, M_{j2j3}, M_{j1j3}, m_{j1}, m_{j2}, m_{j3} \; \Big\} \; .
  \label{eq:basis_6d}
\end{align}
The unfolded 3-jet mass distributions
are shown in Fig.~\ref{fig:6d_biased}, corresponding to 
the 4-dimensional case in Fig.~\ref{fig:4d_biased}. While the unfolded peak in
$M_{jjj}$ is a bit worse than for the easier 4-dimensional
case when unfolding the same value of $m_t$ as used in the training, 
the bias from the training remains in spite of the fact that we 
are weakening the expressive
power of the unfolding network by adding distributions that are mildly affected by the peak position.

\begin{figure}[b!]
    \includegraphics[width=0.45\textwidth, page=13]{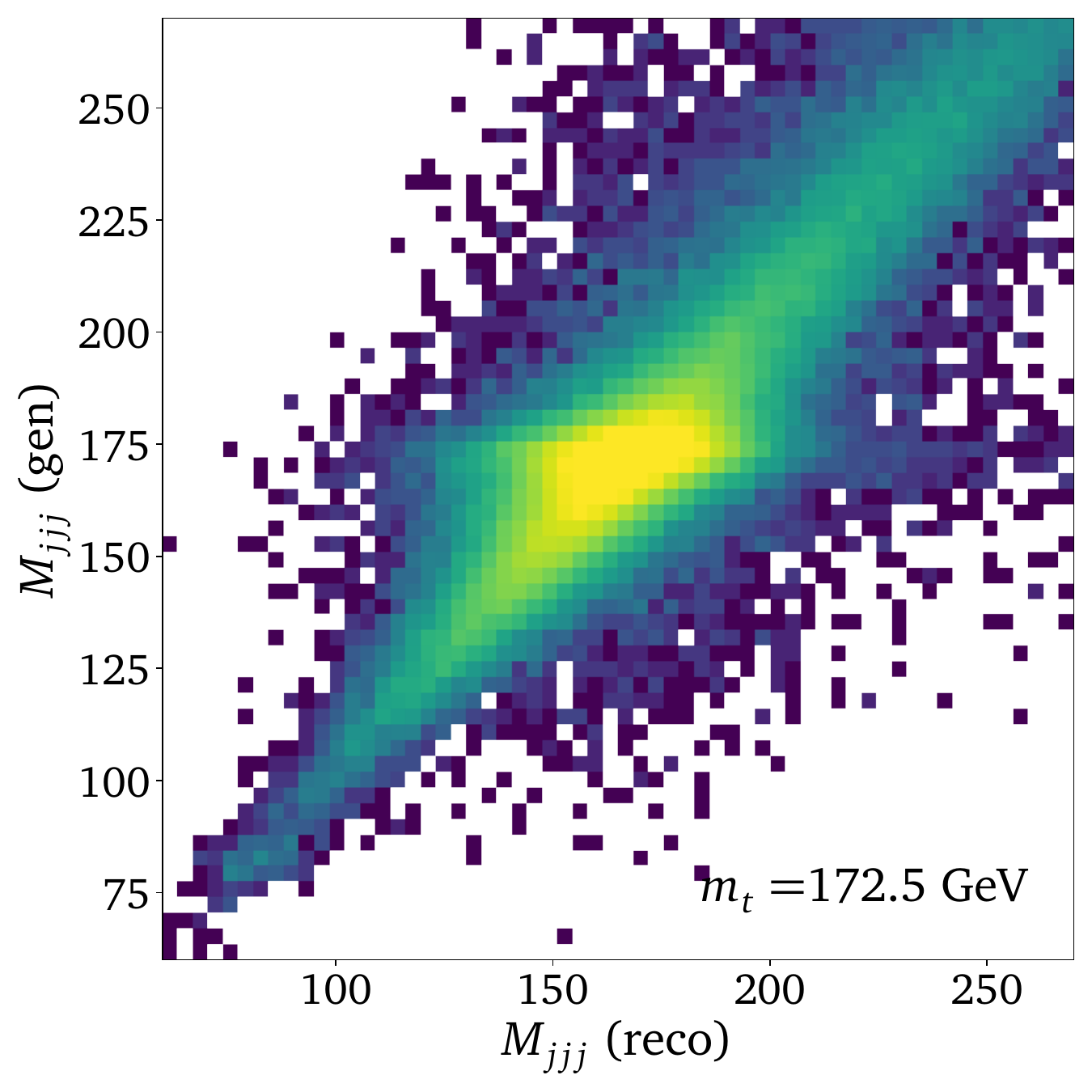}
    \hfill
    \includegraphics[width=0.45\textwidth, page=14]{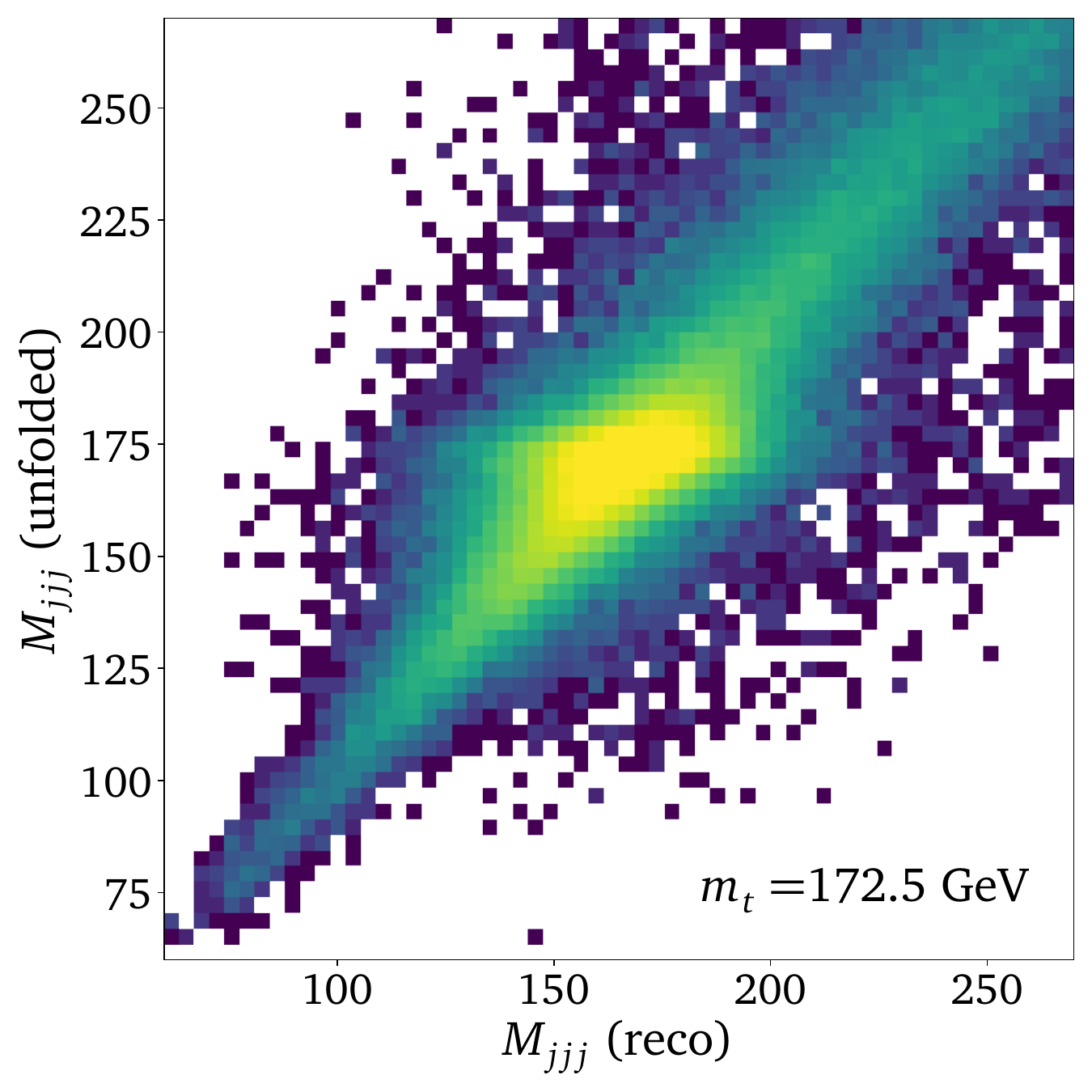}
    \caption{True and learned migrations in the $M_{jjj}$ distribution
      between reco-level and gen-level.}
    \label{fig:migration}
\end{figure}

Finally, it is instructive to study the true and learned migrations between the
reco-level and the gen-level 3-jet distribution. These are shown in
Fig.~\ref{fig:migration}, where in the left panel we see that the forward
simulation maps the sharp peak at gen-level to a broader peak at
reco-level. The problem with the central ellipse describing this
physical migration by detector effects is that it does not indicate
any correlation between the $M_{jjj}$-values at reco-level and at
gen-level. The learned migration in the right panel reproduces the
forward migration exactly. 

For the generative unfolding this means
that small differences at reco-level will always be unfolded to the
same sharp region at gen-level, independent of the information
contained in the reco-level data.
Following Sec.~\ref{sec:ana_cfm} and 
Eq.\eqref{eq:scheme2} the
unfolded distribution $\punf(\xg)$ is entirely determined 
by the training choice $m_s$ and shows practically no dependence on 
the value $m_d$ encoded in the actual data.

\subsection{Taming the training bias}
\label{sec:gen_bias}

The next question is how we can improve the situation where, $m_s$ 
being the top 
mass value used for the simulation and $m_d$ the actual top mass in the 
data, Eq.\eqref{eq:scheme2} turns into
\begin{alignat}{9}
  & \psim(\xg|m_s)
  && \punf(\xg|m_s,\cancel{m_d})
  \notag \\
  & \hspace*{-9mm} {\scriptstyle p(\xr|\xg)} \Bigg\downarrow 
  && \hspace*{+6mm} \Bigg\uparrow {\scriptstyle \pmd(\xg|\xr,m_s)}
  \notag \\
  & \psim(\xr|m_s) 
  \quad \xleftrightarrow{\text{\; correspondence \;}} \quad 
  && \pd(\xr|m_d) \,.
  \label{eq:scheme3} 
\end{alignat}
In the unfolded distribution, the training information 
$m_s$ completely overwrites $m_d$. 
Moreover, even if there was enough sensitivity, a classifier comparing two 
shifted mass peaks learns weights far away from unity, leading to 
numerical challenges. This means we cannot use
the usual iterative methods to remove the bias from the training data. 

Following the strategy from Sec.~\ref{sec:ana}, we first 
increase the sensitivity on $m_d$. For this, we 
pre-process the data such that $m_d$ is directly 
accessible by adding an estimator of $m_d$ 
to the representation of $\xr$. 
Ideally, this estimator would be inspired by an optimal 
observable. Such a one-dimensional observable with sufficient statistical precision 
should exist, and we know how to construct it. For the top mass 
we just use the weighted median of the 3-jet masses at reco-level, 
$M_{jjj}^\text{batch} = \frac{1}{N_\text{batch}} \sum_i^{N_\text{batch}} M_{jjj,i} $, 
where the sum runs over all, possibly weighted, events in one batch.
For a batch size around $10^4$ events, this information will 
be strongly correlated with the top mass,
\begin{align}
\label{eq:batch_wise_condition}
 M_{jjj}^\text{batch} \approx m_d \equiv m_t \Bigg|_\text{data} \; .
\end{align}
This batch-wise kinematic information can be extracted at the 
level of the loss evaluation, and it goes beyond the usual single-event
information, similar to established MMD loss modifications of 
GAN training~\cite{Butter:2019cae,Bellagente:2019uyp}.

Second, we weaken the bias from the training data by 
combining training data with different top masses, but without an additional label,
\begin{align}
m_t = \left\{ 169.5, 172.5, 175.5 \right\}\,\unit{GeV}
\qqquad \text{(combined training).}
\label{eq:training_masses}
\end{align}
It turns out that it is sufficient to cover a range of top masses with 
separate, unmixed training batches.
The range ensures that top masses in the actual data are within 
the range of the training data.
We ensure a balanced training by enlarging the event samples 
with $m_t=169.5$ and $175.5\,\unit{GeV}$ to match the size of the largest 
sample. This is done by repeating and shuffling the input data, 
which effectively uses these events 
several times per epoch. 
we avoid overfitting using an appropriate regularization.  
The limited number of simulated events for the eventual analysis makes 
this training strategy sub-optimal. We expect larger and additional 
$m_t$ simulations, unavailable at this time, to improve the results.
As shown in App.~\ref{app:bias}, both steps need to be included to ensure precise, unbiased results.

Obviously, this strategy
of strengthening the dependence on $m_d$ and reducing the training bias 
is not applicable to all problems, and it does not lead to the endpoint of 
the Bayesian iterative method, but for our combined inference-unfolding 
strategy it works, and this is all we need. 

\subsubsection*{Transfusion architecture}

\begin{figure}[t]
    \centering
    \begin{tikzpicture}[node distance=2cm, scale=0.6, every node/.style={transform shape}]

\node (part1) [txt] {$x_{\text{reco},1}$};
\node (part2) [txt, right of=part1, xshift=-0.5cm] {$...$};
\node (part3) [txt, right of=part2, xshift=-0.5cm] {$x_{\text{reco},n}$};

\node (emb_part1) [embed, below of=part1, yshift=-0.3cm, rotate=90]{Emb};
\node (emb_part2) [txt, below of=part2, yshift=-0.3cm] {$...$};
\node (emb_part3) [embed, below of=part3,yshift=-0.3cm,  rotate=90]{Emb};

\node (TE) [transformer, below of=emb_part2, yshift=-0.7cm, text width=4cm,
text depth=1.5cm, align=center] {Transformer-Encoder};
\node (TE_att) [attention, below of=TE, yshift=1.6cm] {Self-Attention \\
Reco-level correlations};

\node (reco1) [txt, right of=part3, xshift=1.75cm] {$x_{\text{gen},1}(t)$};
\node (reco3) [txt, right of=reco1] {$...$};
\node (reco6) [txt, right of=reco3] {$x_{\text{gen},N}(t)$};

\node (t) [txt, right of=reco6, xshift=0.5cm] {$t$};

\node (emb_reco1) [embed, below of=reco1, yshift=-0.3cm, rotate=90]{Emb};
\node (emb_reco3) [txt, below of=reco3, yshift=-0.3cm] {$...$};
\node (emb_reco6) [embed, below of=reco6, yshift=-0.3cm, rotate=90]{Emb};

\node (TD) [transformer, right of=TE, xshift=5.3cm, yshift=-0.8cm, text width=4cm,
text depth=3.1cm, align=center, minimum height=4cm] {Transformer-Decoder};
\node (TD_att) [attention, right of=TE_att, xshift=5.3cm] {Self-Attention \\
Gen-level correlations};
\node (TD_crossatt) [attention, below of=TD_att, yshift=0.4cm] {Cross-Attention \\
Combinatorics};

\node (inn1) [small_cinn, below of=reco1, yshift=-8.2cm, rotate=90]{Linears};
\node (inn3) [txt, below of=reco3, yshift=-8.2cm]{$...$};
\node (inn6) [small_cinn, below of=reco6, yshift=-8.2cm, rotate=90]{Linears};


\node (prob1) [txt, below of=inn1, yshift=-0.3cm]{$\Big( v_\theta(c_1, t ),$ };
\node (prob2) [txt, below of=inn3, yshift=-0.3cm]{$...$};
\node (prob6) [txt, below of=inn6, yshift=-0.3cm]{$, \; v_\theta(c_{N}, t ) \Big)$};
\node (prob) [txt, left of=prob1, xshift=-2 cm]{$v_\theta(x_\text{gen}(t), t , x_\text{reco}) = \;$};

\draw [arrow, color=black] (part1.south) -- (emb_part1.east);
\draw [arrow, color=black] (part3.south) -- (emb_part3.east);

\draw [arrow, color=black] (emb_part1.west) -- (TE.north -| emb_part1.west);
\draw [arrow, color=black] (emb_part3.west) -- (TE.north -| emb_part3.west);

\draw [arrow, color=black] (TE.south -| emb_part1.west) --  ([yshift=-1cm]TE.south -| emb_part1.west) -- ([yshift=-1cm]TE.south -| TD.west) ; 
\draw [arrow, color=black] (TE.south -| emb_part3.west) --  ([yshift=-0.7cm]TE.south -| emb_part3.west) -- ([yshift=-0.7cm]TE.south -| TD.west);

\draw [arrow, color=black] ([xshift=-0.2cm]reco1.south) -- ([xshift=-0.2cm]emb_reco1.east);
\draw [arrow, color=black] ([xshift=-0.2cm]reco6.south) -- ([xshift=-0.2cm]emb_reco6.east);

\draw [arrow, color=black] (emb_reco1.west) -- (TD.north -| emb_reco1.west);
\draw [arrow, color=black] (emb_reco6.west) -- (TD.north -| emb_reco6.west);

(A) (B);

\draw [arrow, color=black] ([xshift=-0.2cm]TD.south -| emb_reco1.west)  -- node [text width=1.5cm, pos=0.3, font=\Large, right] {$c_{1}$} ([xshift=-0.2cm]inn1.east -| emb_reco1.west);
\draw [arrow, color=black] ([xshift=-0.2cm]TD.south -| emb_reco6.west)  -- node [text width=1.5cm,pos=0.3, font=\Large, right] {$c_{N}$}  ([xshift=-0.2cm]inn6.east -| emb_reco6.west);

\draw [arrow, color=black] (inn1.west -| emb_reco1.west) -- (prob1.north -| emb_reco1.west);
\draw [arrow, color=black] (inn6.west -| emb_reco6.west) -- (prob6.north -| emb_reco6.west);

\draw [arrow, color=black] (t.south) --  ([yshift=-0.5cm]t.south) -- ([yshift=-0.5cm, xshift=0.2cm]t.south -| reco1.center) -- ([xshift=0.2cm]emb_reco1.east); 
\draw [arrow, color=black] ([yshift=-0.5cm, xshift=0.2cm]t.south -| reco6.center) -- ([xshift=0.2cm]emb_reco6.east); 
\draw [arrow, color=black] (t.south) --  ([yshift=-8.1cm]t.south) -- ([yshift=-8.1cm, xshift=0.2cm]t.south -| reco1.center) -- ([xshift=0.2cm]inn1.east); 
\draw [arrow, color=black] ([yshift=-8.1cm, xshift=0.2cm]t.south -| reco6.center) -- ([xshift=0.2cm]inn6.east); 

\end{tikzpicture}
    \caption{Schematic representation of a parallel transfusion network, adapted from~\cite{Huetsch:2024quz}.}
    \label{fig:transfusion}
\end{figure}
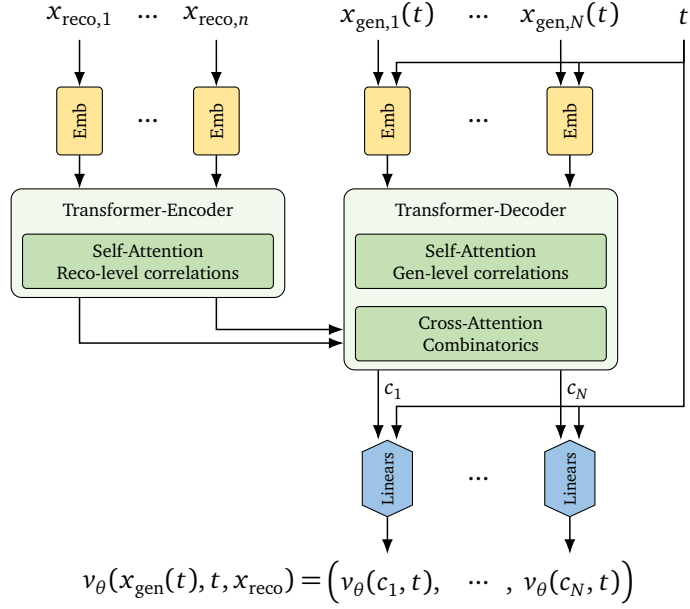

As the network task becomes significantly more difficult we replace the simple 
dense architecture with a transfusion network, described in detail in 
Refs.~\cite{Heimel:2023mvw,Huetsch:2024quz} and visualized in Fig.~\ref{fig:transfusion}.

Each component of the $n$-dimensional condition as well as of the time-dependent $N$-dimen\-sional 
input $x(t)$ are individually embedded by concatenating positional information and zero padding. 
The embedded conditions are passed through the encoder part of a transformer, while the embedded 
input is passed through the decoder counterpart.
In both transformer parts, we apply self-attention to learn the correlations in the condition 
and in the input. 
The network is complemented by a cross-attention between encoder and decoder outputs, to learn the 
correlations between conditions and inputs. These are crucial for the unfolding task.
For every component of the input, the transformer returns one high-dimensional 
embedding vector $c_i$, which is mapped back to a one-dimensional component 
of the velocity field by a shared dense linear network. 
This way, we express the learned $N$-dimensional velocity 
field of Eq.\eqref{eq:velocity_field} as
\begin{align}
    v_\theta (x_\text{gen}(t),t,x_\text{reco} ) = \left(v_\theta (c_{1}, t), \dots, v_\theta (c_{N}, t) \right).
\end{align}
The hyperparameters of the network can be found in Appendix~\ref{app:hyper}.

\begin{figure}[!b]
    \includegraphics[width=0.495\textwidth, page =5] {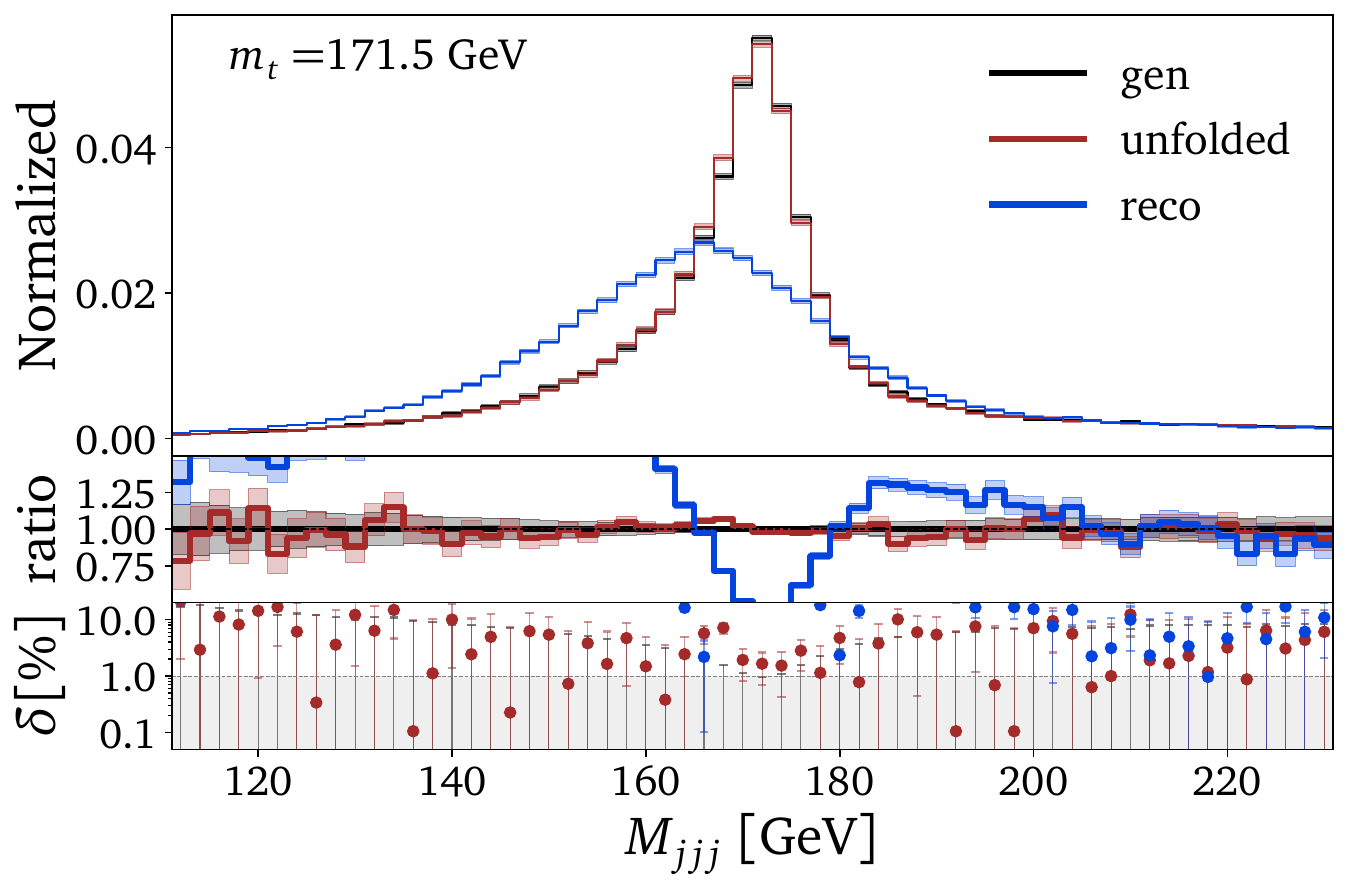}
    \includegraphics[width=0.495\textwidth, page =5] {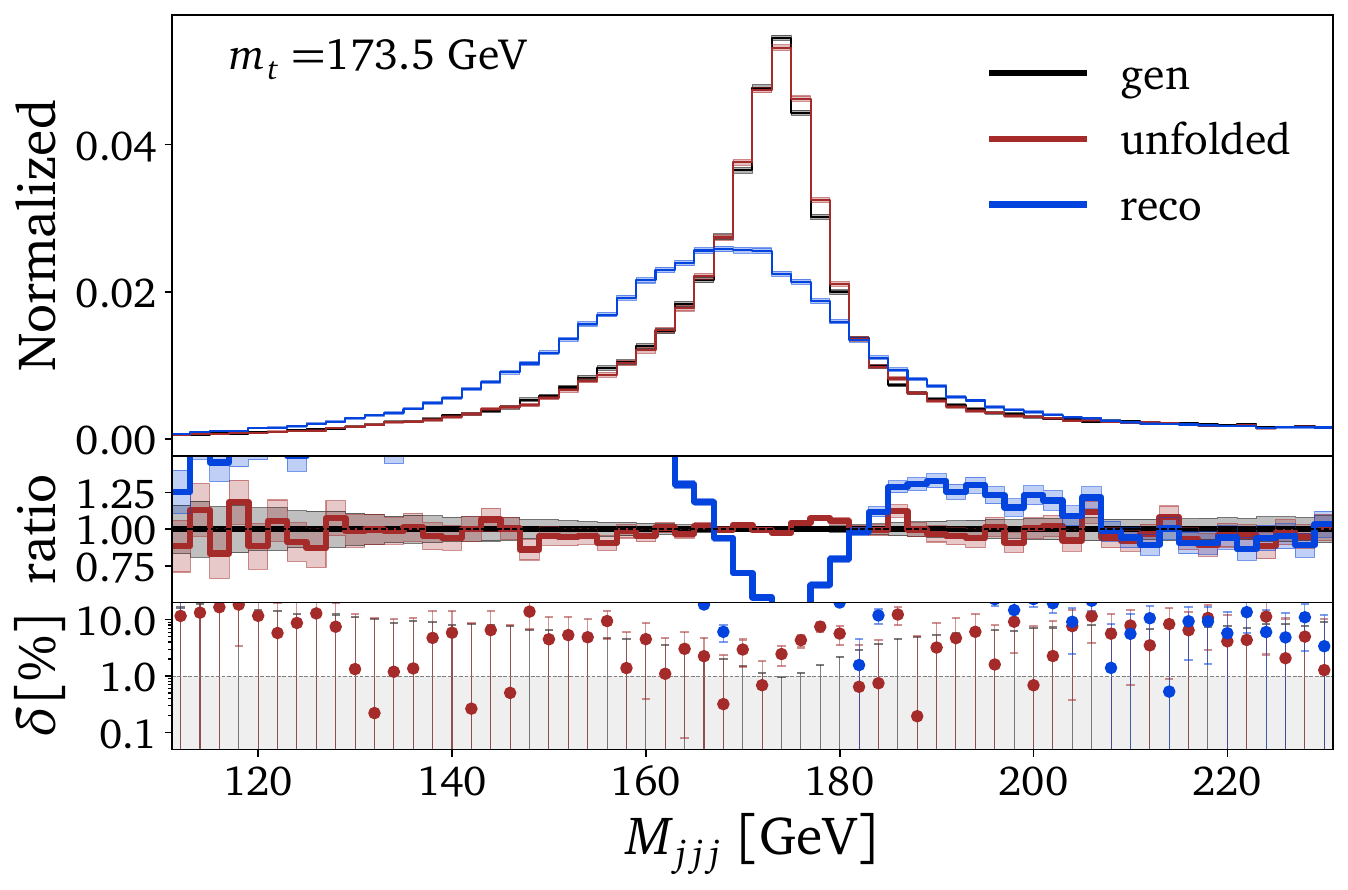}
    \includegraphics[width=0.495\linewidth, page=7]{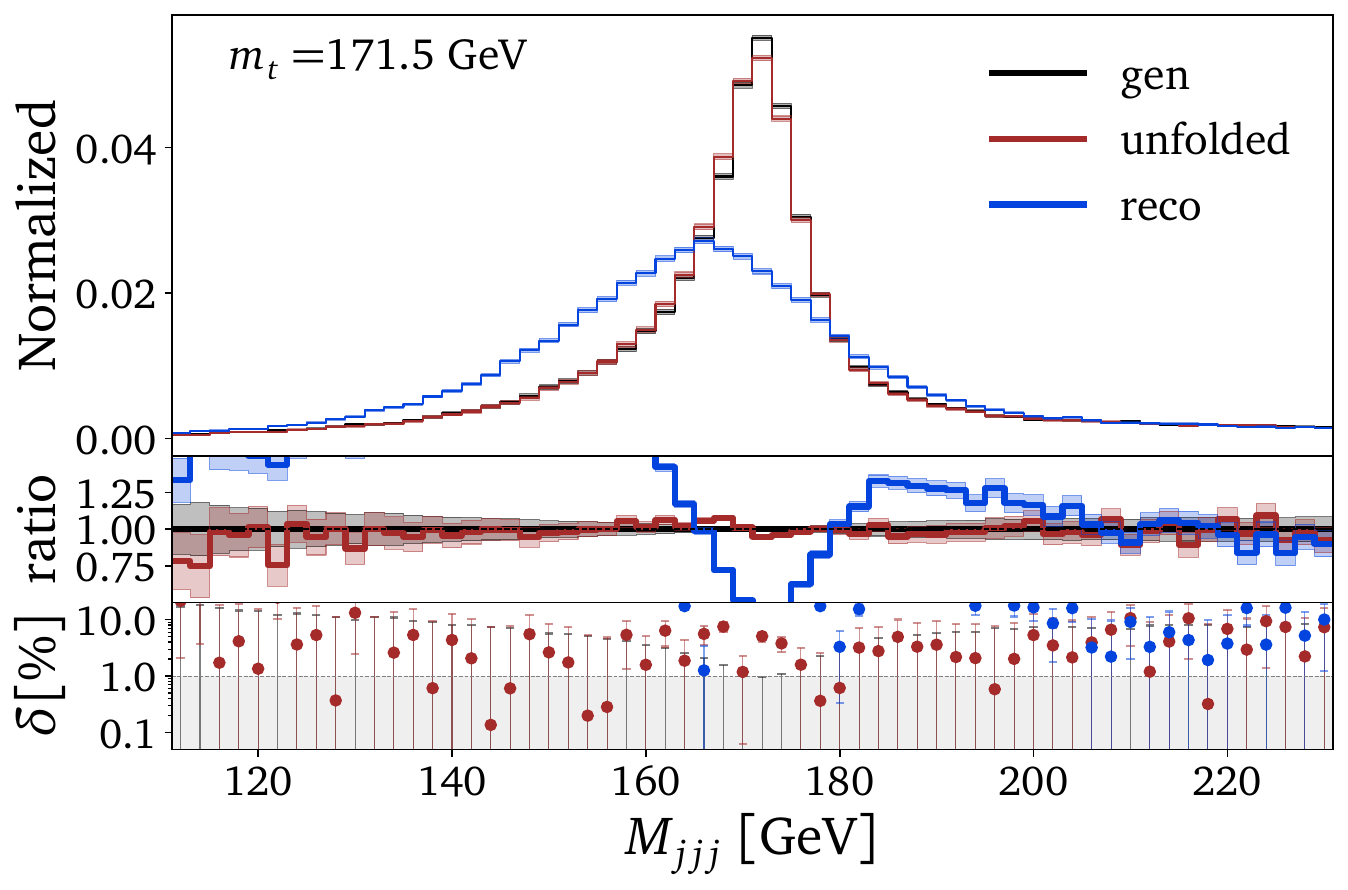}
    \includegraphics[width=0.495\linewidth, page=7]{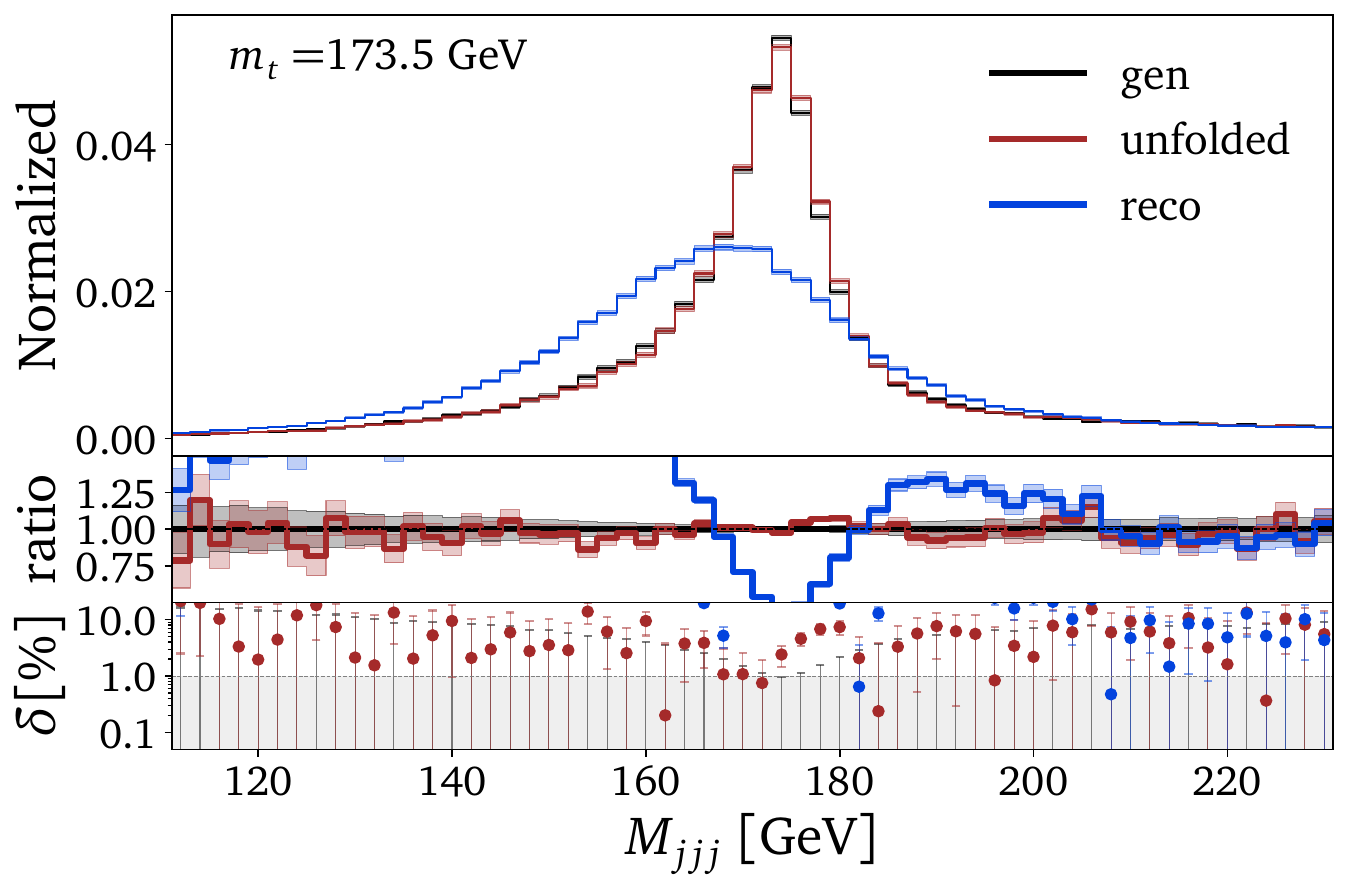} 
    \caption{$M_{jjj}$-distributions from the 4-dimensional (top row) and 
    6-dimensional (bottom row) unfolding of data with 
    $m_t=171.5\,\unit{GeV}$ (left column) and $m_t=173.5\,\unit{GeV}$ (right column). 
    We train the network combining samples with three top masses, Eq.\eqref{eq:training_masses}.
    \label{fig:unbiased_results_4d6d}
    }
\end{figure}
Using the transfusion network we unfold the 4-dimensional phase space from 
Eq.\eqref{eq:basis_4d}. The results are shown in Fig.~\ref{fig:unbiased_results_4d6d} (top row).
We unfold data generated with two different top masses, $m_t = 171.5$ and $173.5\,\unit{GeV}$. 
Neither of these two values are present in the training data. 
We observe in both cases that the top mass as the main kinematic feature is 
reproduced well, without a significant deviation from the gen-level distributions. 
The fitted peak values of the distributions are
$m_\text{peak} = (172 \pm 1)\,\unit{GeV}$ when unfolding data with $m_t = 171.5\,\unit{GeV}$, and 
$m_\text{peak} = (174 \pm 1)\,\unit{GeV}$ when unfolding data with $m_t = 173.5\,\unit{GeV}$. 
While the bias might not have vanished entirely, it is well contained within the 
numerical uncertainties. We will extract the unfolded top mass value properly in
Sec.~\ref{sec:gen_mass}.

\subsubsection*{Dual network}

Given the more complicated training task, we observe a drop in performance 
when we increase  the dimensionality to unfold the 6-dimensional phase space 
\begin{align}
  x = \big( \; \{ m_i \}, \{ M_{ik} \} \big) \; ,
\end{align}
defined in Eq.\eqref{eq:basis_6d} using the transfusion network.
Inspired by Refs.~\cite{Butter:2021csz, Butter:2023fov}, we 
factorize the phase space density into two parts, each encoded in a generative network:
the first network learns the individual jet mass directions in phase space, which are
universal and do not depend on the value of $m_t$; the second network generates the 
2-jet masses conditioned on the individual jet masses,
\begin{align}
p(x_\text{gen}| x_\text{reco}) =
\underbrace{ p\left(\{m_{i,\text{gen}}\}\big|\; x_\text{reco},M_{jjj}^\text{batch}\right)}_\text{network 1}
\underbrace{ p\left(\{M_{ik,\text{gen}}\} \big| \; \{m_{i,\text{gen}}\}, x_\text{reco}, M_{jjj}^\text{batch}\right).}_\text{network 2}
\end{align}
Both CFM-transfusion networks also receive $M_{jjj}^\text{batch}$
calculated for a full batch using Eq.\eqref{eq:trijetmass}. 
For the event generation we first generate the unfolded jet masses $\{ m_i \}$, 
pass them as a condition to the second network,
and then generate the unfolded 2-jet masses $\{ M_{ik} \}$.

Looking at the 6-dimensional correlation giving $M_{jjj}$ in 
Fig.~\ref{fig:unbiased_results_4d6d} (bottom row), 
we observe a hardly visible drop in performance, but still no bias from the training data. 
As before, we observe peak values at
$m_\text{peak} = (172 \pm 1)\,\unit{GeV}$ when unfolding data with $m_t = 171.5\,\unit{GeV}$ 
and at 
$m_\text{peak} = (174 \pm 1)\,\unit{GeV}$ when unfolding data with $m_t = 173.5\,\unit{GeV}$.


\subsection{Mock top quark mass measurement}
\label{sec:gen_mass}

We estimate the benefit from generative unfolding by repeating the 
top quark mass measurement from Ref.~\cite{CMS:2022kqg}, but with 
a large number of bins in the $M_{jjj}$ histogram.
The top mass is extracted from the binned unfolded distributions using a fit based on $\chi^2 = d^T V^{-1} d$, 
where $d$ is the vector of bin-wise differences between the normalized unfolded distribution 
and the normalized prediction from the simulated data. 
The covariance matrix $V$ contains the uncertainties and corresponding bin-to-bin correlations.
A parabola fit provides the central value of $m_t$ and the standard deviation.
Experimental systematics and simulation uncertainties have to be propagated to 
the top mass measurements~\cite{CMS:2022kqg}, combined with an in-situ jet calibration 
using the known $W$-mass peak.
Crucially, these uncertainties do not lead to an 
uncontrolled bias of the unfolding, but will typically manifest themselves as
noise.

\subsubsection*{Statistical and model uncertainties}

First, this fit requires the covariance matrix describing statistical uncertainties~\cite{Backes:2023ixi}.
We sample $N$ times from the latent space, conditional on the reco-level events.
This means we generate $N$ unfolded distributions
from the posterior $\pmd(\xg|\xr)$. 
We then use a Poisson bootstrap, where we assign a weight from a Poisson distribution with unit mean.
The size of one replica is 52,000 events,
corresponding to the approximate number of real data events.
The number of events follows a Poisson distribution, with the mean given by the nominal sample size.

For the measurement, we create $N_\text{rep} = 1000$ replicas by selecting the nominal number of reco-level 
events from the test dataset with $m_t=172.5\,\unit{GeV}$ and the full datasets for the simulations at different top masses.
We unfold each replica, calculate $M_{jjj}$, and use the histogram entries $u_i^{(n)}$ to compute the 
correlation matrix of statistical fluctuations as
\begin{align}
    \text{cov}_{ij} &= \frac{1}{N_{\text{rep}}} \sum_{n=1}^{N_{\text{rep}}} (u_i^{(n)} - \bar{u}_i)(u^{(n)}_j - \bar{u}_j)   
    \qquad  \text{with} \qquad \bar{u}_i = \frac{1}{N_{\text{rep}}} \sum_{n=1}^{N_{\text{rep}}} u^{(n)}_i \notag \\
    \rho_{ij} &= \frac{\text{cov}_{ij}}{\sqrt{\text{cov}_{ii}}\sqrt{\text{cov}_{jj}}} \; .
\label{eq:covariance}
\end{align}
This procedure also takes into account the uncertainties due to the statistical fluctuations of $M_{jjj}^{\text{batch}}$. 
The training of the network itself introduces correlations which are at least one order of magnitude smaller and therefore ignored in the measurement.

\begin{figure}[t]
    \includegraphics[width=0.45\linewidth]{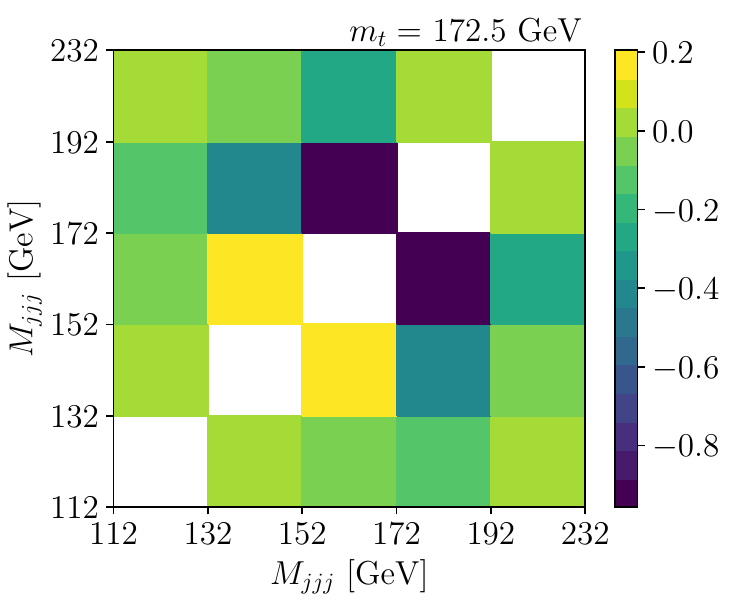}
    \hfill
    \includegraphics[width=0.45\linewidth]{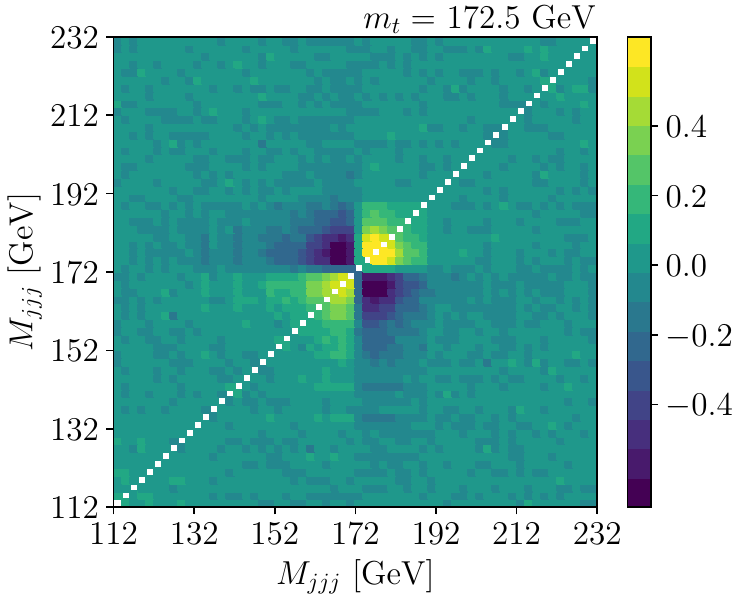}
    \caption{Correlation matrices obtained from $N_{\text{rep}}=1000$ replicas 
    for 5 bins (left) and 60 bins (right) in the 4-dimension unfolding
    with $m_t=172.5\,\unit{GeV}$.
    \label{fig:corr_all}
    }
\end{figure}

The $5 \times 5$ and $60 \times 60$ correlation matrices $\rho_{ij}$ from the 
4-dimensional unfolding using the largest sample generated with $m_t=172.5\,\unit{GeV}$ 
are shown in Fig.~\ref{fig:corr_all}. 
We see two distinct sources of bin-to-bin correlations.
In general, an event migrating from bin $i$ to bin $j$ gives rise to negative 
correlations in $\rho_{ij}$ between the two bins. 
Additionally, unbiasing the unfolding
ensures that a shift in the batch-wise condition also shifts the unfolded peak.
This effect, accounted for in the bootstrapping method, introduces an 
additional contribution to the bin-to-bin correlations. It causes positive
correlations between bins on the same side of the peak and anti-correlations
otherwise.
In our case, both effects are most apparent
in the peak region and its neighboring bins.

We follow Ref.~\cite{CMS:2022kqg} to estimate the uncertainty 
from the choice of $m_t$ in the simulation used for the unfolding. 
We evaluate the difference in each bin $i$ between the unfolded distribution and the 
corresponding simulated gen-level distribution. From 
the differences $d_i$, we construct a covariance matrix 
\begin{align}
    \text{cov}^{\text{model}}_{ij} =  \rho_{ij} d_i d_j \; ,
\end{align} 
where $\rho_{ij}$ are the correlations between bins $i$ and $j$.
Because the bin-to-bin correlations are not known and 
we do not observe any systematic pattern, 
we choose a diagonal covariance matrix with 
$\rho_{ij} = 1$ for $i=j$ and $\rho_{ij} =0$ otherwise. 
It was verified that other choices do not alter the results.
To estimate the impact of this model uncertainty, we perform the $m_t$ extraction twice. 
First, we only include the statistical covariance matrix corresponding to 52,000 
available events at the reco-level.
Second, we repeat the same measurement also including the model uncertainty.

\subsubsection*{Improvement}

\begin{figure}[b!]
    \centering
    \includegraphics[width=0.5\linewidth]{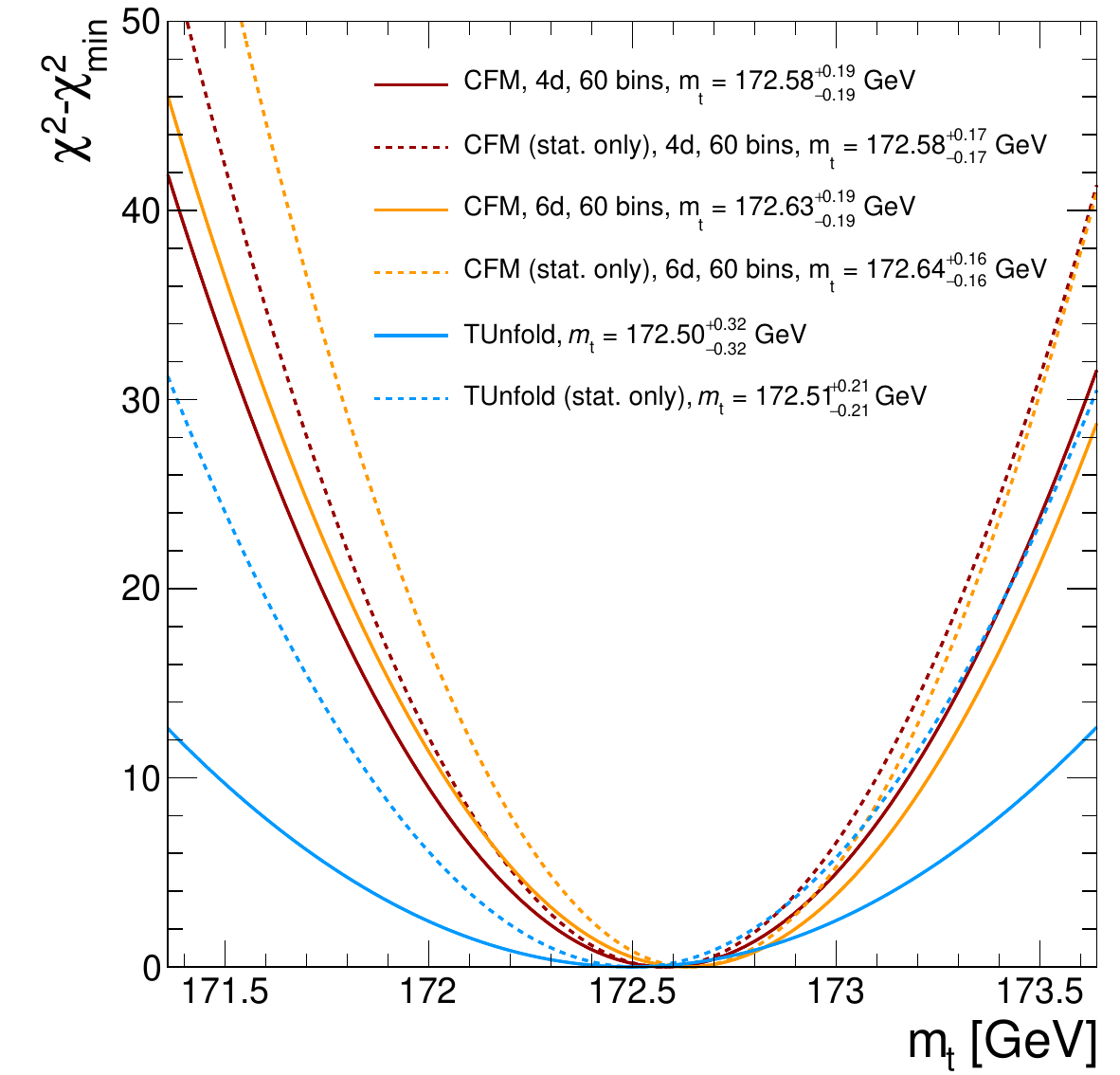}
    \caption{Extraction of $m_t$ with a $\chi^2$ test. The dotted lines include 
    only statistical uncertainties, while the solid lines also include the 
    model uncertainty from the choice of $m_t$.}
    \label{fig:chi2}
\end{figure}

To compare our new unfolding technique to the existing TUnfold results~\cite{CMS:2022kqg}, 
we repeat the extraction using the simulated data set with 172.5\,\unit{GeV} 
and using the statistical covariance matrix from the measured data, 
published in HEPData~\cite{hepdata.130712}.
The $\chi^2$-curves and the corresponding results are displayed in Fig.~\ref{fig:chi2}, 
where we show the 4-dimensional and 6-dimensional 
unfoldings with 60~bins and the TUnfold result.
We see that the uncertainty in the
choice of $m_t$ is reduced from being a leading model uncertainty in the CMS measurement to 
a much smaller level. The statistical uncertainty in the TUnfold result was already small relative to the systematic uncertainties. Both the 4-dimensional and 6-dimensional unfoldings exhibit comparable statistical uncertainties with the 5-bin configuration. However, increasing the number of bins leads to a reduction in statistical uncertainty, as demonstrated below.

\begin{figure}[t]
    \includegraphics[width=0.45\linewidth]{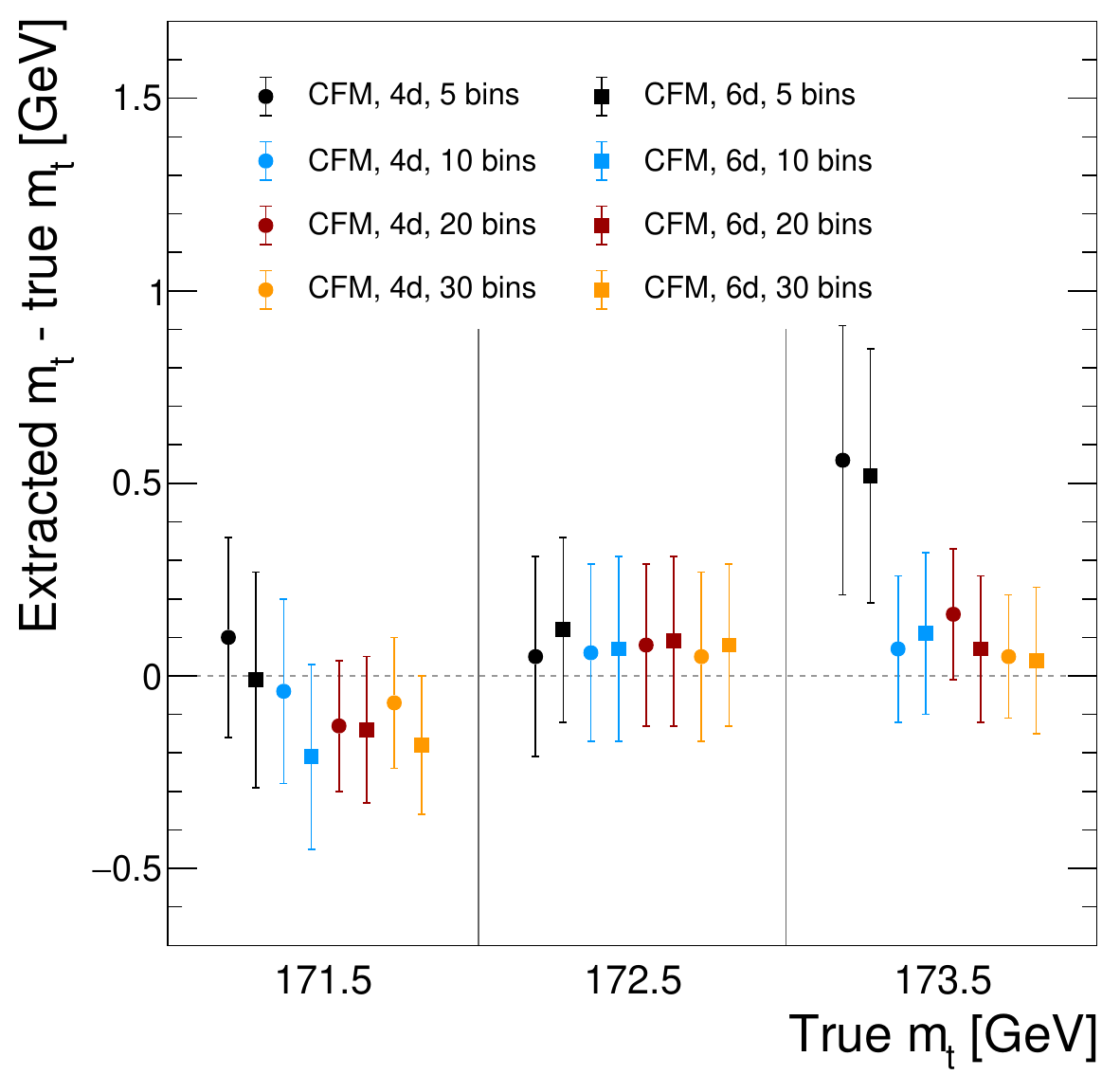}
    \hfill
    \includegraphics[width=0.45\linewidth]{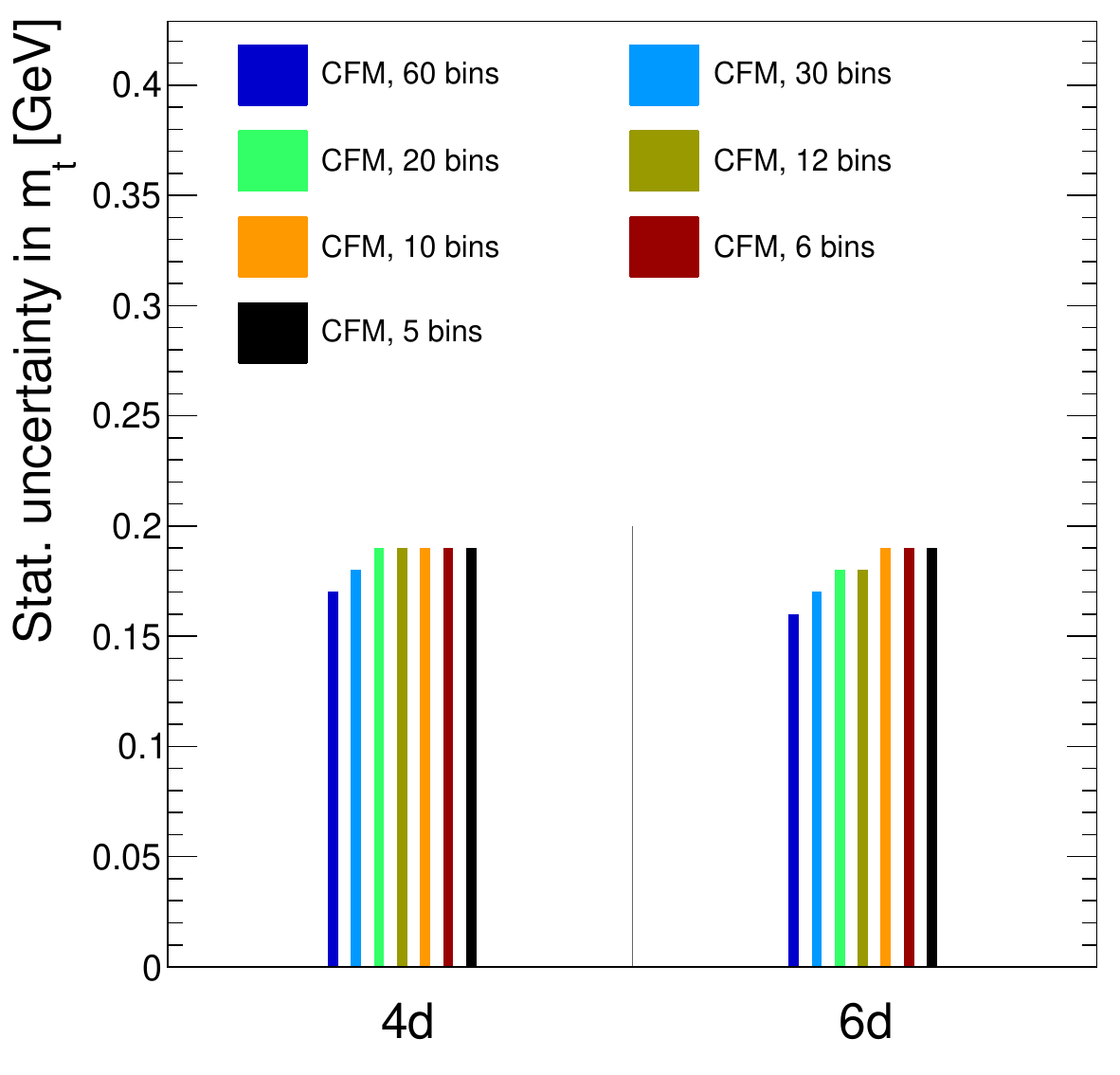}
    \caption{Deviation of the extracted top mass from the reco-level truth, 
    employing 4-dimensional and 6-dimensional unfoldings with different numbers 
    of measurement bins for $m_t = 171.5$, 172.5, and 173.5\,\unit{GeV} (left) .
    The size of the statistical uncertainties in $m_t$ from the 4-dimensional and 6-dimensional 
    unfoldings with different binnings, assuming $m_t = 172.5\,\unit{GeV}$ (right).
    \label{fig:summary}
    }
\end{figure}

To confirm that the choice in $m_t$ does not leave a residual bias, we repeat the 
top quark mass extraction for unfolded data obtained from reco-level data 
simulated with different top masses.
The results are shown in the left panel of Fig.~\ref{fig:summary}. 
For a top mass of $m_t = 173.5\,\unit{GeV}$, we observe a bias of about 0.5\,\unit{GeV} 
when using a measurement with 5~bins. This is not surprising as the exact binning has 
been optimized for a minimal model dependence in the CMS measurement, which we did not do here. 
While the bin width in the unfolding with TUnfold is limited by the jet mass resolution, 
we test various binning schemes for the unbinned unfolding. The bias gets reduced 
when using more bins in the measurement, as expected because the binning introduces 
a regularization in the unfolding which leads to a model dependence. With 10 and more 
measurement bins, we observe that the bias from the model dependence is removed. 
For more measurement bins than 60, the comparably coarse grid of gen-level 
distributions with $m_t = \{169.5, 171.5, 172.5, 173.5, 175.5\}\;\text{GeV}$ 
leads to an unstable closure test.

To circumvent this limitation, we interpolate the gen-level distributions for 
$m_t$-values close to $172.5\,\unit{GeV}$, 
where three samples with a separation of 1\,\unit{GeV} are available and 
a linear dependence of the bin content as a function of $m_t$ represents a valid approximation. 
Now, we can compare the resulting values of $m_t$ from the generative unfolding 
with 5 to 60 bins in terms of the statistical uncertainty.
The results are displayed in the right panel of Fig.~\ref{fig:summary}, 
indicating an increase in the statistical precision in $m_t$ 
due to the improved resolution.

\subsection{Full phase space unfolding}
\label{sec:gen_full}

As a last step of our unfolding program, we unfold the full 12-dimensional 
phase space given the measured top mass. This has the advantage that the 
leading source of training bias is removed. 
Following the same precision arguments as before, we keep the 
mass basis of Eq.\eqref{eq:basis_6d} for the first 6 of the 12 
phase space dimensions. This ensures that the 2-jet and 3-jet masses 
are reproduced well, albeit not at the level of the dedicated first 
unfolding step. 
\begin{figure}[b]
    \includegraphics[width=0.495\linewidth, page=13]{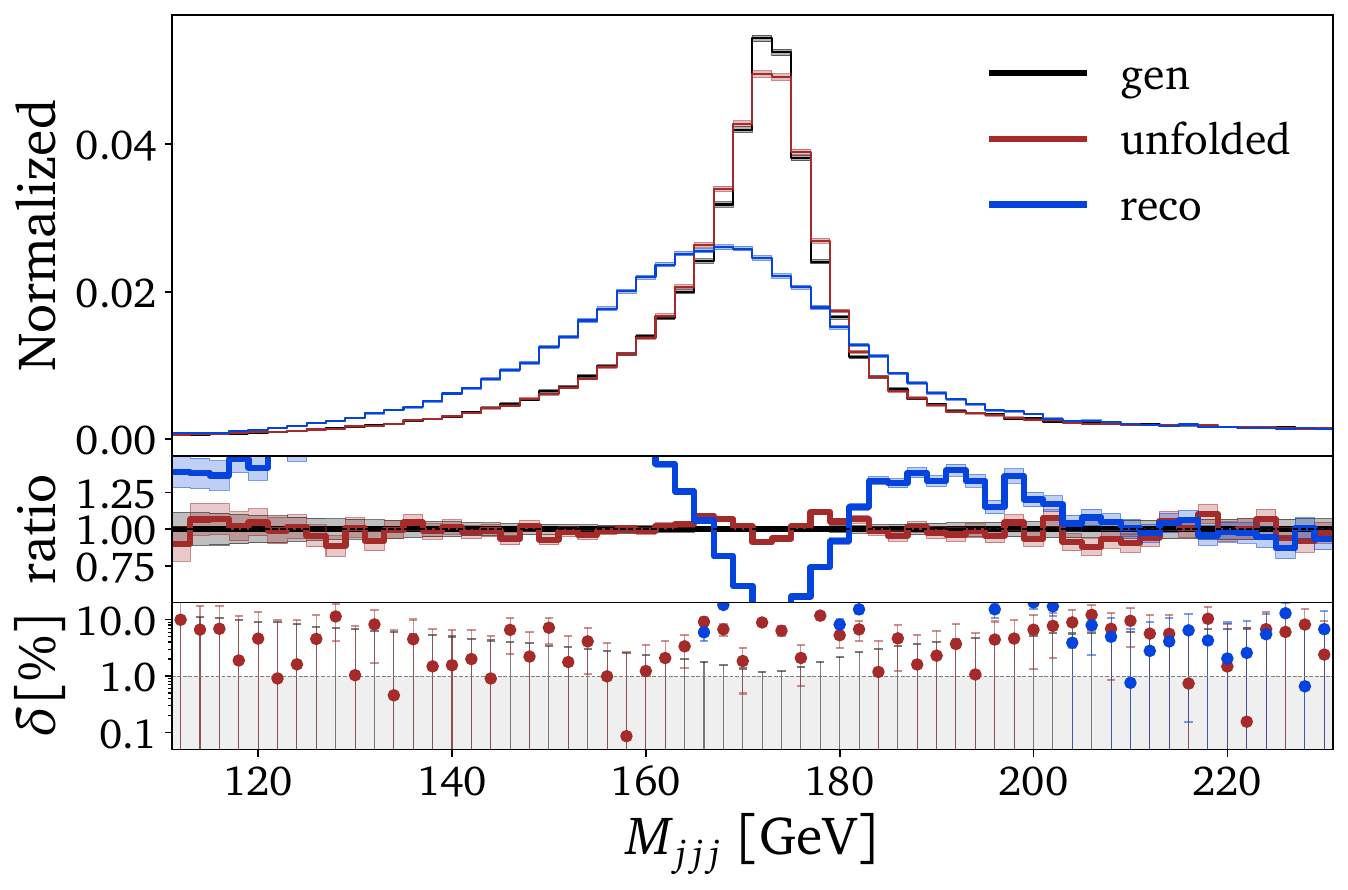}
    \includegraphics[width=0.495\linewidth, page=15]{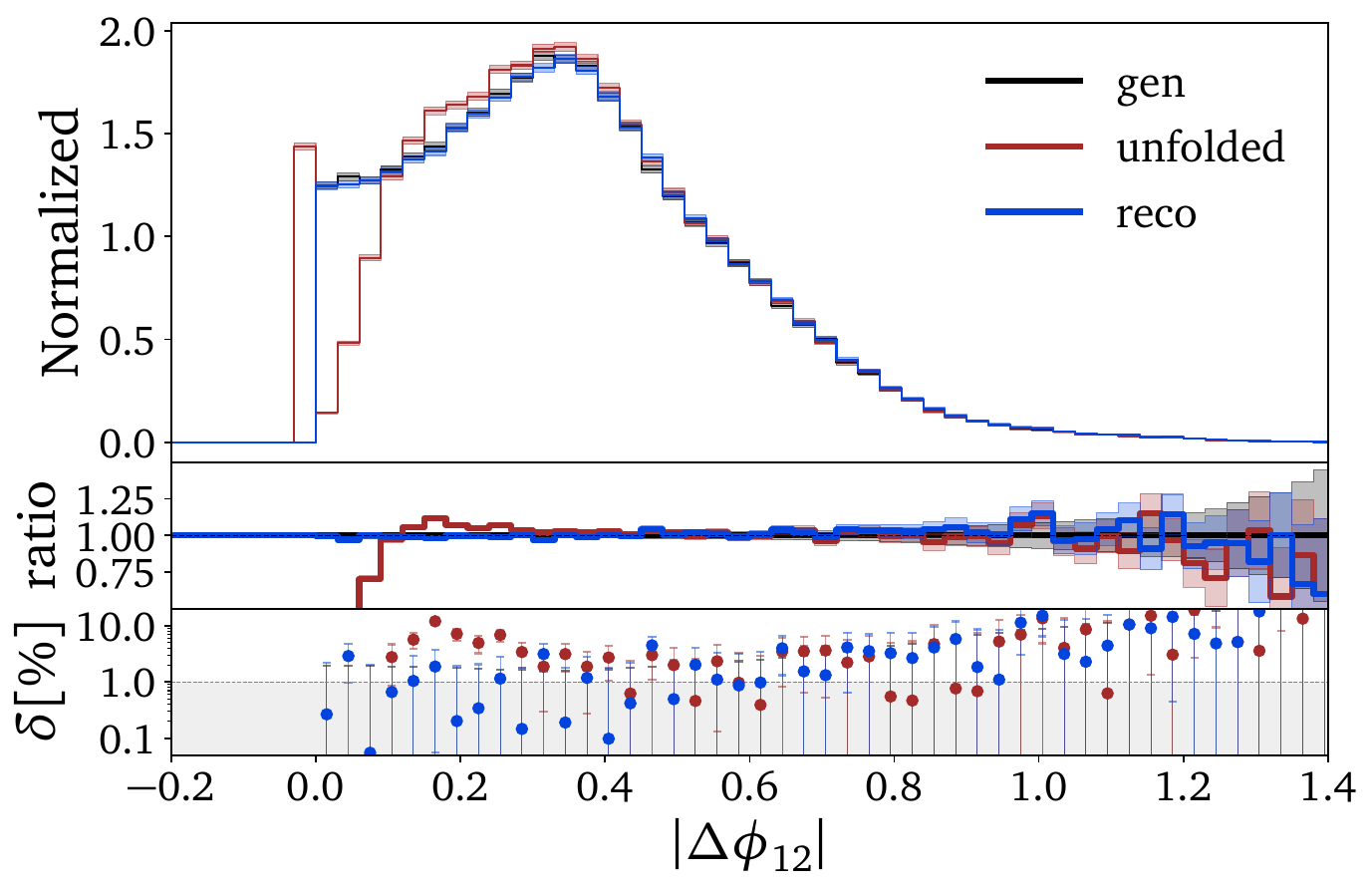}
    \caption{Kinematic distributions from full, 12-dimensional unfolding. We show the 3-jet mass as 
    well as the azimuthal angle between the two leading jets.}
    \label{fig:12d_full_phase_space}
\end{figure}

The remaining phase space dimensions are
\begin{align}
  x = \big( \; \{ m_i \}, \{ M_{ik} \}, \{p_{T,i}\}, \{\eta_i\} \; \big) 
  \qquad i,k=1,2,3 \; ,
\end{align}
all other kinematic observables can be computed from these
basis directions.
For the 12-dimensional unfolding we use a single transfusion network,
after checking that the dual network does not lead to an improvement. 
The hyperparameters are given in Appendix~\ref{app:hyper}. 
\begin{figure}[b!]
    \includegraphics[width=0.495\linewidth, page=15]{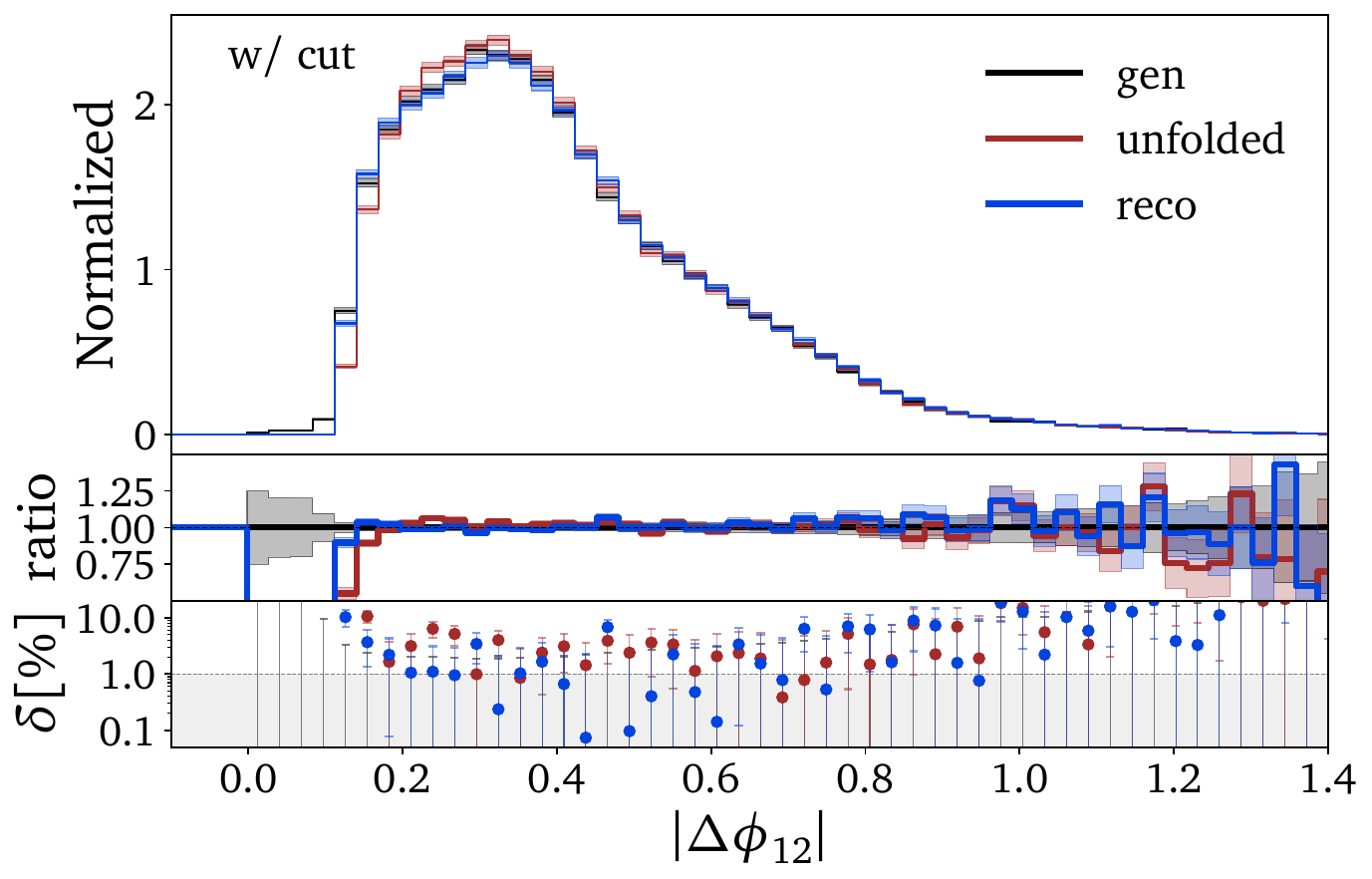} 
    \includegraphics[width=0.495\linewidth, page=13]{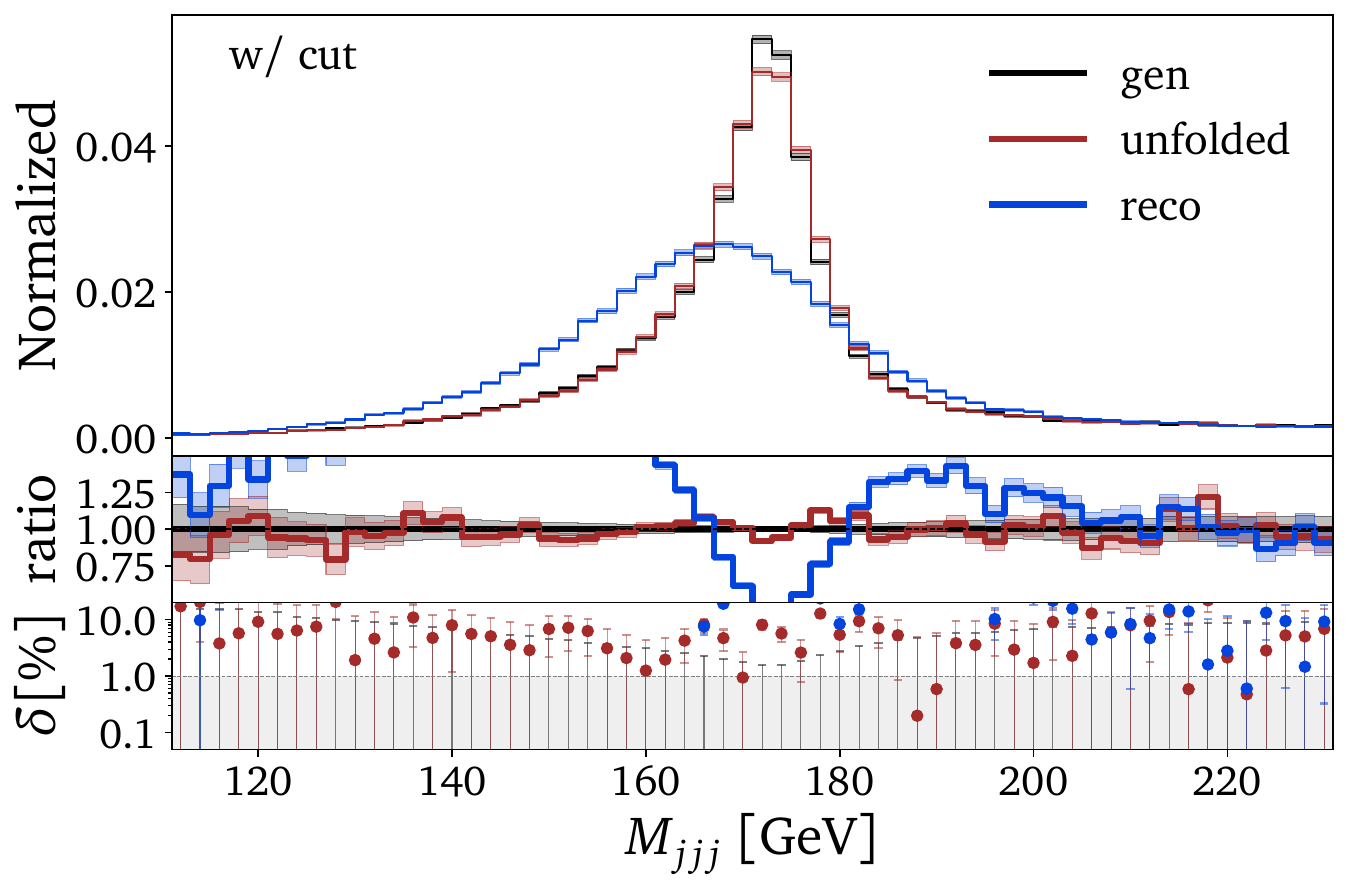} \\
    \includegraphics[width=0.495\linewidth, page=5 ]{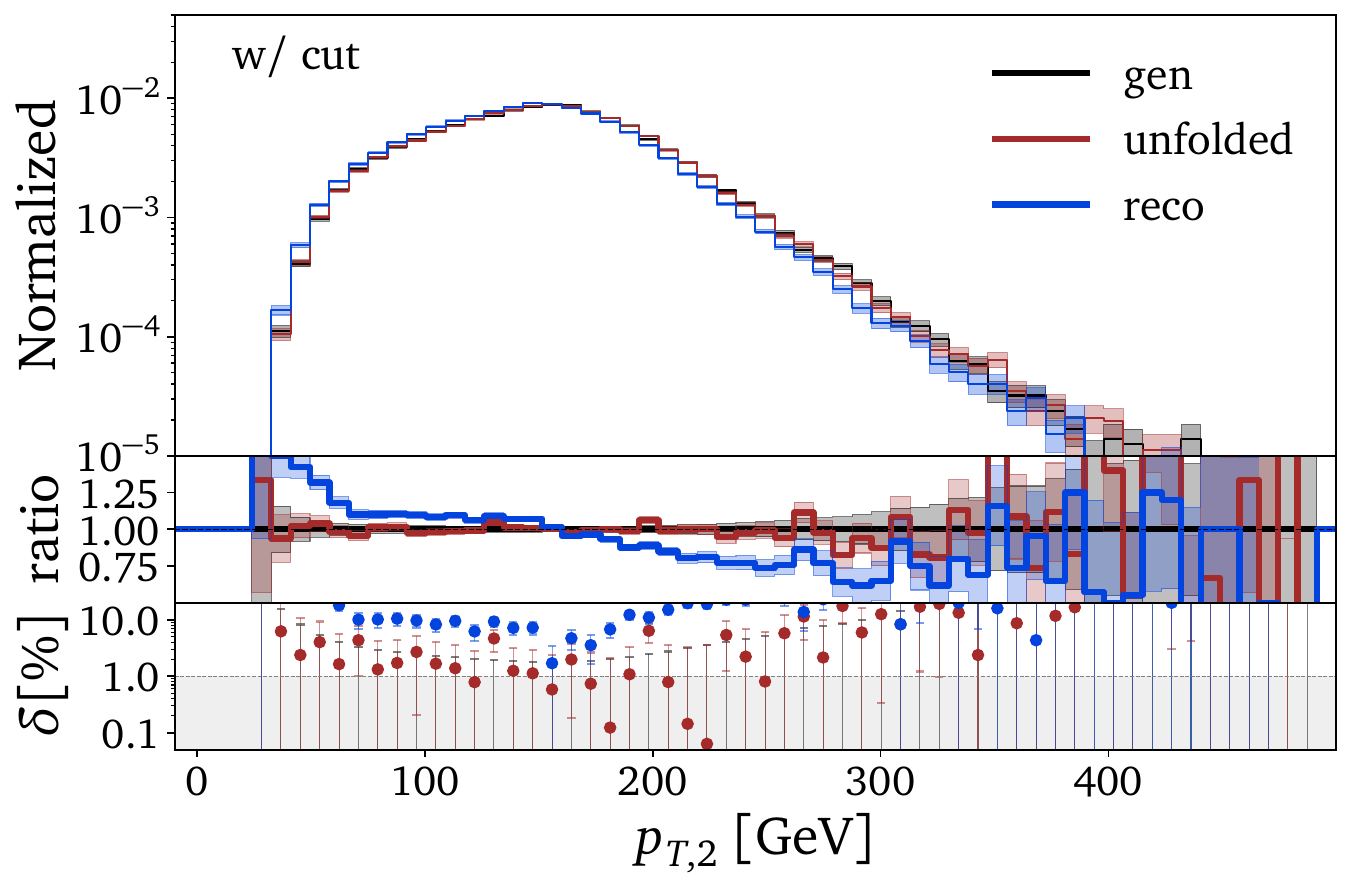} 
    \includegraphics[width=0.495\linewidth, page=38]{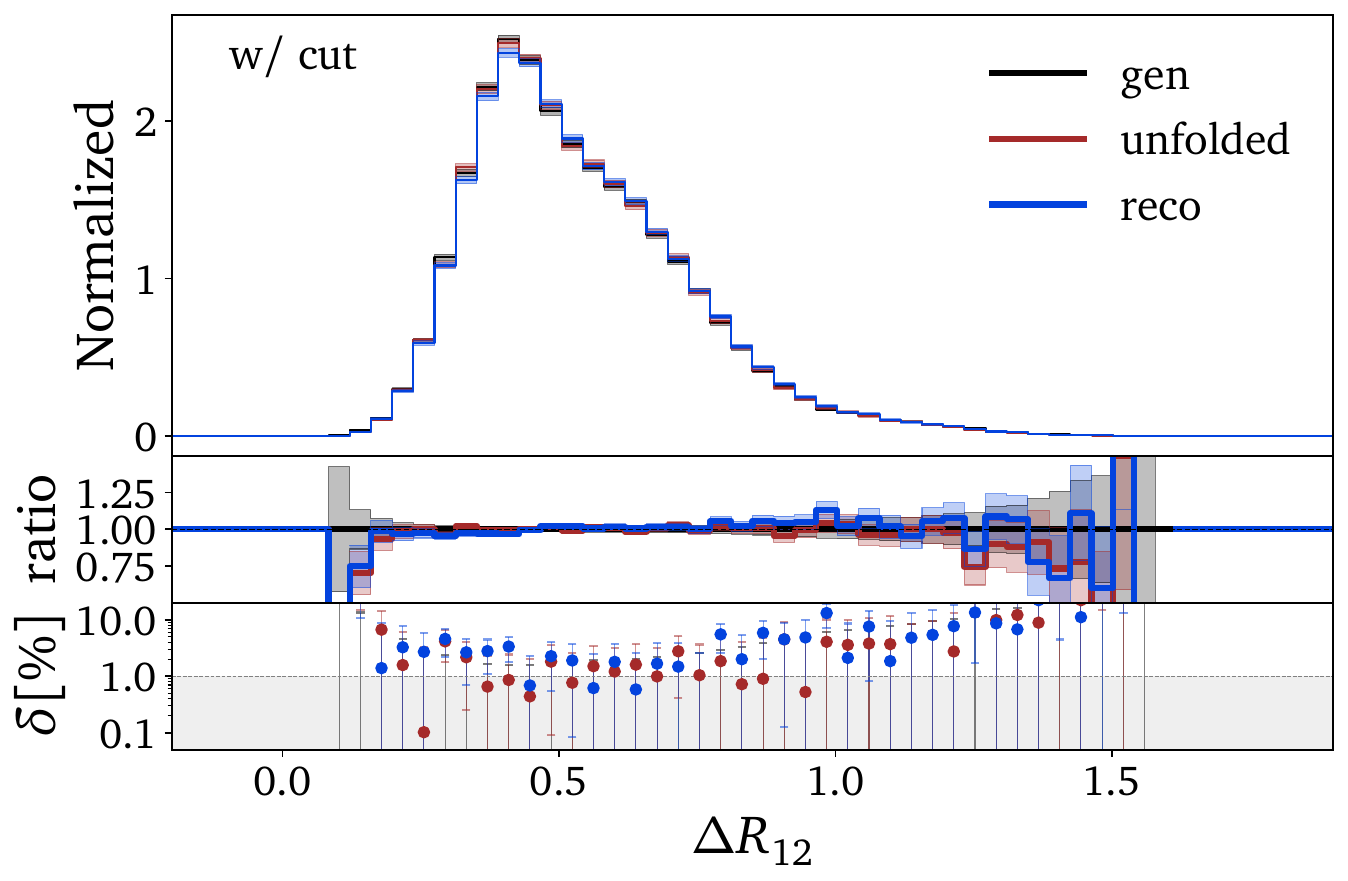} \\
    \includegraphics[width=0.495\linewidth, page=35]{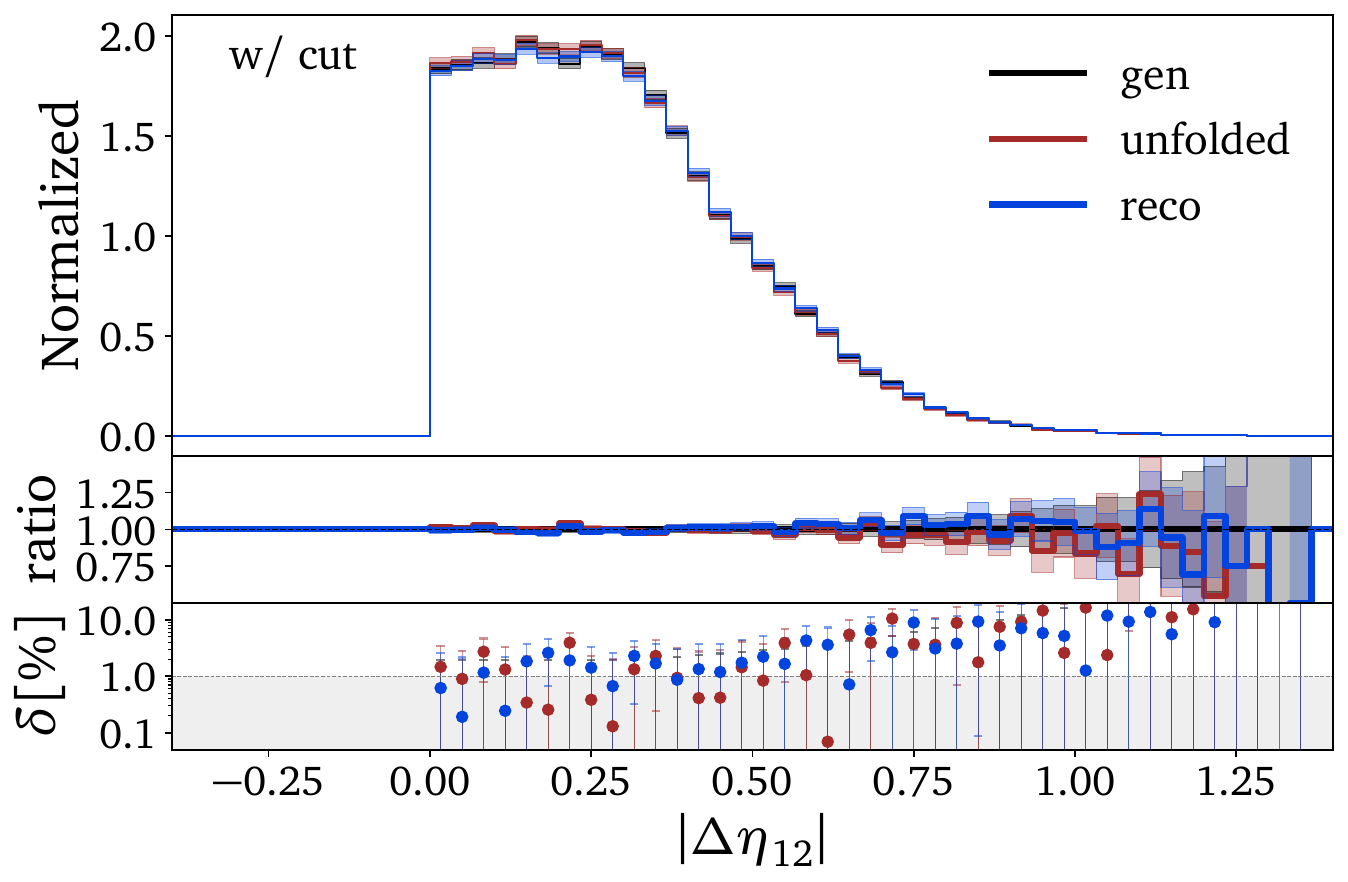} 
    \includegraphics[width=0.495\linewidth, page=2 ]{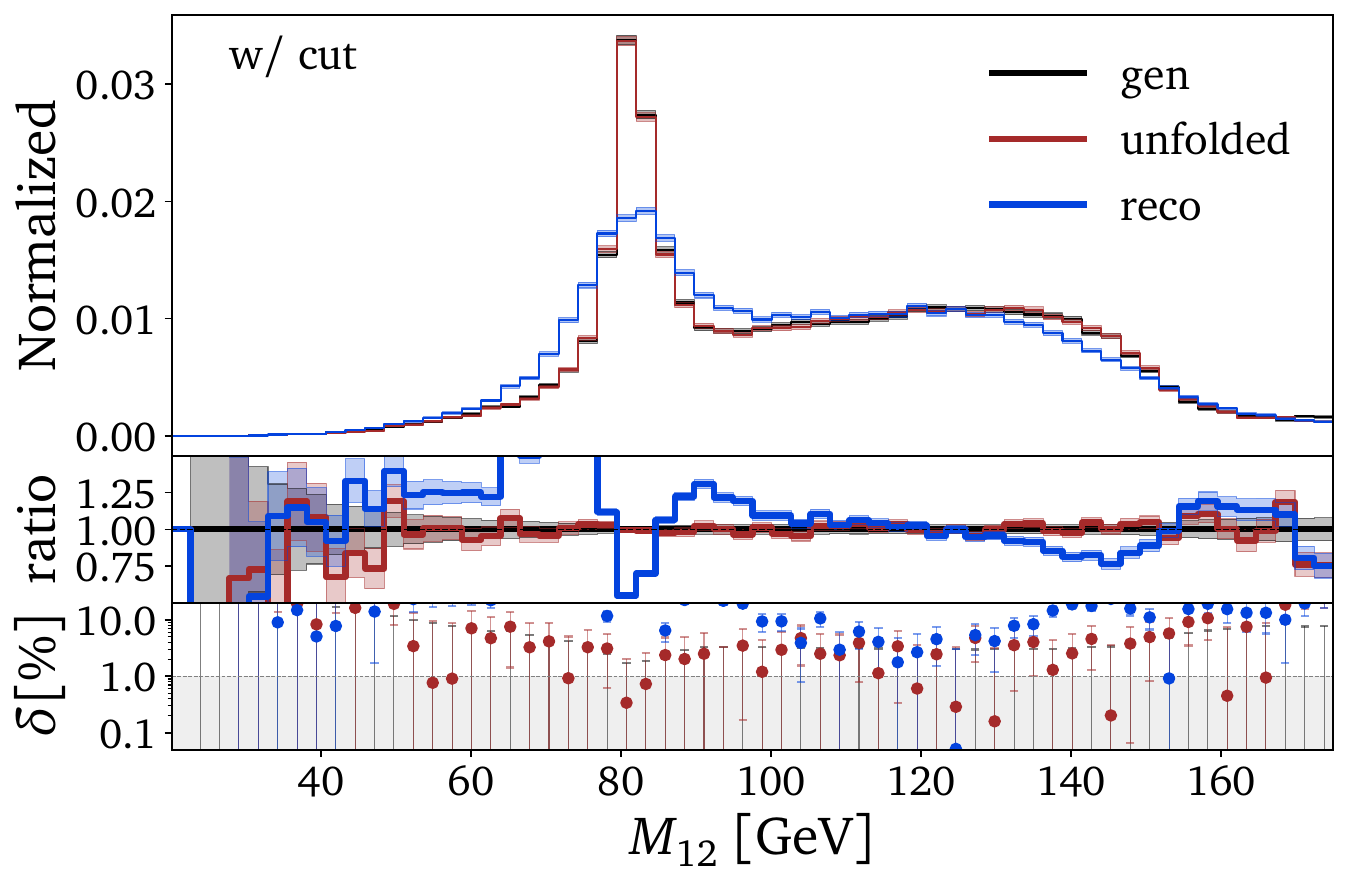} \\
    \includegraphics[width=0.495\linewidth, page=34]{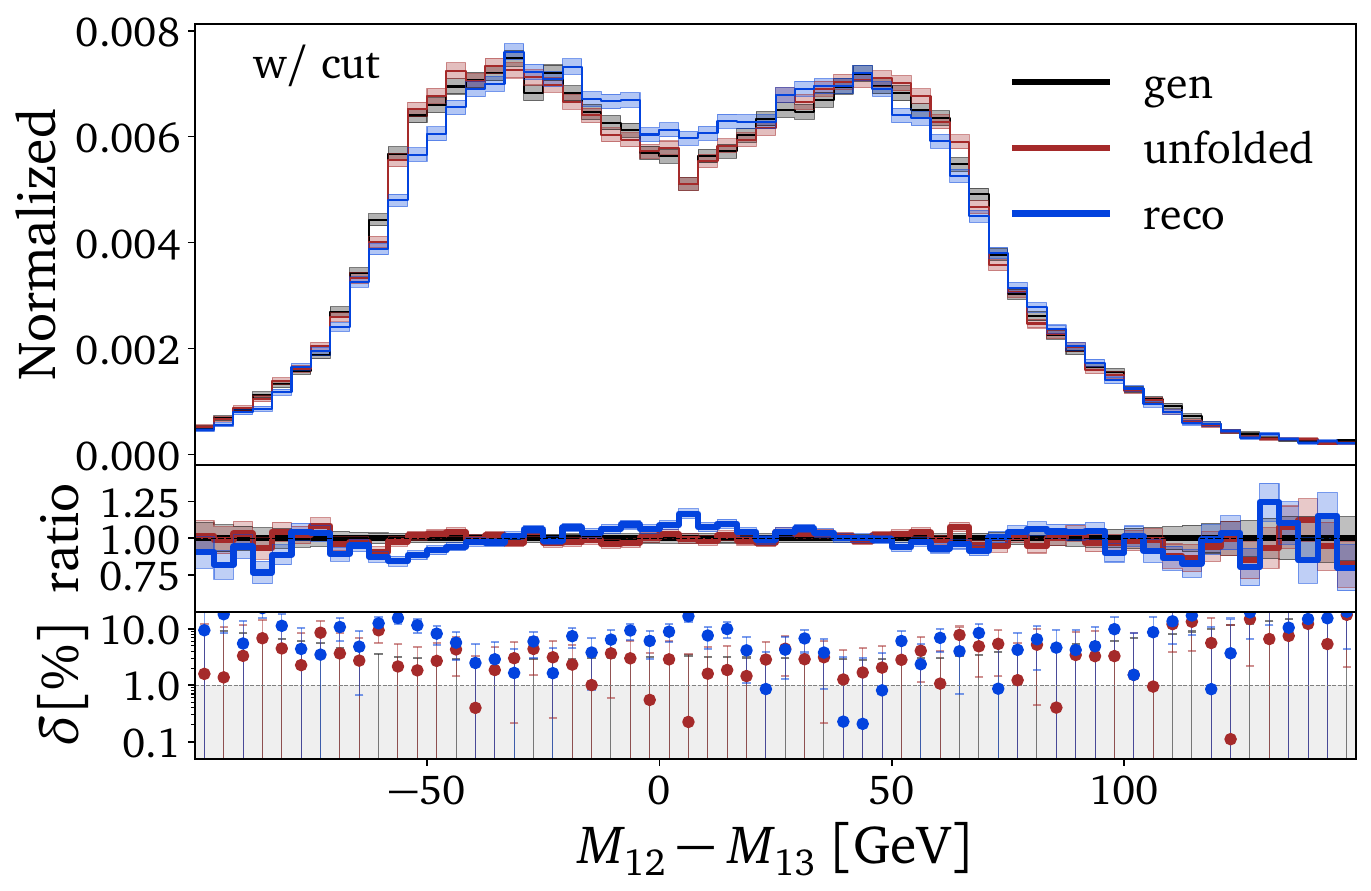} 
    \includegraphics[width=0.495\linewidth, page=22]{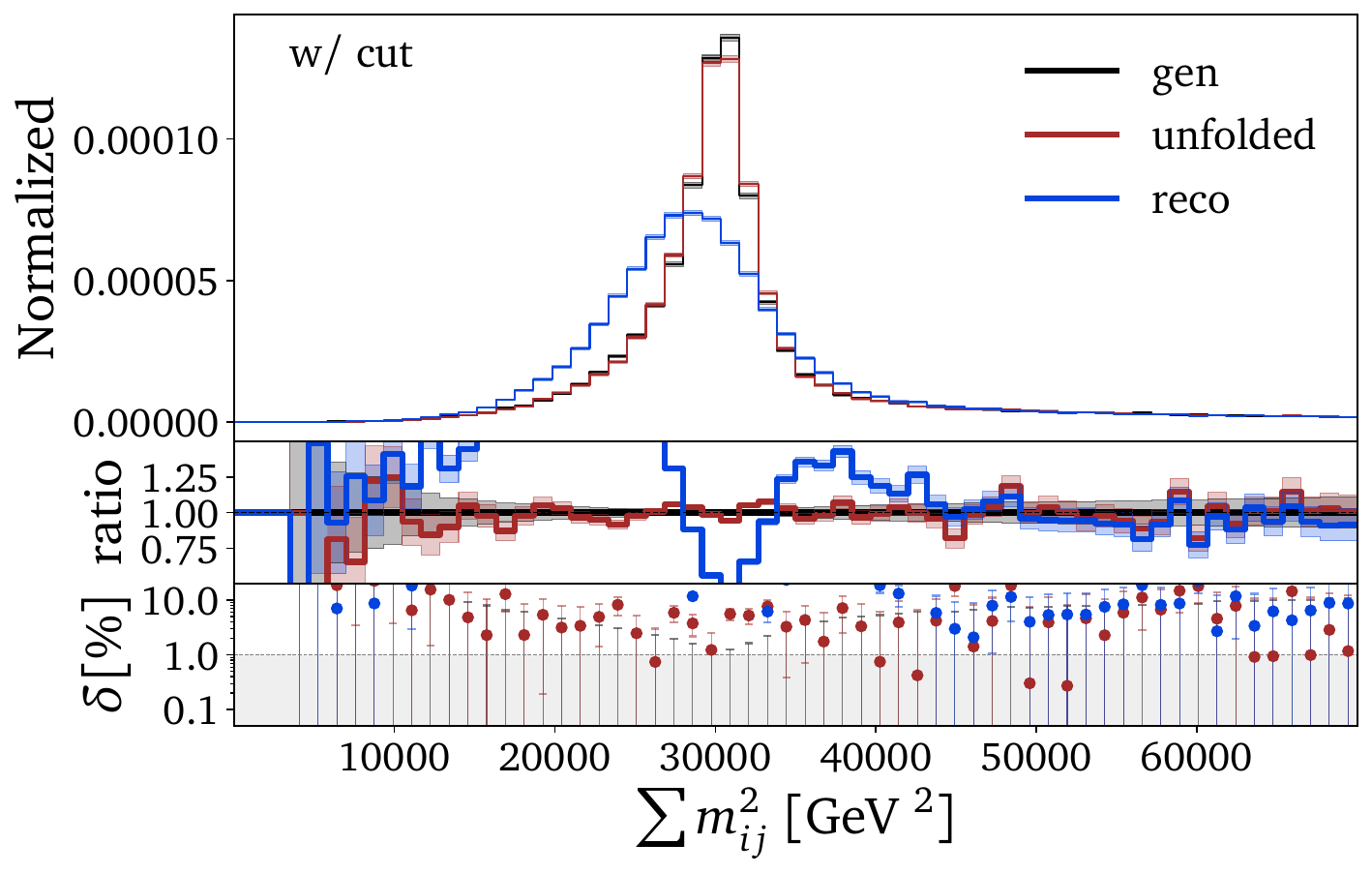} 
    \caption{Kinematic distributions from full, 12-dimensional unfolding. We show the target 3-jet
    distribution, the azimuthal angle between the jets after cut, and a set of single-jet observables, 2-jet correlations, and 3-jet correlations (top to bottom).}
    \label{fig:12d_deltaphi}
\end{figure}
Two kinematic distributions are shown in Fig.~\ref{fig:12d_full_phase_space}.
In the left panel, we see that the top mass peak is learned almost as well
as for the 4-dimensional and 6-dimensional cases. Indeed, this is the case for 
all jet masses and 2-jets masses, which are combined to the 3-jet mass with the 
top peak. 

A serious issue arises from the azimuthal angle between the two leading 
jets, $|\Delta \phi_{12}|$. According to Eq.\eqref{eq:dijetmass} this angle is learned
as a correlation of 7 phase space directions. Moreover, we do not have access 
to the azimuthal angles, only to the cosine of differences between angles.  
Here the problem arises that the network does not ensure that 
this cosine comes out in the physical range $-1\ldots1$. 
We enforce the physical range by clipping the cosine for small angles to one, 
which causes a mis-modelling of the small-$|\Delta \phi_{12}|$ regime, shown in 
the right panel of Fig.~\ref{fig:12d_deltaphi}. 

A simple way to improve this mis-modelling is to require $\cos \Delta \phi_{12} < 1$.
However, from Fig.~\ref{fig:12d_full_phase_space} we know that this does not 
solve the problem. Instead, we accept the fact that for unfolding the masses 
well we might have to pay a prize in the coverage of the angular correlations, 
and we apply an additional acceptance cut
\begin{align}
\Delta \phi_{ik} > 0.1 
\end{align}
both, at the reco- and gen-levels in our simulated events.
This reduces the size of the unfolded dataset by 30\%. 
An extended set of unfolded kinematic distribution after this cut 
are shown in Fig.~\ref{fig:12d_deltaphi}.

\begin{figure}[t]
    \includegraphics[width=0.45\linewidth, page=1]{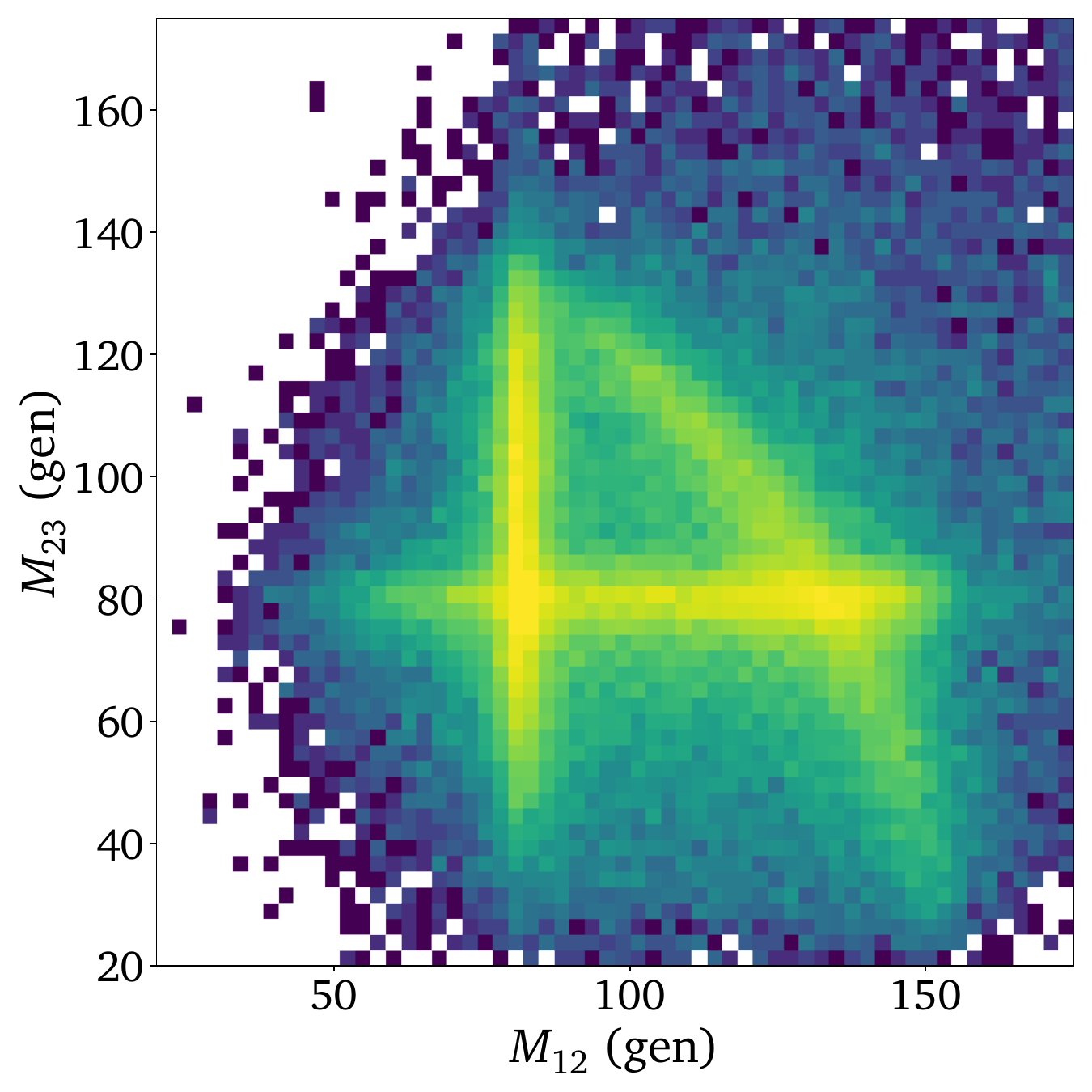}
    \hfill
    \includegraphics[width=0.45\linewidth, page=2]{figs/full_phasespace/12d_dijetmass_correlation_delphes.pdf} 
    \caption{Correlation of two 2-jet masses at gen-level truth (left) and 
    after unfolding (right).}
    \label{fig:12d_correlations}
\end{figure}

We know that our unfolding method covers correlations between
the original phase space directions well, because many of the 
kinematic observables shown in Fig.~\ref{fig:12d_deltaphi} are built from 
complex correlations of our phase space basis. 
However, to end with a nice figure and to drive home the message 
that high-dimensional unfolding using conditional generative networks
does learn the corresponding correlations well, we show one of our 
favorite correlations in Fig.~\ref{fig:12d_correlations}. Indeed, 
there is literally no difference in the correlations between two of the 
three 2-jet masses. This correlation also confirms that the condition 
$M_{ik} \approx m_W$ leads to three distinct lines in phase space, where
close to the crossing points it is impossible to reconstruct which two of the 
jets come from the $W$ decay.

\section{Outlook}
\label{sec:outlook}

Unfolding is one of the ways modern machine learning is transforming the 
way we can do LHC physics. Employing an inverse simulation, it allows for the 
efficient analysis of LHC data by the LHC collaborations, to combine 
analyses between different experiments, and even make unbinned, unfolded 
data accessible to researchers outside the experimental collaborations. 
Unfolding has been used in particle physics frequently, but modern neural networks
allow us to unfold a high-dimensional phase space without a choice of binning. This 
technical advance will turn multi-dimensional and unbinned unfolding 
into a standard analysis method at the LHC and future experiments.

For our study, we unfold detector effects from boosted top quark decay data using
state-of-the-art conditional generative networks. 
Unfolding decay kinematics is especially challenging 
because we expect a large model dependence 
and even systematic bias from the choice of the top mass in the simulated 
training data. Our study shows that generative unfolding with a new methods for prior removal solves this 
problem and provides a first milestone towards incorporating generative unfolding in
an existing CMS analysis. 

First, we showed that for an appropriate phase space parametrization, a 
combination of diffusion network and transformer can reliably unfold 
a 4-dimensional and 6-dimensional subspace of the full top-decay phase space
at the percent level precision. This included the 3-jet mass as a proxy to the
top mass. 
The problem in this unfolding is a strong bias from the 
top mass used to generate the training data. To compensate this bias 
we added a global estimate of the top mass to the representation of the
measured data and weakened the training bias by including a range of top 
masses there. As a result of these two structural modifications, the 
top mass bias was essentially removed. 

Using this setup we showed how to extract the top mass along the lines 
of a recent CMS analysis~\cite{CMS:2022kqg}. We included two 
covariance matrices, one describing all statistical uncertainties and one
covering the model uncertainty from the training data.
We found that, indeed, the impact of the model uncertainty is becoming 
irrelevant, and that the error in the top mass can be reduced when using 
the kind of fine binning allowed by the unbinned unfolding method.

Finally, we unfolded the full, 12-dimensional phase space for a given top 
mass. One failure mode in reproducing the angular distributions was induced
by our phase space parametrization. However, a simple lower cutoff on the 
azimuthal angular separations of the top decay jets allowed for 
an excellent reproduction of all correlations. 

This study serves as a blueprint for an actual CMS analysis, 
both, for a top mass measurement and for a wider use of the unfolded data. Results for full CMS simulations cannot be shown in this publications, but are available from the CMS members on the author team. Their performance is slightly better than for the fast simulation shown here. 

\section*{Acknowledgements}

Most importantly, we would like to thank the organizers and experts at the 
2024 Terascale Statistics School for pointing out that nobody in their 
right mind would ever attempt to use unfolding for a mass measurement. We 
completely agree with that highly motivating point of view. 

Moreover, we like to thank Henning Bahl, Anja Butter, Theo Heimel, Nathan Huetsch and Nikita Schmal for many valuable discussions, and Andrea Giammanco and Anna Benecke for useful discussions on the Delphes detector simulation.
This research is supported 
through the KISS consortium (05D2022) funded by the German Federal Ministry of Education and Research BMBF 
in the ErUM-Data action plan,
by the Deutsche
Forschungsgemeinschaft (DFG, German Research Foundation) under grant
396021762 -- TRR~257: \textsl{Particle Physics Phenomenology after the
  Higgs Discovery}, and through Germany's Excellence Strategy
EXC~2181/1 -- 390900948 (the \textsl{Heidelberg STRUCTURES Excellence
  Cluster}).
We would also 
like to thank the Baden-W\"urttem\-berg Stiftung for financing through
the program \textsl{Internationale Spitzenforschung}, pro\-ject
\textsl{Uncertainties – Teaching AI its Limits} (BWST\_ISF2020-010). 
LF is supported by the Fonds de la Recherche Scientifique - FNRS under Grant No. 4.4503.16.
SPS
is supported by the BMBF Junior Group Generative Precision Networks for Particle Physics
(DLR 01IS22079). The research work of DS has been funded by the Austrian Science Fund~(FWF, grant P33771). The authors acknowledge support by the state of
Baden-W\"urttemberg through bwHPC and the German Research Foundation
(DFG) through grant no INST 39/963-1 FUGG (bwForCluster NEMO).

\appendix
\section{Bias removal methods}
\label{app:bias}
\begin{figure}[t!]
    \includegraphics[width=0.495\linewidth, page=5]{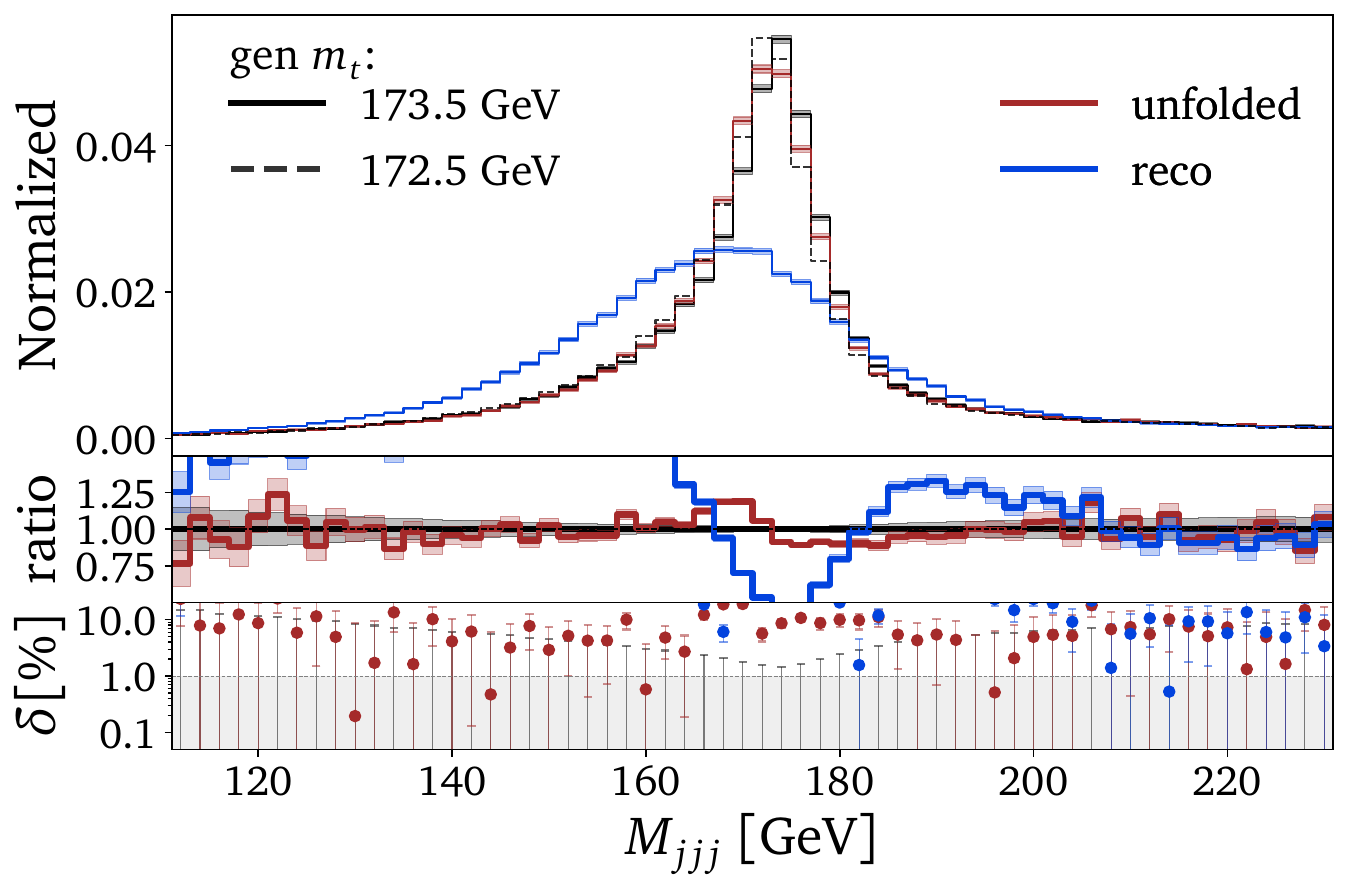}
    \includegraphics[width=0.495\linewidth, page=5]{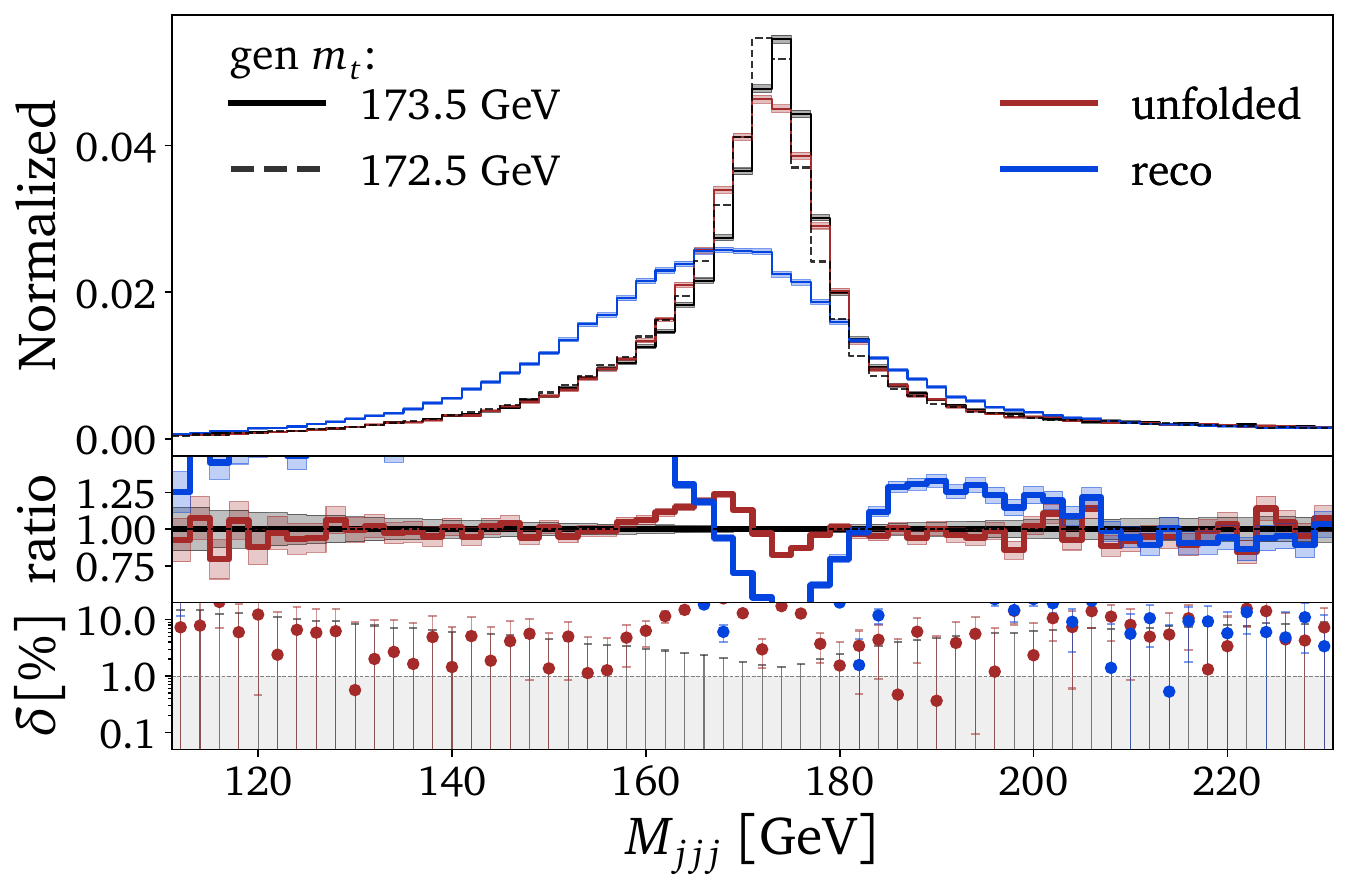}
    \caption{Kinematical distributions from 4-dimensional unfolding. We compare $M_{jjj}$ for $m_t = 172.5 \,\unit{GeV}$ to generated unfolding for $m_t = 173.5 \,\unit{GeV}$, not seen during training. On the left panel we show results where the batch-wise condition of Eq.\eqref{eq:batch_wise_condition} is included into the training pipeline but no augmentation. On the right panel we show results where the training data was augmented with samples of different top masses, but no batch-wise conditioning was included. }
    \label{fig:4d_cross_checks}
\end{figure}
As stated in Sec.~\ref{sec:gen_bias}, we rely on both, batch-wise conditioning and data augmentation, to unfold the triple jet mass without bias. In Fig.~\ref{fig:4d_cross_checks} this is demonstrated by showing unfolding results, where we train either without augmentation or without batch-wise conditioning. For both setups, we observe a clear drop in performance when compared to Fig.~\ref{fig:unbiased_results_4d6d}, although all results are produced with the same hyperparameters of Tab.~\ref{tab:unbiased_hyperparameters}. 
\begin{figure}[b]
    \centering
    \includegraphics[width=0.45\linewidth, page=1]{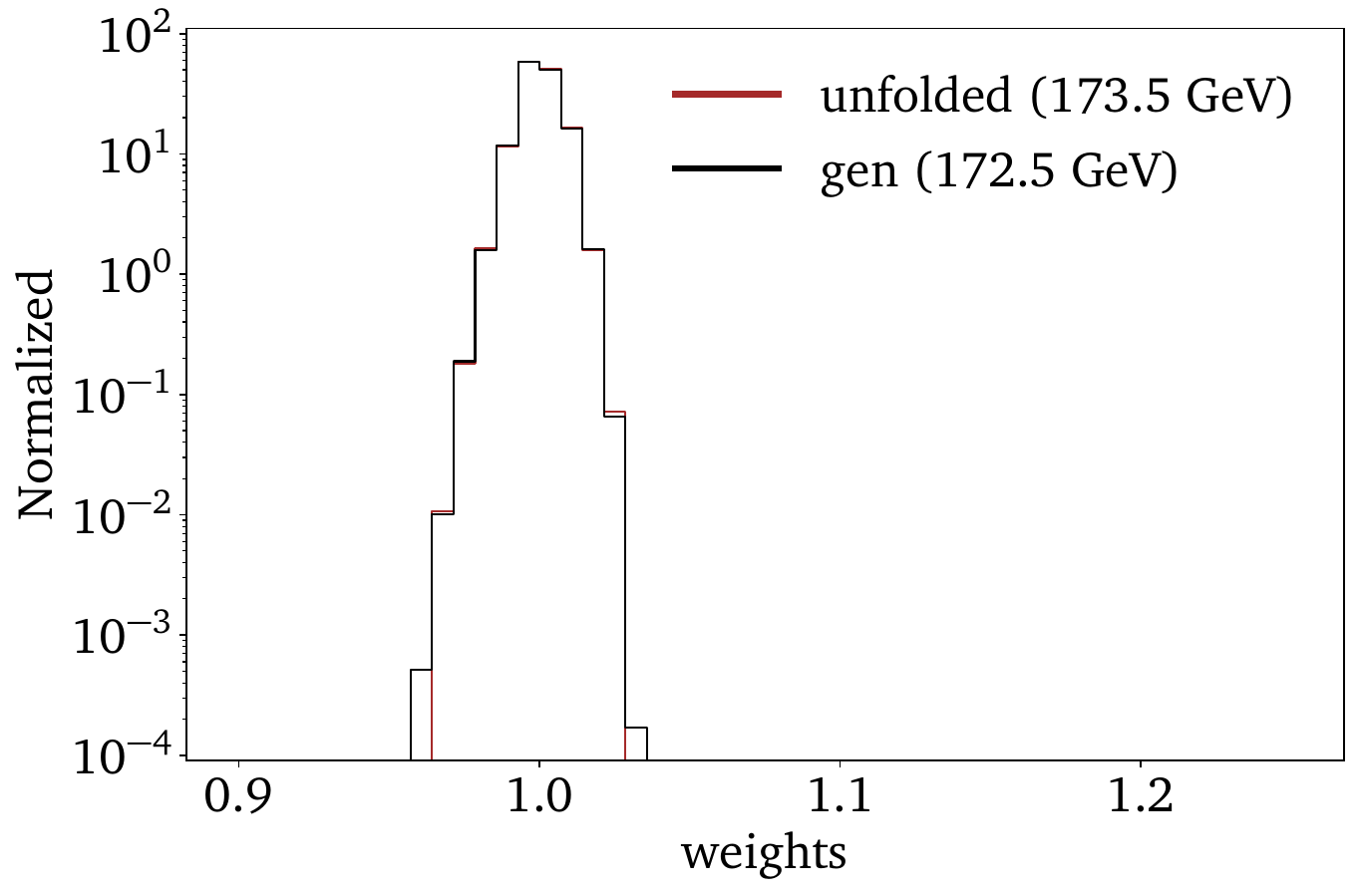}
    \caption{Classifier weights to reweight MC gen-level simulation to unfolded pseudo-data in the 4-dimensional parametrization, as part of the second step in iterative generative unfolding. The top masses are given in parentheses.}
    \label{fig:4d_iterative_weights}
\end{figure}
Iterative generative unfolding can ensure prior independence~\cite{Backes:2022vmn}, 
but does not succeed for the triple jet mass in our application.
For iterative generative unfolding~\cite{Backes:2022vmn} the first step consists of a generative network to learn the posterior distribution of Eq.\eqref{eq:posterior}. For the 4-dimensional unfolding scenario, we look at the unfolding results of Fig.~\ref{fig:4d_biased} as our first step. We train the generative network on MC simulations with $m_t = 172.5 \, \unit{GeV}$ and try to unfold reco-level pseudo-data with a corresponding top-quark mass of $m_t=173.5 \, \unit{GeV}$. 
In a second step we learn a reweighting between the unfolded pseudo-data and the MC simulation used during training. We see in the lower right panel of Fig.~\ref{fig:4d_biased} that the unfolded results collapse back to the distribution of the prior MC simulation. The learned reweighting will barely correct the MC simulation, which is confirmed when looking at the learned classifier weights in Fig.~\ref{fig:4d_iterative_weights}. 
They are sharply centered around unity, so we do not gain from the iterations as the MC simulation from the fist iteration matches the MC simulation from the second iteration. 

\begin{figure}[t]
    \includegraphics[width=0.495\linewidth, page=1]{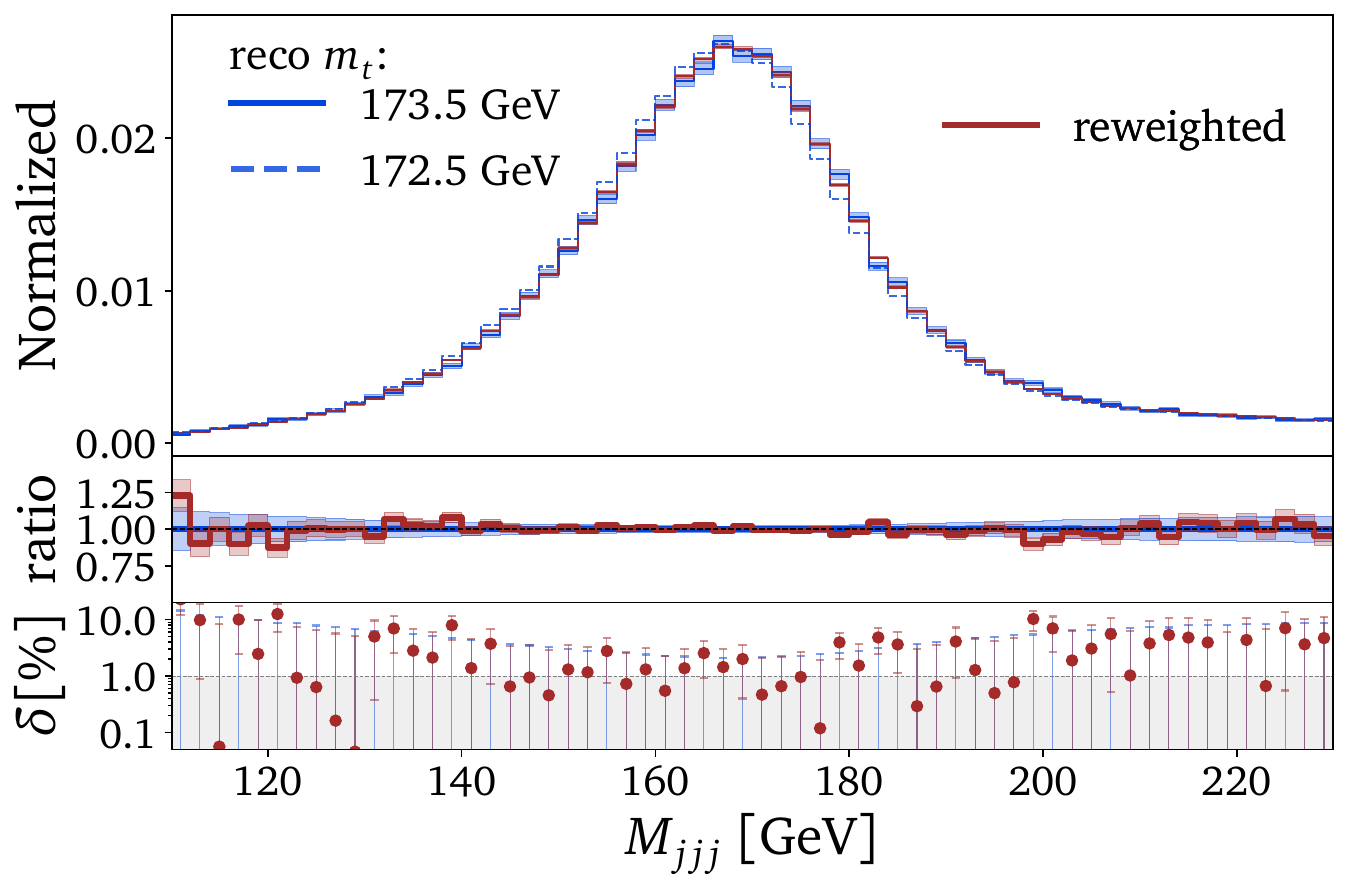}
    \includegraphics[width=0.495\linewidth, page=1]{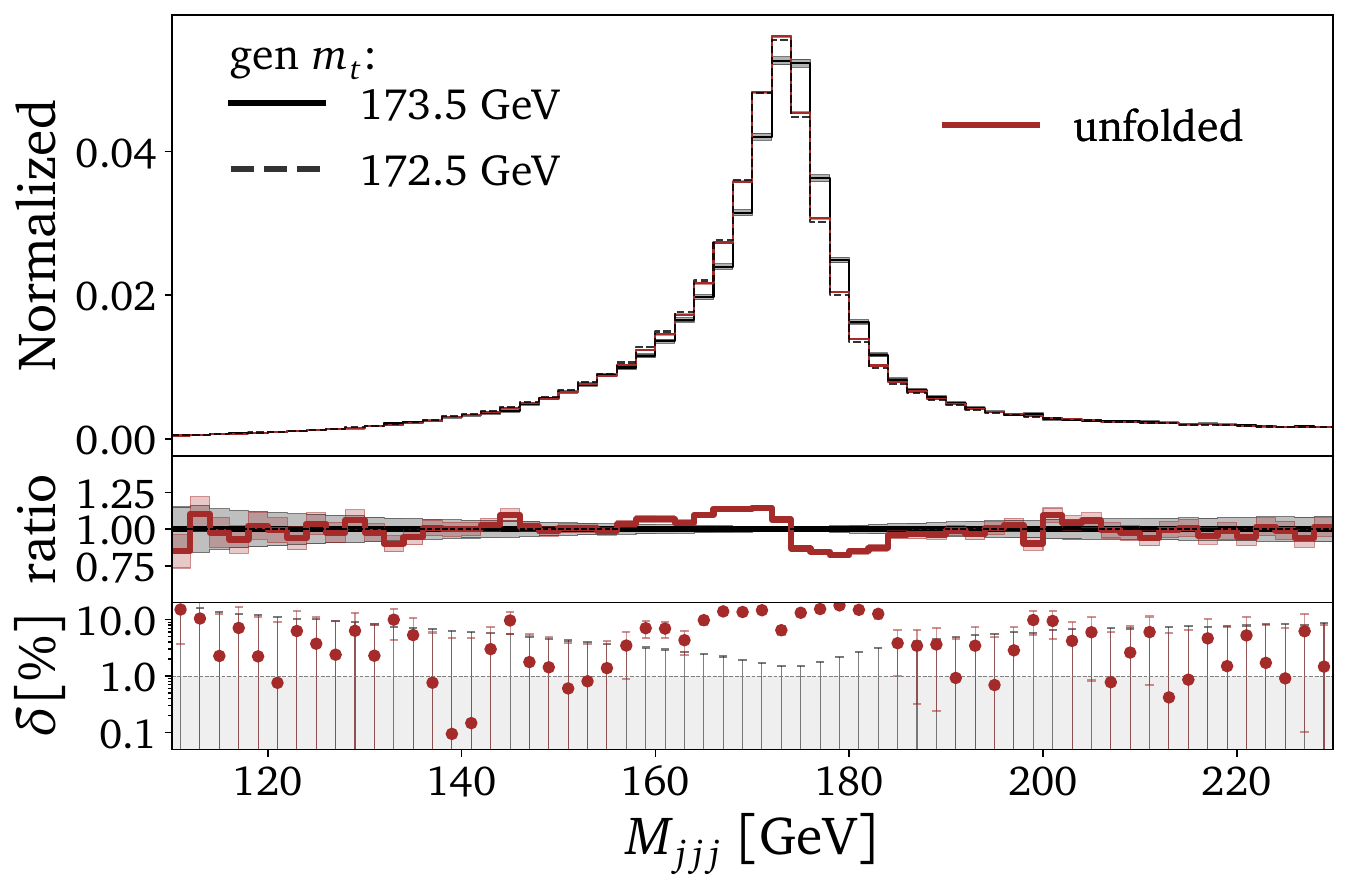}
    \caption{Kinematical distributions from (4+1)-dimensional unfolding. On the left panel, we compare the reco-level $M_{jjj}$ distribution for $m_t = 172.5 \,\unit{GeV}$ to pseudo-data for $m_t = 173.5 \,\unit{GeV}$ and the reweighted version computed with the first step of the omnifold algorithm. On the right panel, we make the same comparison on gen-level where unfolded is now the reweighted MC simulation on gen-level.}
    \label{fig:4d_omnifold}
\end{figure}

OmniFold~\cite{Andreassen:2019cjw} learns a classifier-based reweighting between the pseudo-data and the MC simulation on reco-level. In a second step, the OmniFold algorithms pulls the learned reco-level weights to gen-level, event by event, and learns a second classifier-reweighting between the reweighted gen-level distribution and the initial MC gen-level distribution. The procedure can be repeated iteratively.  However, for shifted resonances such as the triple jet mass the correction is not learned correctly. This can be confirmed when looking at Fig.~\ref{fig:4d_omnifold}. Here, we train OmniFold on the 4-dimensional parametrization plus the triple jet mass. The first step correctly reweights the reco-level kinematical distribution of the triple jet mass of the MC simulation ($m_t=172.5 \, \unit{GeV}$) to our pseudo-data ($m_t=173.5 \, \unit{GeV}$). When we pull the learned weights to gen-level, they are not sufficient to reweight the peaked distributions of the gen-level triple jet mass. The reweighted distribution collapses back to the prior MC distribution, indicating again that we cannot remove the prior using iterations.
These findings motivate the use of our novel unfolding strategy resulting in Fig.~\ref{fig:unbiased_results_4d6d}.

Although the standard OmniFold approach fails in our unfolding tasks, it does not mean that similar adaptions to the algorithm could not lead to unbiased results. However, we leave a concrete investigation of the matter to the OmniFold authors.
\section{Hyperparameters}
\label{app:hyper}

\begin{table}[h!]
    \centering
    \begin{small} \begin{tabular}{lc} \toprule
        Parameter     &\\ \midrule
        LR sched. & cosine\\
        Max LR    & $10^{-3}$   \\
        Optimizer & Adam \\
        Batch size & 16384\\
        \midrule
        Network & Resnet \\
        Dim embedding  &  64  \\
        Intermediate dim   & 512 \\
        Num layers   & 8  \\
        \bottomrule
    \end{tabular} \end{small}
    \caption{Parameters for the 4-dimensional and 6-dimensional networks in Sec.~\ref{sec:gen_naive}.}
    \label{tab:biased_hyperparameters}
\end{table}

\begin{table}[h!]
    \centering
    \begin{small} \begin{tabular}{lcc} \toprule
        Parameter     & 4D  &  6D \\ \midrule
        Epochs    & 800    & $500 (+ 1000)$     \\
        LR sched. & cosine & cosine\\
        Max LR    & $10^{-3}$   & $10^{-3}$\\
        Optimizer & Adam & Adam\\
        Train batch size & 10000 & 10000\\
        Inference batch size & 50000 & 50000\\
        Dropout & 0.1 & 0.1 \\
        \midrule
        Network & Transfusion & Transfusion\\
        Dim embedding  &  64 & 64 \\
        Intermediate dim   & 512 & 512\\
        Num layers   & 4 & 4 \\
        Num heads & 4 & 4 \\
        \bottomrule
    \end{tabular} \end{small}
    \caption{Parameters for the 4-dimensional and 6-dimensional networks in Sec.~\ref{sec:gen_bias}.}
    \label{tab:unbiased_hyperparameters}
\end{table}

\begin{table}[h!]
    \centering
    \begin{small} \begin{tabular}{lc} \toprule
        Parameter     & 12D  \\ \midrule
        Epochs    & 500       \\
        LR sched. & cosine \\
        Max LR    & $10^{-3}$ \\
        Optimizer & Adam \\
        Batch size & 16384\\
        Dropout & 0.1\\
        \midrule
        Network & Transfusion \\
        Dim embedding  &  128\\
        Intermediate dim   & 512 \\
        Num layers   & 6 \\
        Num heads & 4  \\
        \bottomrule
    \end{tabular} \end{small}
    \caption{Parameters for the 12-dimensional network in Sec.~\ref{sec:gen_full}.}
    \label{tab:full_hyperparameters}
\end{table}
\clearpage
\bibliography{tilman,refs, literature}
\end{document}